\documentclass[lineno]{jfm}

\makeatletter
\gdef\@underjournal{}
\makeatother

\usepackage{graphicx}
\usepackage{newtxtext}
\usepackage{newtxmath}
\usepackage{natbib}
\usepackage{hyperref}
\usepackage{comment}

\hypersetup{
    colorlinks = true,
    urlcolor   = blue,
    citecolor  = black,
}

\newcommand{\RomanNumeralCaps}[1]
\linenumbers

\usepackage{multirow}
\usepackage{ragged2e}
\usepackage{float}
\usepackage{subcaption}
\usepackage{booktabs}

\title{A Spherical Multipole Expansion of Acoustic Analogy for Propeller Noise}

\author{Felice Fruncillo\aff{1,2}
  \corresp{\email{ffruncillo@unisa.it}},
  Paolo Luchini\aff{1}
 \and Flavio Giannetti\aff{1}}

\affiliation{\aff{1}University of Salerno, Via Giovanni Paolo II, 132, Fisciano, 84084
\aff{2}CIRA, Italian Aerospace Research Center, Via Maiorise, Capua, 81043}

\usepackage[dvipsnames,svgnames,x11names]{xcolor}

\usepackage[final,commandnameprefix=ifneeded]{changes}

\definechangesauthor[color=BrickRed]{FF}

\setauthormarkup{}
\setauthormarkuptext{}

\definecolor{myaddcolor}{RGB}{0,0,255} 

\setaddedmarkup{\textcolor{myaddcolor}{#1}}

\newcommand{\stkout}[1]{\ifmmode\text{\sout{\ensuremath{#1}}}\else\sout{#1}\fi}
\setdeletedmarkup{\stkout{#1}}

\makeatletter
\newif\iffinal
\@ifpackagewith{changes}{final}{\finaltrue}{\finalfalse}

\newcommand{\inlinecomment}[1]{%
    \iffinal
    \else
    \textcolor{Green}{[#1]}%
    \fi
}
\makeatother

\begin{document}
\maketitle

\begin{abstract}
This work develops a spherical-multipole expansion of \added{the surface terms of an acoustic-analogy formulation}\deleted[id=FF]{Goldstein’s acoustic analogy}, for the prediction of tonal noise from rotating propellers. The acoustic field is expressed through spherical multipoles, which separate source integrals from the observer dependence. This decoupling leads to computational efficiency: once the multipole coefficients are computed from blade geometry and aerodynamics, the sound field at any observer location is obtained by a simple evaluation of spherical harmonics and radial propagation factors, avoiding repeated integrations for each observer point. Moreover, this enables a straightforward radiated power calculation, without resorting to far-field pressure integrals. For hovering subsonic propellers, the results show a rapid convergence of the expansion. For each harmonic, the dominant radiation is accurately captured by the first two non-zero multipoles, corresponding to the leading symmetric and antisymmetric contributions with respect to the plane of rotation. To interpret the physical content of these leading terms, two simplified descriptions of the source integral are developed. The first is a lifting-surface formulation, suited to blades at small incidence, in which the thin-airfoil approximation allows to separate \deleted[id=FF]{lift}\added{thrust}-like loading, \deleted[id=FF]{drag}\added{torque}-like loading, and thickness contributions. The second is a lifting-line formulation, suited to high-aspect-ratio blades, in which the surface integral is reduced to spanwise integrals of compact sectional moments. The validity of the two formulations is assessed through comparisons of directivity, power distribution over harmonics and time-domain waveforms. \deleted[id=FF]{The results show good accuracy in their respective regimes of validity, together with substantial computational savings.} Finally, both models are validated against \deleted[id=FF]{two} experimental datasets taken from the literature, showing excellent agreement for tonal directivity and blade-passing harmonics.
\end{abstract}

\section{Introduction}
\label{sec:Introduction}

The prediction of tonal noise from rotating blades is a long-standing problem in aeroacoustics. Early successful work by \cite{Gutin1936} represented the propeller as a distribution of concentrated forces acting on the rotor disk, yielding analytical expressions for the noise harmonics. In the following year, \cite{deming1937} identified the contribution of thickness, thereby correcting Gutin's results for blades operating at low incidence. A few years later, \cite{Garrick1954} extended these findings to account for the presence of a uniform mean flow. 

As in many acoustic problems, progress was strongly influenced by the development of acoustic analogies. The classic theories initiated by \cite{Lighthill1952} and \cite{Curle1955} recast the fluid-dynamic equations into an inhomogeneous wave equation and represent the effect of solid boundaries through equivalent sources. From the outset, these approaches provided a prolific framework for propeller noise predictions \citep{FfowcsWilliams1969.2,morfey1972,Hawkings1974}. The Ffowcs Williams-Hawkings (\deleted[id=FF]{F}FW-H) equation \citep{FfowcsWilliams1969} generalized this framework to arbitrarily moving \deleted[id=FF]{and permeable} surfaces. \deleted[id=FF]{Together with the time-domain integral formulations developed by Farassat, FFW-H became a standard reference for rotor-noise modelling.} \added{This equation became a standard reference for rotor-noise modelling, particularly through its time-domain integral formulations derived by Farassat \citep{Farassat1977,Farassat1981}.} Time-domain methods are powerful for high-fidelity prediction \added{in complex unsteady scenarios}, but they can be computationally demanding when the full-sphere directivity is desired, since they require numerical \added{interpolations} \deleted[id=FF]{differentiation} and calculation of retarded blade locations \citep{hanson1980b}. \deleted[id=FF]{Moreover, frequency-domain provides direct insight into how specific propeller design features influence the radiated noise characteristics.} \added{By contrast, frequency-domain formulations are naturally suited to periodic tonal problems and are less straightforward to apply to transient or strongly unsteady flows. Their analytical expressions can, however, relate individual harmonics directly to geometric and operating parameters, thereby facilitating the interpretation of how specific design features influence the radiated noise \citep{hubbard1991}.} For these reasons, in parallel with \deleted[id=FF]{these} developments \added{in the time domain}, in the 1970s Hanson pursued frequency-domain approaches aimed at \deleted[id=FF]{efficient evaluation and} physical interpretability. His helicoidal-surface theory \citep{Hanson1980}, based on Goldstein's analogy \citep{Goldstein1976}, offered analytical expressions for propeller far-field tones. Relative to earlier models, this theory represented a significant generalization: it accounts for finite chordwise extent, forward flight, and a more realistic geometric description by replacing the disk idealization with a helicoidal surface. The final expressions follow from a far-field expansion of the Green's function, in a spirit similar to that employed by \cite{Gutin1936}. Hanson later extended these results to the near field \citep{Hanson1985}, building on a\added{n} \deleted[id=FF]{cylindrical-coordinate} expansion of the Green's function \citep{Hanson1983}. He noted that, relative to the far-field case, the near-field formulation introduces an additional integration that can be interpreted as a Fourier transform in the wavenumber \added{space}. More complete reviews of these early developments can be found in \cite{morfey1973} and \cite{Farassat1980}. \added{With these tools at their disposal, other investigations addressed a range of complementary aspects, including the validity of linearized models in reproducing measured waveforms and directivities up to supersonic tip Mach numbers \citep{Hanson1976}, the effect of particular blade geometries \citep{Hanson1979,Dobrzynski1993}, and the role of scattering by the hub and nearby airframe structures \citep{Glegg1991}. For an in-depth review of these works, the reader is referred to \cite{metzger1995}, who also highlights that, from the 1990s onward, research efforts increasingly shifted toward computational aeroacoustics codes rather than analytical studies of acoustic analogies.} \inlinecomment{For clarity, this paragraph has been moved up from the bottom.}

Many early studies were \deleted[id=FF]{therefore} rooted in planform-based blade representations on which source distributions were prescribed. Further simplifications were often obtained through a ring approximation of the propeller disk \citep{Gutin1936,Garrick1954,lowson1969}, in which the distributed sources are assumed to be concentrated at a particular radius. Subsequently, lifting-line descriptions for high-aspect-ratio propellers showed that the tonal sound field can be predicted directly from the spanwise lift distribution, with good agreement against more detailed computations and experimental data \citep{clark1974,brouwer1989,Brouwer1992}. \added{Such approaches are computationally attractive because they avoid a full surface description of the blade sources. Their accuracy, however, is naturally limited when chordwise non-compactness becomes important, especially for highly-swept propellers and at higher harmonics \citep{Schulten1984}. Related efforts have sought to retain part of the chordwise source structure without resorting to full blade-surface integration. \cite{Brentner1994} replaced the distributed chordwise pressure by an equivalent loading preserving the sectional lift. \cite{Bernardini2016} developed a spectral compact-source formulation in which analytical expressions for the chordwise pressure distribution are reconstructed from sectional loads, thereby reducing the amount of aerodynamic data required by the acoustic prediction. Computational savings have also been pursued through numerical and algorithmic improvements of acoustic-analogy solvers \citep{Brentner1997}. In the time domain, \cite{Casalino2003} introduced an advanced-time formulation for FW-H predictions, avoiding the iterative solution of retarded-time equations. Source-mode formulations \citep{Roger2019} follow yet another route, allowing rotating tonal sources to be represented by spinning modes, which can be efficiently coupled with Helmholtz solvers to account for scattering by nearby solid surfaces.} \deleted[id=FF]{With these tools at disposal, other investigations addressed a range of complementary aspects, including the validity of linearized models in reproducing measured waveforms and directivities up to supersonic tip Mach numbers, the effect of particular blade geometries, and the role of scattering by the hub and nearby airframe structures. For an in-depth review of these works, the reader is referred to Metzger (1995), which also highlights that, from the 1990s onward, research efforts increasingly shifted toward computational aeroacoustics codes rather than analytical studies of acoustic analogies.} \inlinecomment{For clarity, this paragraph has been moved earlier.} In parallel, developments have also focused on efficient numerical evaluation of radiation integrals. As already \deleted[id=FF]{noted in} \added{employed by} \cite{Gutin1936} and frequently exploited thereafter, far-field approximations \added{can} offer \deleted[id=FF]{the simplest} \added{a} reduction in computational complexity. This is due to the simplified structure of the radiated field in the \added{acoustic} far zone, which allows the observer dependence to be treated in a much simpler form. In the more general case, cylindrical-coordinate expansions have often been adopted \citep{Schulten1984,Glegg1991,lewy1994} to alleviate the cost of the underlying integrals, which are also frequently evaluated directly in their unexpanded form \citep{Hanson1976,Chapman1993,Carley1999}. \deleted[id=FF]{Inevitably, both near and far-field formulations in the literature lead to expressions in which the observer location appears inside the source integrals.} 

\added{The methods discussed above provide important reductions in the cost of evaluating radiation integrals, either by simplifying the Green's function, or by reducing the source description. However, they reduce the cost of each individual source integration. In many classical near- and far-field formulations, the observer location remains embedded within the integrand, so that changing the microphone position changes the integral to be evaluated. Predictions at many observer locations may therefore require repeated evaluations of the radiation integrals. The far}\deleted[id=FF]{Far}-field assumption reduces the number of observer coordinates embedded in the integration, but it does not remove the dependence entirely. This becomes an important limitation when predictions are required at a large number of microphone locations, as is very common in practical applications. \added{The practical difficulty associated with many observer locations was also addressed by \cite{Long2000}, who showed the benefit of parallelized calculations to alleviate costs.} Contributions in this direction were provided by \cite{Parry1989} for subsonic conditions and by \cite{Crighton1991} for supersonic one\added{s}; however, these results are restricted to the limit of an infinite number of blades, in which Hanson's integrals can be approximated asymptotically. \added{The extension of these studies to near-field observers was considered by \cite{peake1991}.} Despite the substantial body of work outlined above, there is therefore a need for a formulation in which the source integrals do not have to be recomputed for each observer position.

This paper develops a formulation to address this need. Starting from \deleted[id=FF]{Goldstein’s} \added{an} acoustic analogy, we retain its two surface source terms, namely the aerodynamic force per unit area and the surface normal velocity. The core of the approach is a spherical expansion of the free-space Green’s function \citep{jackson1962}, which leads to a spherical-multipole representation of the acoustic pressure valid in both the near and far field. A central feature of the resulting expression is the complete separation between source and observer dependence: the multipole coefficients depend only on blade geometry and aerodynamics, whereas the observer coordinates appear only through spherical harmonics and radial propagation factors. This separation makes the method computationally attractive, since the source-dependent coefficients are computed once and can then be reused to evaluate the acoustic field rapidly at arbitrary observer locations. \added{This advantage is complementary to the simplifications introduced in the works discussed above. Far-field approximations, asymptotic treatments or reduced source descriptions may still be employed to decrease the cost of evaluating the source integrals. In the present formulation, however, these integrals are independent of the observer location. The source integration is thus performed once and may itself benefit from any available approximation or acceleration technique. Related spherical-wave expansions have previously been developed for canonical radiation and diffraction problems involving circular pistons \citep{Hasegawa1983,Hasegawa1984,Wittmann1991,Douglas2005}. An extension to rotating-source Rayleigh integrals \citep{Rayleigh1896} was considered by \citet{Carley2006}.} \deleted[id=FF]{In recent work} \added{More recently}, \cite{fruncillo2025} applied the \deleted[id=FF]{same idea} \added{source-observer separation} to a Kirchhoff-Helmholtz formulation \citep{Williams2000}\added{. In that work, thin-airfoil theory was used to prescribe pressure and normal-velocity boundary conditions from the blade geometry, reducing the acoustic problem to a half-plane boundary integral. The resulting multipole representation was shown to} \deleted[id=FF]{and showed that it can} reduce the computational cost of far-field multi-microphone predictions by up to two orders of magnitude relative to Hanson’s formulation \citep{Hanson1980}. The present work differs from that study, since it is based on an acoustic-analogy formulation rather than on a potential-flow boundary integral. As a result, the source description retains explicitly the aerodynamic\added{-}force \added{and surface-normal-velocity} term\added{s} appearing in \deleted[id=FF]{Goldstein’s} \added{the acoustic} analogy. This makes it possible to examine the separate roles played by loading, thickness, and drag-related contributions in the multipole expansion of propeller noise, whereas the latter effects cannot be considered within the framework adopted previously. A further advantage emphasized here is that the spherical representation yields a direct expression for the radiated power, without the need for additional pressure integrations\deleted[id=FF]{, which is especially attractive for design and optimization purposes.}\added{. This makes the approach particularly useful for interpreting the distribution of acoustic energy among harmonics and multipoles, and for design or optimization applications.}

The spherical-multipole formulation is first\deleted[id=FF]{ly} examined \deleted[id=FF]{in its exact form, i.e.} by evaluating the resulting coefficients on the true blade geometry. In particular, the results show that, for the hovering subsonic conditions considered here, the radiated pressure field is largely governed by the first two non-zero multipoles, which already capture the essential directivity and power content of the tonal field. \added{Such a limited multipole content is expected for acoustically compact sources, for which the product of the wavenumber and the maximum source dimension is much smaller than unity. It is not, however, an obvious result under typical propeller-noise conditions, where this compactness parameter may be of order unity or larger.} Once this feature has been established, it becomes useful to seek approximate descriptions of the surface integral to reveal which geometric and aerodynamic mechanisms feed the dominant multipole contributions. To this end, two complementary simplified descriptions are developed. The first is a lifting-surface (LS) formulation, in which the blade is assumed at small incidence. Adopting a thin-airfoil approximation \citep{Ashley1965}, the resulting source terms can be decomposed into contributions associated with the pressure jump and the pressure sum over the blades, and the thickness distribution. Because the chordwise phase is retained inside the spanwise integrals, this formulation remains accurate when large-chord effects are important, in particular at higher frequency. The second is a lifting-line (LL) formulation, in which each blade section is assumed \deleted[id=FF]{compact} \added{slender} and collapsed onto a spanwise curve. A local expansion of the Green’s function and its gradient is then performed about each radial station, so that the resulting multipole coefficients reduce to one-dimensional spanwise integrals of compact sectional moments. In this case the chordwise phase is neglected, but large incidence and induced-velocity effects are naturally retained. Geometric compactness arguments of this kind have been discussed in several contexts \citep{lowson1969,hubbard1991}. Here\deleted[id=FF]{, however,} they are exploited within the spherical-multipole framework to clarify the physical content of the leading terms, and in particular to identify how lift, drag, and thickness contribute to the dominant multipoles of the expansion. To compare the two models on a consistent basis, both sets of multipole coefficients are assessed against \deleted[id=FF]{exact} surface integration on the true blade geometry. This comparison shows that the two approximations perform well in their respective regimes of validity. In particular, once the multipole expansion is truncated to its dominant terms, both approximations retain the relevant acoustic content \deleted[id=FF]{of the exact solution} while \deleted[id=FF]{yielding substantial computational savings relative to exact} \added{reducing the cost of the} surface integration. The theoretical developments are complemented by analyses of far-field directivity, power distribution over harmonics and time-domain waveforms. In addition, validation against two experimental datasets taken from the literature \citep{Grande2022b,Bensignor2026} for different propellers is presented in the final part of the paper. In both cases, the agreement with the measurements is excellent, confirming that the proposed formulations provide not only computational efficiency, but also accurate prediction of the dominant tonal radiation. \added{A separate comparison is carried out under a high tip Mach number condition, using the experimental data of \citet{Boxwell1978}. In this case, the present results are also consistent with previous linear predictions.}

The paper is organized as follows. Section \ref{sec:Formulation} introduces and discusses the spherical-multipole formulation of \deleted[id=FF]{Goldstein’s} \added{the retained surface terms of an acoustic} analogy. Section \ref{sec:Approximations} derives the lifting-surface and lifting-line approximations. Experimental validation is reported in Section \ref{sec:Experimental Validation}. Final remarks are given in Section \ref{sec:Conclusions}.

\section{Multipole Expansion Formulation}
\label{sec:Formulation}

Acoustic analogies are exact identities obtained by recasting the compressible-flow equations as a linear wave equation forced by terms that collect the remaining physics \citep[Eq.~4]{Lighthill1952}. Their practical use, however, hinges on the crucial assumption to regard the source term as known a priori, even though\deleted[id=FF]{,} there is no exact way to determine the acoustic pressure and the source term independently of one another. In practice, the analogy becomes predictive only under a one-way coupling hypothesis, whereby the flow field is provided by independent calculations or measurements and the acoustic field is computed as the corresponding solution of the linear wave equation. This limitation was recognized from the outset. In Lighthill’s original analogy \citep{Lighthill1952}, no solid surfaces were included, and the forcing term was the Lighthill\deleted[id=FF]{'s} stress tensor. This tensor involves fluctuating density (or pressure) and therefore contains, in principle, the very unknown of the wave equation. To decouple the forcing from the unknown acoustic field, it was proposed to approximate Lighthill's stress tensor by its Reynolds-stress contribution, which can be estimated from the incompressible part of the flow without explicitly resolving the acoustic component. The validity of this approximation was examined by \cite{Crow1970} through an asymptotic expansion, showing that it is reasonable for sufficiently low Mach numbers and when the characteristic eddy size of the flow is comparable with a representative dimension of the vortical region. The inclusion of solid surfaces as additional source terms is due to the developments of \cite{Curle1955} and \cite{FfowcsWilliams1969}. In the context of propeller noise, early studies demonstrated that, at low speeds and for the frequencies relevant to tonal radiation, \deleted[id=FF]{the} Lighthill's contribution is \deleted[id=FF]{of higher order} \added{marginal} with respect to the surface ones \citep{FfowcsWilliams1969.2} and is therefore commonly neglected in tonal-noise analyses \citep{Hawkings1974,Hanson1979}. 

\added{\cite{goldstein1974} derived a general integral representation for aerodynamic sound in the presence of moving solid boundaries. In the present work, we specialize this representation by neglecting both convection effects and the volume quadrupole contribution.} Under th\deleted[id=FF]{is}\added{ese} assumption\added{s}, in a reference frame where the medium is at rest\added{,} the acoustic pressure radiated by a moving surface $S(t)$, at source position $\mathbf{x}$ and time $t$, and observed at position $\mathbf{x}_0$ and time $t_0$, is written as 
\begin{equation}
    p(\mathbf{x}_0,t_0)
    =\int_{-\infty}^{\infty}\int_{S(t)}
    \left[ \mathbf{f}(\mathbf{x},t)\cdot\nabla_{\mathbf{x}}G(\mathbf{x}_0,t_0 ; \mathbf{x},t)
    - \rho_0 V_n(\mathbf{x},t)  \partial_{t}G(\mathbf{x}_0,t_0 ; \mathbf{x},t) \right] dS   dt
    \label{eq:goldstein-time}
\end{equation}
where the free-space Green's function \citep{jackson1962} is
\begin{equation}
        G = \frac{\delta\left(t_0-t-|\mathbf{x}_0-\mathbf{x}|/a_s\right)}{4\pi|\mathbf{x}_0-\mathbf{x}|}
\end{equation}
and vector \(\mathbf{f}(\mathbf{x},t)\) is the force per unit area exerted by the body on the fluid and \(V_n(\mathbf{x},t) = \mathbf{V}(\mathbf{x},t)\cdot \hat{\mathbf{n}}(\mathbf{x},t)\) is the normal velocity of the surface. We adopt the convention of \cite{hanson1993}, in which the unit normal $\hat{\mathbf{n}}$ has outward direction from the blade. In Eq. \eqref{eq:goldstein-time}, \(\nabla_{\mathbf{x}}\) denotes the gradient with respect to the source coordinates, while \(\partial_t\) denotes differentiation with respect to the source time. The constants \(\rho_0\) and \(a_s\) are the ambient density and speed of sound, respectively. \added{As clarified by \cite{goldstein1974}, adopting the free-space Green's function reduces Eq. \eqref{eq:goldstein-time} to the surface-source terms of the FW-H formulation.}

\deleted[id=FF]{In the acoustic analogy, the Eq. \eqref{eq:goldstein-time} is a closed-form expression to obtain acoustic pressure. This constitutes the key distinction from a classical Kirchhoff-Helmholtz boundary-integral formulation adopted in acoustics. For the wave equation, the pressure and its normal derivative on a surface cannot be prescribed independently, and their relation is not known a priori, leading to an integral equation to be solved. This overspecification difficulty can be avoided by looking for tailored Green’s functions that render one of the boundary conditions unnecessary. A recent example of this approach, in the context of propeller noise, has been presented by Fruncillo (2025).}

In the common terminology of aeroacoustics, the source terms \added{associated with the distributions of} \(V_n(\mathbf{x},t)\) and \(\mathbf{f}(\mathbf{x},t)\) in Eq. \eqref{eq:goldstein-time} are usually referred to as monopole and dipole \added{contributions}. Indeed, those are the far-field behaviours they would originate if they were present as isolated point sources. By the same convention, the Lighthill\deleted[id=FF]{'s} term neglected here is called a quadrupole. It should be emphasized, however, that this nomenclature pertains only to the character of an elementary point source: when the entire radiating surface is observed from distances much larger than its characteristic size, the resulting total field may display complicated features that differ from those of the individual elemental sources. The aim of the present work is precisely to analyze and characterize the \deleted[id=FF]{far-field multipoles} \added{global multipole content} of the radiated sound \added{by decomposing the acoustic field into outgoing spherical-wave components}.

\begin{figure}
\centering
\def\svgwidth{0.8\linewidth} 
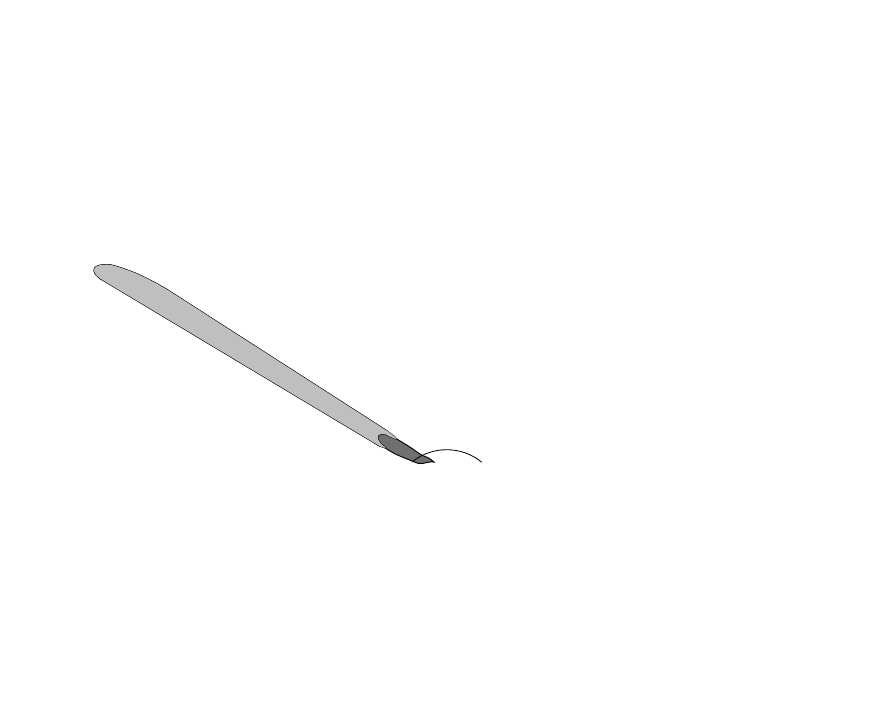
\caption{Reference geometry and observer coordinates for the rotating two-bladed propeller. The rotor spins about the $z$-axis with angular velocity $\Omega$ in the fixed Cartesian frame $(x,y,z)$. The observer is located at $\mathbf{x}_0$ at distance $r_0$ from the origin, with spherical elevation  $(\theta_0)$ and azimuth angles $(\phi_0)$. The source position in the frame attached to the propeller is denoted by $\mathbf{x}'$.}
\label{fig:blade}
\end{figure}

For a propeller rotating at constant angular speed \(\Omega\) without translation (figure \ref{fig:blade}), we describe a source point on the blade by \(\mathbf{x}=(r,\theta,\phi)\) and the observer position by \(\mathbf{x}_0=(r_0,\theta_0,\phi_0)\), both expressed in spherical coordinates of a fixed inertial frame. Under steady operating conditions, the acoustic pressure measured at a fixed observer location is periodic in time and can therefore be expanded in the Fourier series
\begin{equation}
    p(\mathbf{x}_0,t_0)=\sum_{m=-\infty}^{\infty} \hat{p}_m(\mathbf{x}_0)\,e^{-im\Omega t_0}
    \label{pressure fourier}
\end{equation}
If the rotor consists of \(N\) identical blades equally spaced in azimuth, only those harmonics whose index \(m\) is an integer multiple of \(N\) are non-zero.

It is convenient to evaluate \deleted[id=FF]{Goldstein’s} \added{the} time-domain surface integral, Eq. \eqref{eq:goldstein-time}, in a frame co-rotating with the blades. In that frame, a source point is described by \(\mathbf{x}'=(r',\theta',\phi')\added{\added{\in S}}\), where the azimuth \(\phi'\) is fixed to the rotor, and the relation between \deleted[id=FF]{inertial} \added{fixed} and rotating coordinates is
\begin{equation}
r=r',\qquad \theta=\theta',\qquad \phi=\phi'+\Omega t
\end{equation}
Thus, in the rotating frame the blade surface is stationary, and the explicit time dependence associated with the motion of the integration surface can be removed. Following \citet[Eqs.~20 and 28]{hanson1993}, the harmonic coefficients may then be written as \inlinecomment{The notation in this equation has been slightly revised for clarity.}
\begin{equation}
    \hat{p}_m(\mathbf{x}_0)
    =\frac{N}{2\pi}\int_{S} e^{-im\phi'}
    \int_{0}^{2\pi} 
    \left[
    \mathbf{f}(\mathbf{x},\phi)\cdot\nabla_{\mathbf{x}}G_m(\mathbf{x}_0;\mathbf{x})
    - i\rho_0 \omega_m V_n(\mathbf{x})\,G_m(\mathbf{x}_0;\mathbf{x})
    \right]
    e^{im\phi} d\phi  dS
    \label{eq:goldstein-m}
\end{equation}
where \(\omega_m=m\Omega\) is the angular frequency of the \(m\)-th harmonic and the time integration has been replaced by an equivalent integration over the \deleted[id=FF]{inertial} azimuth \added{angle} \(\phi\) swept by the source point as the blade rotates. In this expression, the gradient \(\nabla_{\mathbf{x}}\) is taken with respect to the source coordinates in the \deleted[id=FF]{inertial} \added{fixed} frame\added{.}\deleted[id=FF]{, after which the inner integral is} \added{The resulting integrand is then} evaluated at the source position corresponding to the co-rotating position $\mathbf{x}'$ \added{to perform the surface integration.} \added{This equation is simplified with respect to that derived by Hanson, since we are not considering the moving medium.} 

The function
\begin{equation}
G_m(\mathbf{x}_0;\mathbf{x})=\frac{e^{ik_m|\mathbf{x}_0-\mathbf{x}|}}{4\pi |\mathbf{x}_0-\mathbf{x}|}
\end{equation}
is the free-space Green’s function of the Helmholtz equation at wavenumber \(k_m=\omega_m/a_s=mk\).

The aim of this work is to achieve a separation between source and observer quantities, so that the integrals in Eq. \eqref{eq:goldstein-m} do not need to be recomputed each time the observer position is changed. For observers outside the sphere containing the rotor, this can be done through a spherical expansion of the Green's function \citep[Eq. 16.22]{jackson1962}, which we write as
\begin{equation}
    G_m(\mathbf{x}_0;\mathbf{x})=ik_m \sum^\infty_{n=-\infty}\sum^{\infty}_{\ell=|n|} f_{\ell n}(\mathbf{x}_0)  g_{\ell n}(\mathbf{x})
\end{equation}
where \(f_{\ell n}(\mathbf{x}_0)\) contains all the observer dependence, while \(g_{\ell n}(\mathbf{x})\) contains only the source dependence. In particular, in spherical coordinates one has 
\begin{equation}
    f_{\ell n}(\mathbf{x}_0)=h_\ell(k_m r_0)Y_{\ell n}(\theta_0,\phi_0), \qquad g_{\ell n}(\mathbf{x})=j_\ell(k_m r)Y^*_{\ell n}(\theta,\phi)
    \label{f and g definitions}
\end{equation}
where $h_\ell$ and $Y_{\ell n}$ are spherical Hankel functions (of the first kind) and spherical harmonics of degree $\ell$ and order $n$, whereas $j_\ell$ are spherical Bessel functions and the symbol $^*$ indicates complex conjugation.

Under the simplifying assumption that \deleted[id=FF]{surface fields \(\mathbf{f}\) and} \(V_n\) \added{and the spherical components of \(\mathbf{f}\) are steady in the blade-fixed frame, these source quantities} do not depend explicitly on the azimuth angle \(\phi\)\deleted[id=FF]{, as is true in hovering conditions, the}\added{. The} only explicit dependence on \(\phi\) inside the inner integral is \added{therefore} carried by the function \(g_{\ell n}(\mathbf{x})\propto Y^*_{\ell n}(\theta,\phi) \propto e^{-in\phi}\). \added{In spherical coordinates, all components of \(\nabla_{\mathbf{x}}g_{\ell n}\) retain the same factor \(e^{-in\phi}\), such that the} \deleted[id=FF]{The} inner integration then simply yields 
\begin{equation}
\int_0^{2\pi} e^{-i\left(n-m\right)\phi} d\phi = 2\pi \delta_{n m}
\end{equation}
where $\delta_{nm}$ is the Kronecker delta. The harmonic pressure coefficient becomes \inlinecomment{The notation in this equation has been slightly revised for clarity.}
\begin{equation}
    \hat{p}_m(\mathbf{x}_0)
    = N  ik_m \sum^{\infty}_{\ell=|m|} f_{\ell m}(\mathbf{x}_0)  \int_{S} \left[ \mathbf{f}(\mathbf{x}')\cdot\nabla_{\mathbf{x}'}g_{\ell m}(\mathbf{x}')
    -  i \rho_0 a_s k_m V_{n}(\mathbf{x}')  g_{\ell m}(\mathbf{x}') \right]
      dS
\end{equation}
\added{In this final expression,} \deleted[id=FF]{where} the azimuthal dependence \(e^{-i m \phi'}\) \added{appearing in Eq.~\eqref{eq:goldstein-m} has not been discarded; it} is now contained in the evaluation of function \(g_{\ell m}\) in the co-rotating frame. Only the source-dependent functions appear inside the surface integral, while the observer-dependent part \(f_{\ell m}(\mathbf{x}_0)\) is factored out. The complex multipole coefficients of degree $\ell$ and order $m$ are defined as \inlinecomment{The notation in this equation has been slightly revised for clarity.}
\begin{equation}
    A_{\ell m}
    = N  ik_m \int_{S} \left[ \mathbf{f}(\mathbf{x}')\cdot\nabla_{\mathbf{x}'}g_{\ell m}(\mathbf{x}')
    -  i \rho_0 a_s k_m V_{n}(\mathbf{x}')  g_{\ell m}(\mathbf{x}') \right]
      dS
    \label{eq:Alm-start}
\end{equation}
and depend only on the source geometry and on the surface distributions of \(\mathbf{f}(\mathbf{x}')\) and \(V_n(\mathbf{x}')\). 

The complex acoustic pressure at harmonic \(m\) can then be expressed as a spherical multipole expansion in terms of the source coefficients \(A_{\ell m}\), namely
\begin{equation}
    \hat{p}_m(\mathbf{x}_0)=\sum_{\ell=|m|}^{\infty} A_{\ell m}\, f_{\ell m}(\mathbf{x}_0)
    \label{eq:pressure_formula}
\end{equation}
and the corresponding real acoustic pressure is then recovered from Eq. \eqref{pressure fourier}.

Once the multipole coefficients \(A_{\ell m}\) are known, the acoustic power associated with the harmonic $m$ is proportional to the sum of their squared moduli \citep[Eq. 6.113]{Williams2000}, as
\begin{equation}
\Pi_m = \frac{1}{\deleted[id=FF]{2}\rho_0 a_s k^2_m} \sum^{\infty}_{\ell=|m|} \left|A_{\ell m} \right|^2 \label{eq:power formula}
\end{equation}
\added{Here, unlike in \cite{Williams2000}, $A_{\ell m}$ corresponds to a coefficient of the two-sided Fourier series in Eq. \eqref{pressure fourier}. For a real pressure field, the contributions at $m$ and $-m$ are equal, so that the physical power of the corresponding harmonic is $2\Pi_m$. } This enables fast, straightforward computations without resorting to far-field pressure integrals, as is otherwise required when adopting alternative coordinate systems.

The surface integral in Eq. \eqref{eq:Alm-start} can be evaluated numerically for an arbitrarily complex blade geometry. The practical advantage of the multipole representation becomes apparent when the expansion in the index \(\ell\) converges rapidly. It is worth noting that, since \(\ell\ge |m|\) and the harmonic indices \(m\) are integer multiples of the blade count \(N\), the lowest admissible multipole order is constrained by the rotor periodicity. In particular, for \(N\ge 2\), that is, excluding the single-blade case, monopole (\(\ell=0\)) and dipole (\(\ell=1\)) contributions are not admissible; the first possible far-field multipole is therefore at least a quadrupole (\(\ell=2\)) for \added{a} two-bladed propeller. \added{This approach provides a direct physical interpretation of the radiated field as a superposition of elementary spherical radiation patterns. The coefficients \(A_{\ell m}\) quantify how strongly, and with which phase, each admissible spherical mode is excited by the blade geometry and aerodynamic loading, while the spherical harmonics \(Y_{\ell m}\) determine the corresponding angular structure. The rotating surface excites a hierarchy of admissible spherical multipoles, selected by the rotor periodicity. Therefore, even before evaluating the source integrals, the angular shapes of the radiated field can be inferred from the spherical harmonics associated with the allowed harmonics and degrees.}

\subsection{Far-field pressure}

\inlinecomment{This subsection has been added for clarity.}

The complex pressure amplitudes can be obtained from Eq. \eqref{eq:pressure_formula} at any observer location lying outside the sphere enclosing the propeller. \deleted[id=FF]{In propeller acoustics, the emitted frequency content is conveniently expressed in terms of the tip Mach number $M_t=\Omega R/a_s$ where \(R\) is the propeller radius, namely the distance between the blade tip and the axis of rotation. Equivalently} \added{For a rotating tonal source, the emitted frequencies are locked to the shaft rotational speed. Introducing the rotational tip Mach number $M_t=\Omega R/a_s$, where \(R\) is the propeller radius, namely the distance between the blade tip and the axis of rotation, the rotor Helmholtz number at the $m$-th harmonic is} 
\begin{equation}
k\added{_m}R=\added{m} M_t=\frac{2\pi R}{\lambda\added{_m}}
\end{equation}
where \(\lambda\added{_m}\) is the corresponding \added{acoustic} wavelength. \added{The product $k_m R$ characterizes the acoustic compactness of the propeller at the $m$-th harmonic, and can be rewritten as $mM_t$, providing a direct interpretation in terms of tip Mach number \citep{lowson1969}.}

\deleted[id=FF]{In the present work, the main interest is in the radiation observed at large distance from the source, and the analysis therefore focuses primarily on the acoustic far field.} \added{The expressions derived in the present work are valid in both the near and far fields, provided that the observer lies outside the sphere enclosing the source. Particular attention is devoted to the far-field case and to the corresponding expressions.} In this limit, the spherical representation is especially convenient because the observer-dependent factor \(f_{\ell m}(\mathbf{x}_0)=h_\ell(\deleted[id=FF]{m M_t \bar r_0}\added{k_m r_0})Y_{\ell m}(\theta_0,\phi_0)\) simplifies significantly. For an observer in the far field, the spherical Hankel function admits the large-argument approximation \citep[Eq. 16.13]{jackson1962}
\begin{equation}
h_\ell\!\left(m M_t \bar r_0\right)\approx (-i)^{\ell+1}\frac{e^{i m M_t \bar r_0}}{m M_t \bar r_0},
\qquad
m M_t \bar r_0\gg \ell
\label{hankel FF}
\end{equation}
where \(\bar r_0=r_0/R\) is the observer distance normalized by the propeller radius. Substituting Eq. \eqref{hankel FF} into Eq. \eqref{eq:pressure_formula}, the complex far-field (FF) pressure coefficient of harmonic \(m\) becomes
\begin{equation}
\hat{p}^{(FF)}_m(\mathbf x_0)
=
\frac{e^{i m M_t \bar r_0}}{m M_t \bar r_0}
\sum_{\ell=|m|}^{\infty}
(-i)^{\ell+1}A_{\ell m}Y_{\ell m}(\theta_0,\phi_0)
\label{eq:pressure-formula-FF}
\end{equation}
This expression makes the structure of the far field particularly transparent: the radial decay and propagation are explicit, while the angular dependence is entirely contained in the spherical harmonics, weighted by the multipole coefficients. \added{When the source is observed at a sufficiently large distance, fine spatial details of the near pressure field are progressively filtered out, and the radiated field tends to exhibit a simpler angular structure. In this limit, one may expect that only a few spherical patterns are required to identify the dominant directivity features. In the near field, where the structure is complex \citep{Chapman1993}, the spherical Hankel functions retain their full \(\ell\)-dependent radial behaviour and more terms are needed. Even in that case, however, the expansion still indicates which spherical shapes contribute to the pressure field.}

\subsection{Dominant multipoles}
\label{sec:dominant-multipoles}
We now examine the \added{predictions of the} multipole coefficients \deleted[id=FF]{predictions} obtained from Eq. \eqref{eq:Alm-start}. In particular, we are interested in the decay of the coefficients with respect to the index \(\ell\), i.e. in the number of terms required to reconstruct the \deleted[id=FF]{far-}field accurately at each frequency. The dependence on the azimuthal index \(m\) reflects the frequency content of the radiated field, since each value of \(m\) corresponds to a distinct harmonic of the rotor motion.

\begin{figure}
    \centering
    \subfloat[Propeller A, with large chord-to-diameter ratio $\gamma=c/(2R)=0.20$ and mild twist from $\beta=10^\circ$ at the root to $2^\circ$ at the tip. \label{fig:blade_A}]{
        \begingroup
        \def\svgwidth{1\linewidth}
        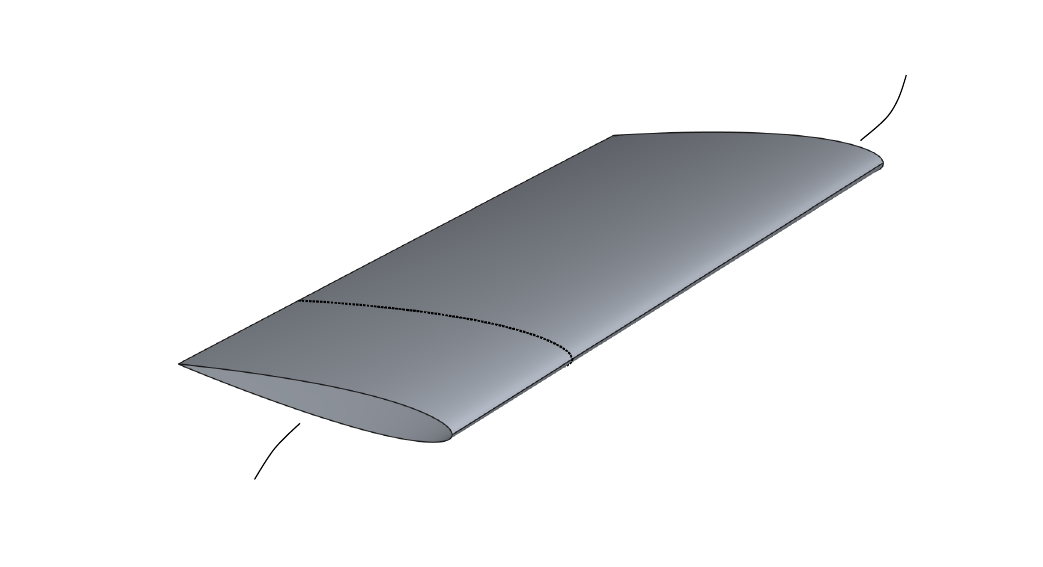
        \endgroup}
    \vspace{0.8em}
    \subfloat[Propeller B, with small chord-to-diameter ratio $\gamma=c/(2R)=0.05$ and stronger twist from $\beta=17^\circ$ at the root to $7^\circ$ at the tip. \label{fig:blade_B}]
        {\begingroup
        \def\svgwidth{1\linewidth}
        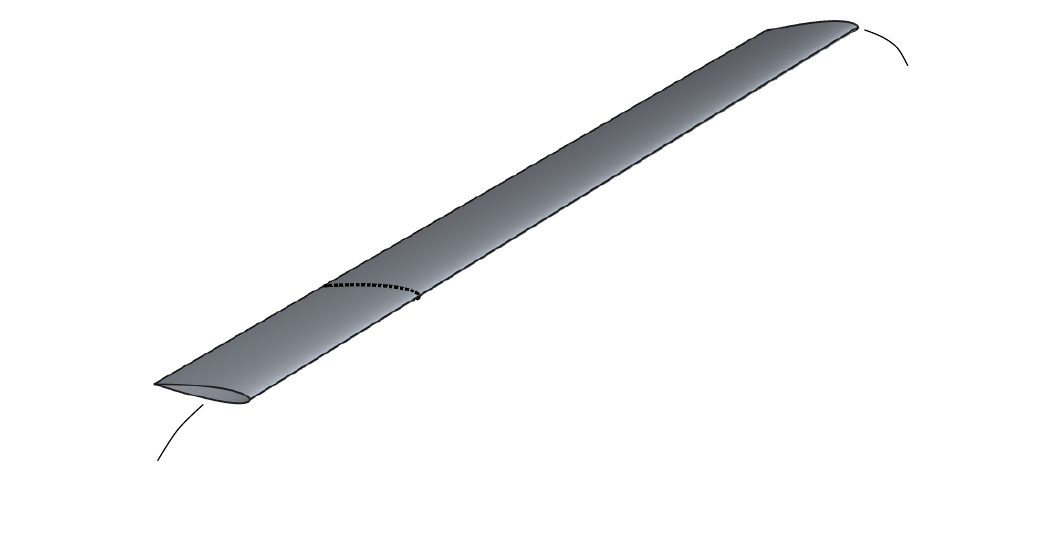
        \endgroup}
    \caption{Geometry and aerodynamic loading of the two propellers. In each panel, the upper insets show the spanwise distributions of the sectional lift and drag coefficients, \(c_l\) and \(c_d\), at \(M_t=0.2\) (blue lines) and \(M_t=0.8\) (red lines). The lower insets report the chordwise pressure-coefficient distributions, \(c_p\), at the representative section \(r/R=0.75\). Solid and dotted lines denote the suction and pressure sides, respectively. \inlinecomment{This figure has been added in the revised manuscript.}}
    \label{fig:blade_geometries}
\end{figure}

To explore these points over a representative range of propeller geometries, two baseline configurations are considered. The choice is also motivated by the developments presented later, since the same configurations will serve as convenient reference cases for discussing the approximate formulations of the surface integral. Propeller A has a large chord-to-diameter ratio, \(\gamma=c/(2R)\), where \(c\) is the local chord length of the blade. For this propeller, \(\gamma=0.20\). Propeller B has a smaller chord-to-diameter ratio, \(\gamma=0.05\). Both propellers are two-bladed rotors, so that the blade count is \(N=2\), with tip radius $R=0.1$ m and hub radius $R_{\mathrm{hub}}=0.01$ m. The blades are unswept, so that their mean chord positions are aligned along a straight line, and are built with NACA~0012 airfoils along the span. \added{A linear twist law \(\beta(r)\), measured with respect to the plane of rotation \((z_0=0)\), is prescribed along the spanwise coordinate \(r\) for each blade. Propeller A has a comparatively mild twist distribution, decreasing from \(10^\circ\) at the hub to \(2^\circ\) at the tip. Propeller B is deliberately more strongly twisted, with \(\beta\) decreasing from \(17^\circ\) to \(7^\circ\). The two geometries are illustrated in figure \ref{fig:blade_geometries}.} 

\begin{figure}
\centering
\subfloat[Propeller A: larger chord length $c=0.04$ m at smaller incidence.]
{\includegraphics[width=.5\textwidth]{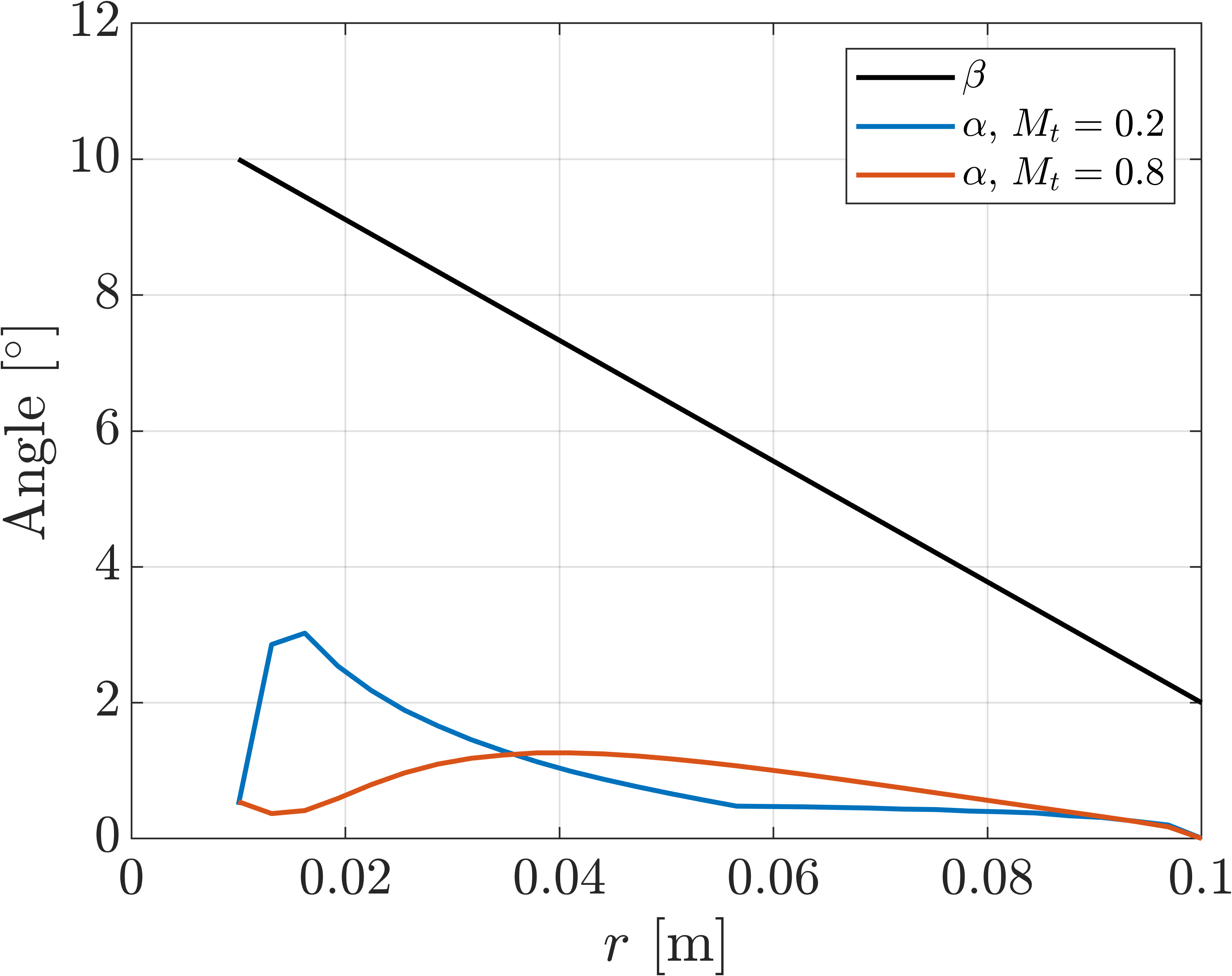}\label{fig:aero-span-propA}}\hfill
\subfloat[Propeller B: smaller chord length $c=0.01$ m at higher incidence.]
{\includegraphics[width=.5\textwidth]{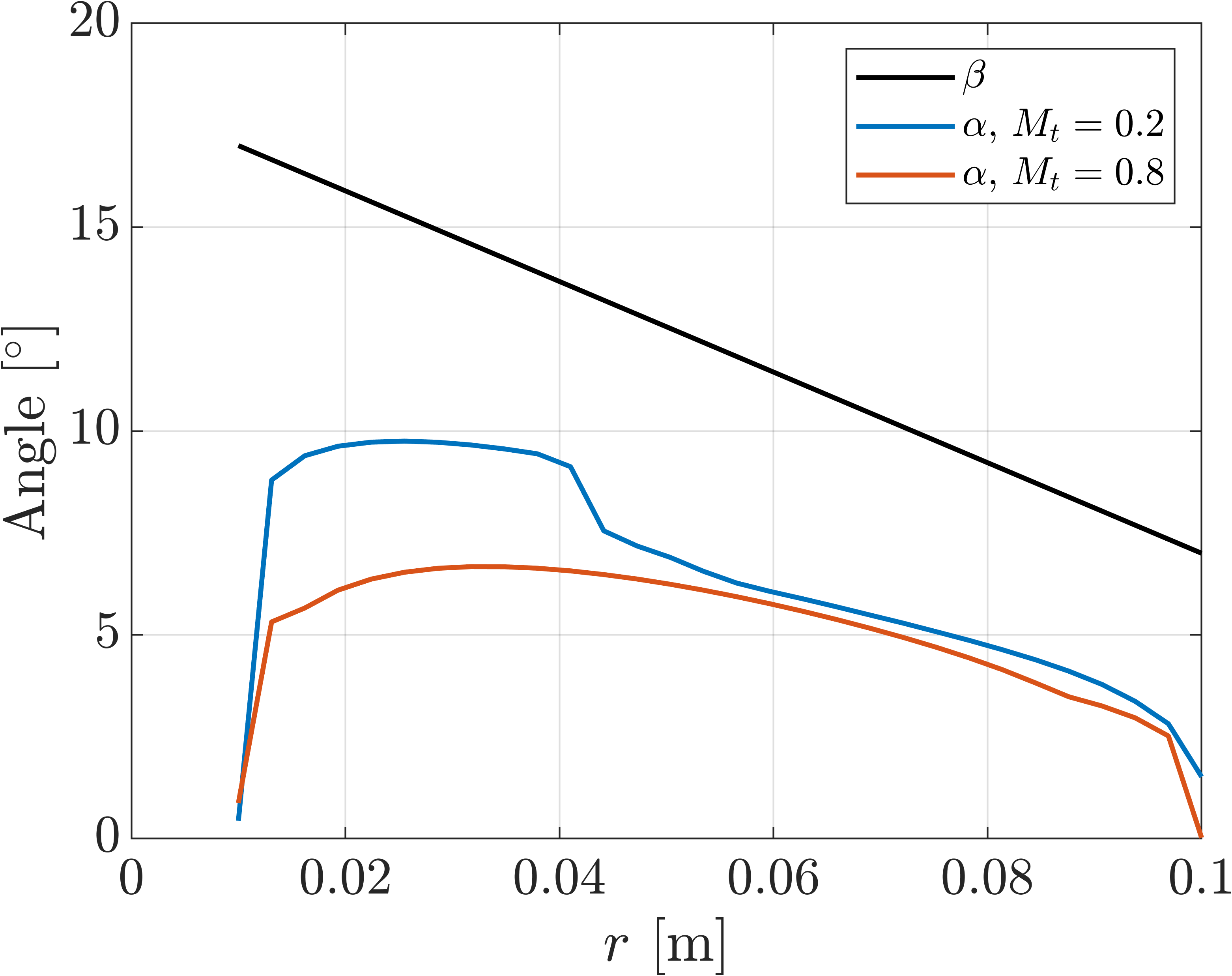}\label{fig:aero-span-propB}}
\caption{Spanwise aerodynamic distributions for both propellers. In each panel, the solid \added{black} line represents the prescribed twist distribution, while the \deleted[id=FF]{dashed} \added{blue and red} lines represent the angle-of-attack \(\alpha\) \deleted[id=FF]{and inflow angle \(\varphi_i\)} distributions at \(M_t=0.2\) and \(M_t=0.8\). \inlinecomment{This figure has been updated in the revised manuscript.}}
\label{AeroSpan}
\end{figure}

Two operating conditions are considered in order to assess the effect\added{s} of \added{loading and} acoustic compactness, namely \(M_t=0.2\) and \(M_t=0.8\). \deleted[id=FF]{An aerodynamic dataset is generated at each tip Mach number from sectional aerodynamic calculations performed with XFOIL, using a reference Reynolds number \(Re=10^5\). Each dataset contains the sectional pressure distributions, lift and drag coefficients. For each propeller, a linear blade-twist distribution \(\beta(r)\), measured with respect to the plane of rotation \((z_0=0)\), is prescribed along the spanwise coordinate \(r\). In particular, Propeller A is characterized by a small twist distribution, whereas Propeller B has a more pronounced one.} \added{For each propeller and tip Mach number, a spanwise aerodynamic database is generated from sectional calculations performed with XFOIL \citep{drela1989}. At each radial station, the local tangential velocity \(\Omega r\) is used to evaluate the corresponding Mach number and chord-based Reynolds number for the aerodynamic calculation. For each radial position and angle of attack, the database stores the sectional lift and drag coefficients, \(c_l\) and \(c_d\), together with the pressure-coefficient distributions on the suction and pressure sides of the airfoil.} The spanwise distributions of angle of attack, \(\alpha(r)\), and inflow angle, \(\varphi_i(r)\), namely the angle between the rotor plane and the local relative velocity vector, are obtained by solving iteratively the blade element momentum theory relation \citep{Bramwell2001}  
\begin{equation}
N c\left(c_l \cos\varphi_i-c_d \sin\varphi_i\right)=8\pi \added{F} r \sin^2\varphi_i 
\label{bemt relation}
\end{equation}
\added{A Prandtl tip- and hub-loss correction $F$ is included in the BEMT calculation to account for the finite number of blades and for the reduction of circulation close to the blade extremities \citep{betz1919,Glauert1935}. This correction is usually retained for propeller and rotor calculations, including recent aeroacoustic applications \citep{Adkins1994,goyal2024}.} For each radial station, the local twist, inflow angle, and angle of attack satisfy the geometric relation $\beta(r)=\alpha(r)+\varphi_i(r)$. \added{The use of BEMT as aerodynamic input for propeller-noise prediction is well established, with comparisons against several experimental datasets showing good agreement \citep{Hambrey2017,Kotwicz2019}. Recent studies have further compared BEMT-based acoustic estimates with experiments and high-fidelity simulations for isolated propellers \citep{Pereira2024}.} The resulting spanwise distributions are shown in figure \ref{AeroSpan}.

\added{Although the spherical expansion removes the need to repeat the blade-surface integration for each observer position, the observer-independent multipole coefficients in Eq. \eqref{eq:Alm-start} must still be evaluated numerically over the source surface. The spatial discretization therefore remains an important aspect of the present formulation. A sufficiently fine mesh is required to resolve both the aerodynamic loading distribution and the acoustic phase variation over the blade. Conversely, an unnecessarily fine discretization increases the computational cost while providing only marginal improvements in accuracy. This trade-off is common to surface-integral formulations for propeller and rotor noise \citep{brentner1996,Brentner1997c}, and recent works have further examined the sensitivity of tonal-noise predictions to the adopted acoustic discretization \citep{tannoury2013,dayde2026}.}
\added{\citet{hanson1993} adopted a maximum deviation of 0.5 dB from a refined solution as the discretization criterion and used it to determine the minimum numbers of spanwise and chordwise source points. The required spatial resolution depends on the blade count, harmonic order and flight Mach number; consequently, mesh adequacy should be verified separately for each propeller geometry and operating condition.} 

\begin{figure}
\centering
\subfloat[Spanwise refinement at fixed \(N_c=100\).]
{\includegraphics[width=.5\textwidth]{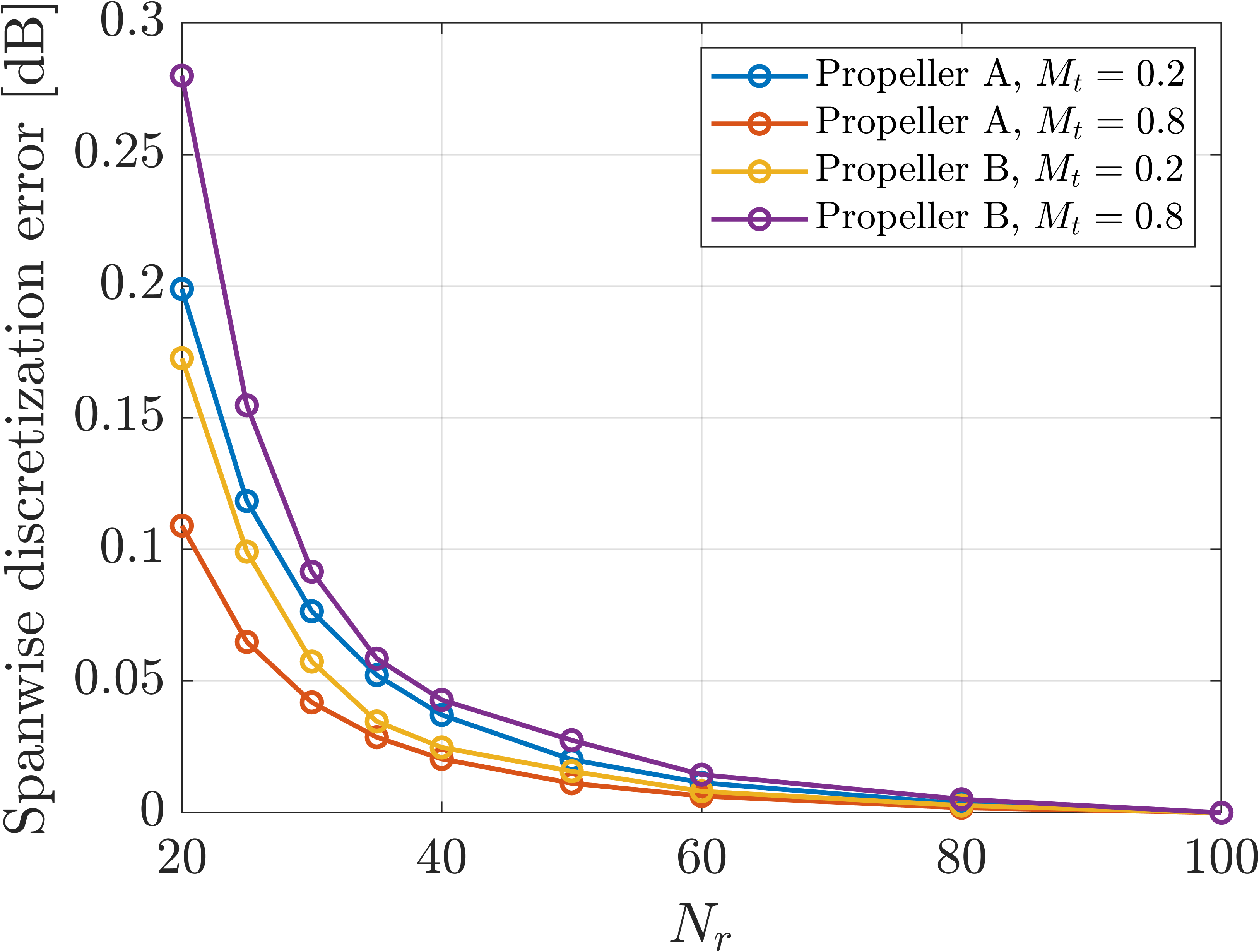}\label{fig:SpanwiseDiscretizationError_all4}}\hfill
\subfloat[Chordwise refinement at fixed \(N_r=100\).]
{\includegraphics[width=.5\textwidth]{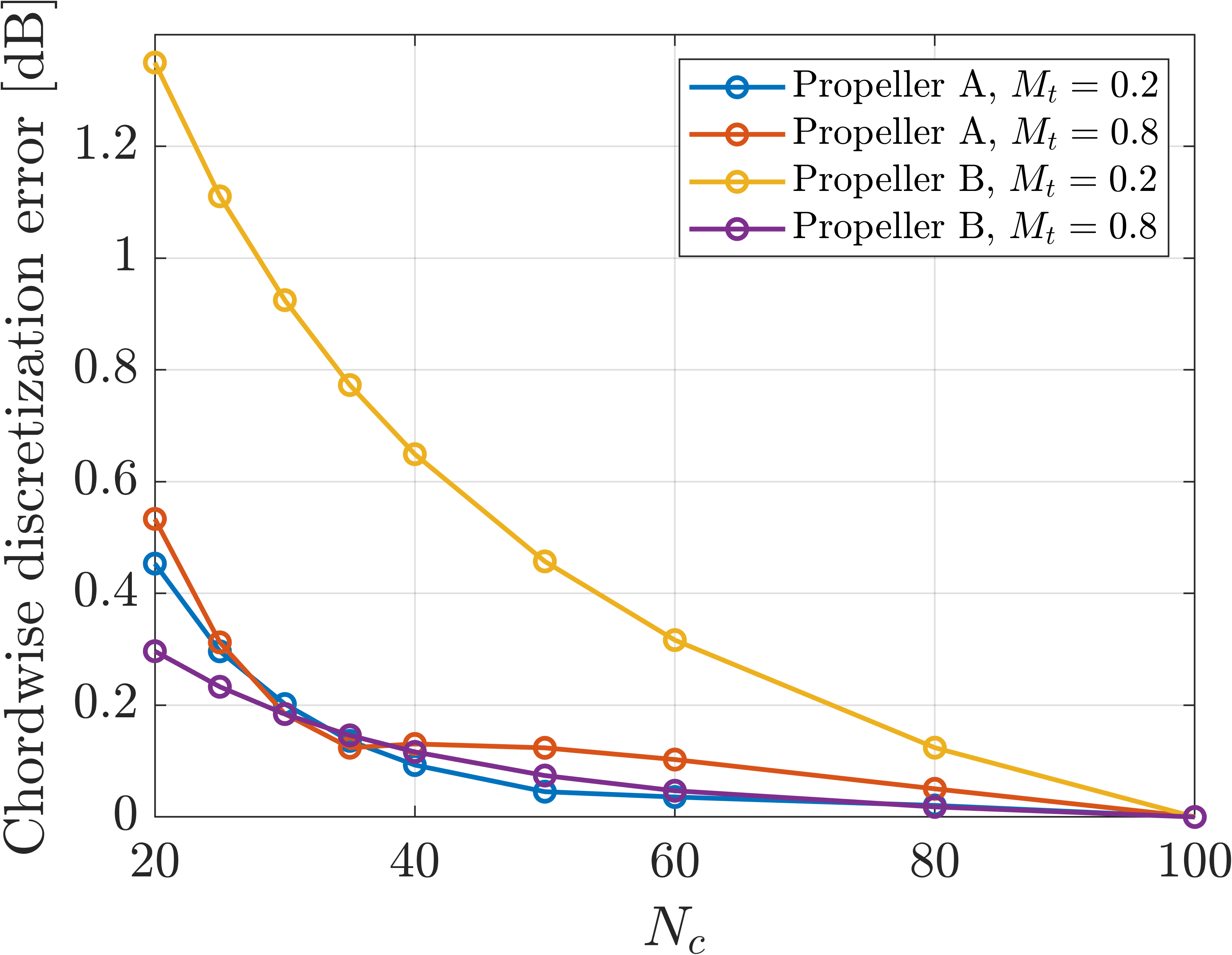}\label{fig:ChordwiseDiscretizationError_all4}}
\caption{Blade-surface discretization error. The error is evaluated relative to the \(100\times100\) reference mesh and corresponds to the maximum harmonic-power difference over the retained tones. \inlinecomment{This figure has been added in the revised manuscript.}}
\label{fig:surface-discretization-convergence}
\end{figure}

\deleted[id=FF]{A preliminary mesh-convergence study was carried out to identify a discretization yielding grid-independent acoustic predictions.} \added{In the present work, the blade-surface discretization is selected through a dedicated convergence study covering both propellers and both tip-Mach-number conditions. For each configuration, the reference surface representation employs $N_c=100$ chordwise points on each side of the airfoil, with clustering near the leading edge, and $N_r=100$ uniformly spaced radial stations between hub and tip. The chordwise reference grid retains the full resolution of the pressure-coefficient distributions stored in the aerodynamic database.  The surface integral in Eq. \eqref{eq:Alm-start} is then evaluated over the corresponding blade surface obtained by connecting the sectional airfoils along the span. The influence of the chordwise and radial resolutions is then examined separately. In the chordwise-refinement study, $N_c$ is progressively reduced while $N_r=100$ is retained. In the radial-refinement study, $N_r$ is varied while $N_c=100$ is fixed. For each mesh, the radiated power is evaluated for all retained tones, from $m=2$ to $m=20$. The discretization error is defined as the absolute difference in harmonic-power level relative to the reference representation. The quantity reported in figure \ref{fig:surface-discretization-convergence} is the largest difference over the complete retained harmonic range, for each discretization.}

\added{The convergence study shows that, for these propellers, the chordwise discretization is generally more restrictive, since it must resolve both the sectional pressure distribution and the acoustic phase variation over the airfoil chord. The convergence cannot be predicted from the acoustic size of the surface elements alone. This is illustrated by Propeller B at \(M_t=0.2\), for which the slowest chordwise convergence occurs despite the source being more compact. In this case, the largest error is found at the first radiating harmonic, \(m=2\), and decreases at higher harmonic orders, indicating that the integration of the aerodynamic loading also plays an important role. Actually, an accurate prediction may require a suitable representation of the chordwise pressure variation, particularly when strong loading gradients make a purely compact sectional description inadequate \citep{Bernardini2016}. For Propeller A, the largest discrepancies tend to occur at the higher retained harmonics, where phase variations over the blade surface are more pronounced. For Propeller B, the maximum error can instead occur at lower harmonics, depending on the operating condition and on whether the chordwise or radial resolution is varied. Based on these results, a \(N_c=40, N_r=20\) surface discretization is adopted in the following, since it is the coarsest mesh that keeps the harmonic-power error below \(0.5~\mathrm{dB}\) over the entire frequency range considered.}

\deleted[id=FF]{The analysis was carried out for the most demanding case, namely Propeller A at \(M_t=0.8\), which combines the largest chord ratio with the shortest acoustic wavelength. For this configuration, the far-field directivity defined by Eq. \eqref{directivity definition} in Appendix \ref{appB} was computed retaining harmonics up to \(m=20\), , and \(100\) values of \(\ell\) for each harmonic. Starting from a very fine mesh, the discretization was progressively reduced until the mean difference in directivity level with respect to the finer mesh, evaluated over the polar angle \(\theta\), remained below \(0.01\) dB. The final surface representation employs \(34\) chordwise points on each side of the airfoil section, for a total of \(68\) points over the suction and pressure surfaces, with clustering near the leading edge, together with \(30\) uniformly spaced radial stations between hub and tip. The surface integral in Eq. \eqref{eq:Alm-start} is then evaluated over the corresponding blade surface obtained by connecting the sectional airfoils along the span.}

For both propellers and operating conditions, the multipole coefficients have been evaluated \added{and the values} for the first harmonics \deleted[id=FF]{and}\added{are} reported in tables \ref{tab:almabs_exact_Mt_0.2_a_2}-\ref{tab:almabs_exact_Mt_0.8_a_10}. \added{For a fixed azimuthal order $m$, the spherical harmonics alternate in symmetry as the degree $\ell$ increases. Indeed, reflection across the rotor plane ($z_0=0$) gives $Y_{\ell m}(\pi-\theta,\phi)=(-1)^{\ell-m}Y_{\ell m}(\theta,\phi)$ \citep[p. 141, Eq. 4]{Varshalovich1988}. Modes with $\ell-m$ even are therefore symmetric with respect to the rotor plane, whereas those with $\ell-m$ odd are antisymmetric. Accordingly, the relative importance of successive multipole coefficients should be assessed within the same symmetry family, and this distinction is taken into account in the following discussion.} A first important outcome of these calculations is the \deleted[id=FF]{very} rapid decay of the expansion with respect to the index \(\ell\). Tables show that, for each azimuthal order \(m\), the dominant part of the acoustic field is concentrated at \deleted[id=FF]{the} most in the first two admissible degrees, namely \(\ell=|m|\) and \(\ell=|m|+1\)\deleted[id=FF]{, while the coefficients associated with \(\ell\ge |m|+2\) are already much smaller}. \added{The following terms, \(\ell\ge |m|+2\), are generally much smaller, although their relative importance may increase at higher tip Mach number and higher harmonic order.} This means that the leading symmetric ($\ell=|m|$) and antisymmetric ($\ell=|m|+1$) contributions with respect to the rotor plane \deleted[id=FF]{($z_0=0$)} are generally sufficient to capture the essential angular structure of the radiation. 

A clear difference emerges between the two propellers. For Propeller A, the multipole content is strongly dominated by the first admissible term, \(\ell=|m|\), at both Mach numbers. At \(M_t=0.2\), table \ref{tab:almabs_exact_Mt_0.2_a_2} shows that the coefficient at \(\ell=|m|\) is systematically the largest one, while the next order, \(\ell=|m|+1\), is already smaller by roughly \deleted[id=FF]{one} \added{two} order\added{s} of magnitude\deleted[id=FF]{, or more}. The higher orders are then much weaker still, so that the decay is essentially monotonic and very rapid with increasing \(\ell\). The same structure remains visible at \(M_t=0.8\), see table \ref{tab:almabs_exact_Mt_0.8_a_2}, although the decay becomes less steep and the contribution of \(\ell=|m|+2\) becomes more noticeable, especially at the higher harmonics. Nevertheless, the leading term \(\ell=|m|\) remains the dominant one throughout. Propeller B exhibits a qualitatively different behaviour. At \(M_t=0.2\), table \ref{tab:almabs_exact_Mt_0.2_a_10} shows that the first two admissible multipoles, \(\ell=|m|\) and \(\ell=|m|+1\), are both important, with \(\ell=|m|+1\) remaining within the same order of magnitude as the leading term. By contrast, the following pair, \(\ell=|m|+2\) and \(\ell=|m|+3\), is much smaller. This establishes a two-by-two structure in the decay with \(\ell\): the first pair \((\ell=|m|,\ell=|m|+1)\) dominates, while the second pair \((\ell=|m|+2,\ell=|m|+3)\) is significantly weaker. The same pattern persists at \(M_t=0.8\), as reported in table \ref{tab:almabs_exact_Mt_0.8_a_10}. In this case the first two multipoles become even more comparable, and for \(m=2\) the coefficient at \(\ell=|m|+1\) is slightly larger than that at \(\ell=|m|\). The next two orders remain clearly smaller.

A second feature visible in the tables is the behaviour with increasing harmonic index \(m\). At fixed \(\ell-|m|\), the coefficients generally decrease as \(m\) increases when \(M_t=0.2\). Conversely, at \(M_t=0.8\), the coefficients grow initially before decreasing at higher frequencies. The dependence on \(m\) reflects the frequency content of the subsonic propeller, a behaviour that is well documented in the experimental literature \citep{Trebble1987a,Trebble1987b}.

\begin{table}
\centering
\def~{\hphantom{0}}
\begin{tabular}{ccccc}
$\ell-|m|$ & $m=2$ & $m=4$ & $m=6$ & $m=8$ \\ \hline
0 & $3.147\times 10^{-1}$ & $6.541\times 10^{-2}$ & $7.889\times 10^{-3}$ & $7.538\times 10^{-4}$ \\
1 & $7.520\times 10^{-3}$ & $6.259\times 10^{-4}$ & $4.367\times 10^{-5}$ & $3.153\times 10^{-6}$ \\
2 & $2.766\times 10^{-4}$ & $1.486\times 10^{-4}$ & $2.546\times 10^{-5}$ & $2.936\times 10^{-6}$ \\
3 & $1.201\times 10^{-5}$ & $2.085\times 10^{-6}$ & $2.034\times 10^{-7}$ & $1.785\times 10^{-8}$ \\
\end{tabular}
\caption{Leading multipoles \(|A_{\ell m}| \added{[\mathrm{Pa}]}\) obtained from Eq. \eqref{eq:Alm-start} for Propeller A at \(M_t=0.2\). \inlinecomment{This table has been updated in the revised manuscript.}}
\label{tab:almabs_exact_Mt_0.2_a_2}
\end{table}

\begin{table}
\centering
\def~{\hphantom{0}}
\begin{tabular}{ccccc}
$\ell-|m|$ & $m=2$ & $m=4$ & $m=6$ & $m=8$ \\ \hline
0 & $3.916\times 10^{-2}$ & $5.702\times 10^{-3}$ & $6.506\times 10^{-4}$ & $6.626\times 10^{-5}$ \\
1 & $1.285\times 10^{-2}$ & $1.251\times 10^{-3}$ & $9.539\times 10^{-5}$ & $6.790\times 10^{-6}$ \\
2 & $4.472\times 10^{-5}$ & $1.236\times 10^{-5}$ & $1.952\times 10^{-6}$ & $2.409\times 10^{-7}$ \\
3 & $2.296\times 10^{-5}$ & $4.485\times 10^{-6}$ & $4.677\times 10^{-7}$ & $3.997\times 10^{-8}$ \\
\end{tabular}
\caption{Leading multipoles \(|A_{\ell m}| \added{[\mathrm{Pa}]}\) obtained from Eq. \eqref{eq:Alm-start} for Propeller B at \(M_t=0.2\). \inlinecomment{This table has been updated in the revised manuscript.}}
\label{tab:almabs_exact_Mt_0.2_a_10}
\end{table}

\begin{table}
\centering
\def~{\hphantom{0}}
\begin{tabular}{ccccc}
$\ell-|m|$ & $m=2$ & $m=4$ & $m=6$ & $m=8$ \\ \hline
0 & $2.868\times 10^{2}$ & $7.573\times 10^{2}$ & $1.108\times 10^{3}$ & $1.254\times 10^{3}$ \\
1 & $2.611\times 10^{1}$ & $3.140\times 10^{1}$ & $3.029\times 10^{1}$ & $2.876\times 10^{1}$ \\
2 & 2.952 & $2.636\times 10^{1}$ & $5.936\times 10^{1}$ & $8.518\times 10^{1}$ \\
3 & $6.931\times 10^{-1}$ & 1.755 & 2.390 & 2.807 \\
\end{tabular}
\caption{Leading multipoles \(|A_{\ell m}| \added{[\mathrm{Pa}]}\) obtained from Eq. \eqref{eq:Alm-start} for Propeller A at \(M_t=0.8\). \inlinecomment{This table has been updated in the revised manuscript.}}
\label{tab:almabs_exact_Mt_0.8_a_2}
\end{table}

\begin{table}
\centering
\def~{\hphantom{0}}
\begin{tabular}{ccccc}
$\ell-|m|$ & $m=2$ & $m=4$ & $m=6$ & $m=8$ \\ \hline
0 & $3.089\times 10^{1}$ & $6.106\times 10^{1}$ & $8.921\times 10^{1}$ & $1.119\times 10^{2}$ \\
1 & $5.786\times 10^{1}$ & $7.479\times 10^{1}$ & $7.290\times 10^{1}$ & $6.479\times 10^{1}$ \\
2 & $4.944\times 10^{-1}$ & 1.905 & 4.219 & 6.827 \\
3 & 1.683 & 4.524 & 6.207 & 6.782 \\
\end{tabular}
\caption{Leading multipoles \(|A_{\ell m}| \added{[\mathrm{Pa}]}\) obtained from Eq. \eqref{eq:Alm-start} for Propeller B at \(M_t=0.8\). \inlinecomment{This table has been updated in the revised manuscript.}}
\label{tab:almabs_exact_Mt_0.8_a_10}
\end{table}

\begin{figure}
\centering
\subfloat[Propeller A, $M_t=0.2$.]
{\includegraphics[width=.5\textwidth]{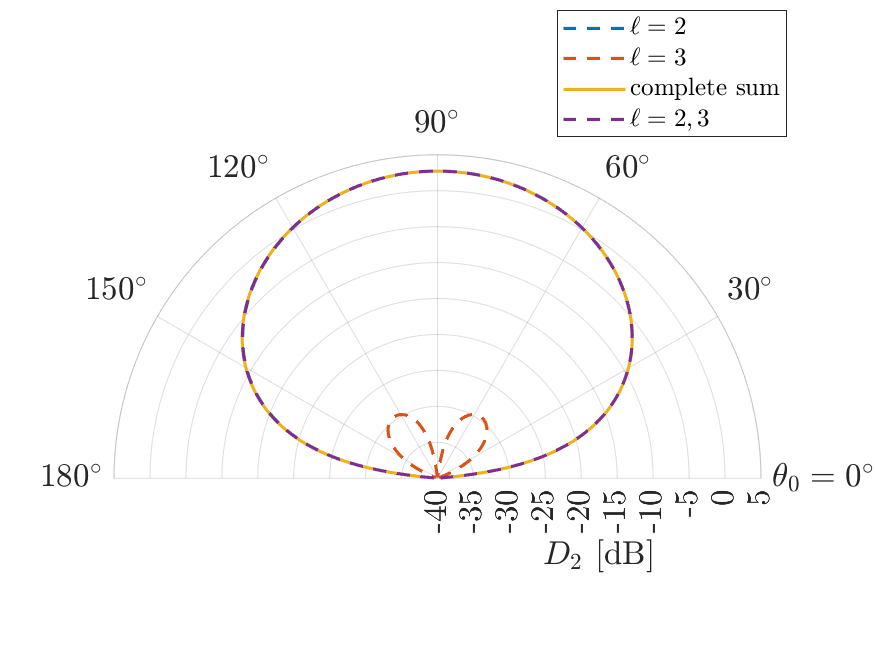}\label{fig:m2_decomp_A02}}\hfill
\subfloat[Propeller B, $M_t=0.2$.]
{\includegraphics[width=.5\textwidth]{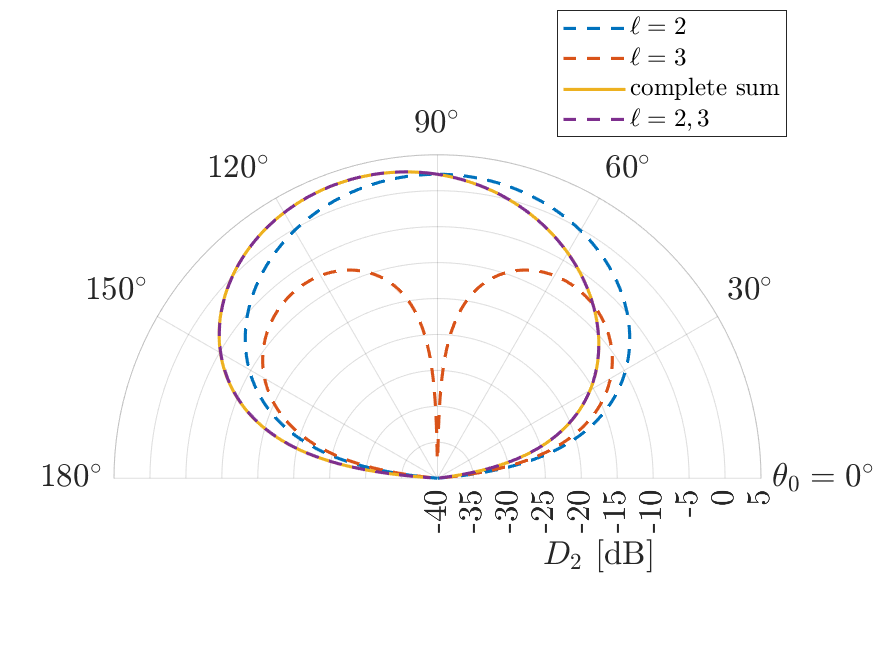}\label{fig:m2_decomp_B02}}\hfill
\subfloat[Propeller A, $M_t=0.8$.]
{\includegraphics[width=.5\textwidth]{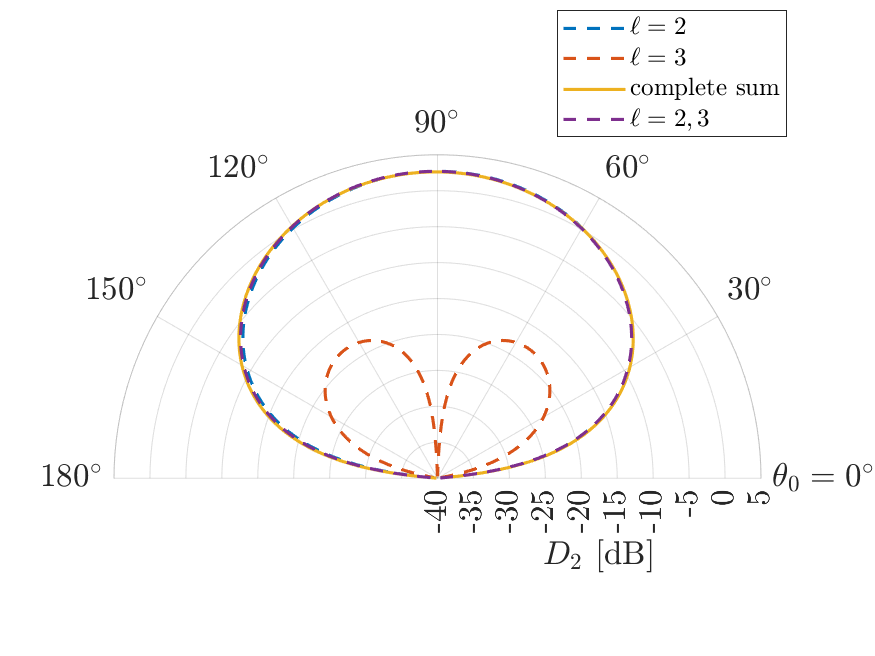}\label{fig:m2_decomp_A08}}\hfill
\subfloat[Propeller B, $M_t=0.8$.]
{\includegraphics[width=.5\textwidth]{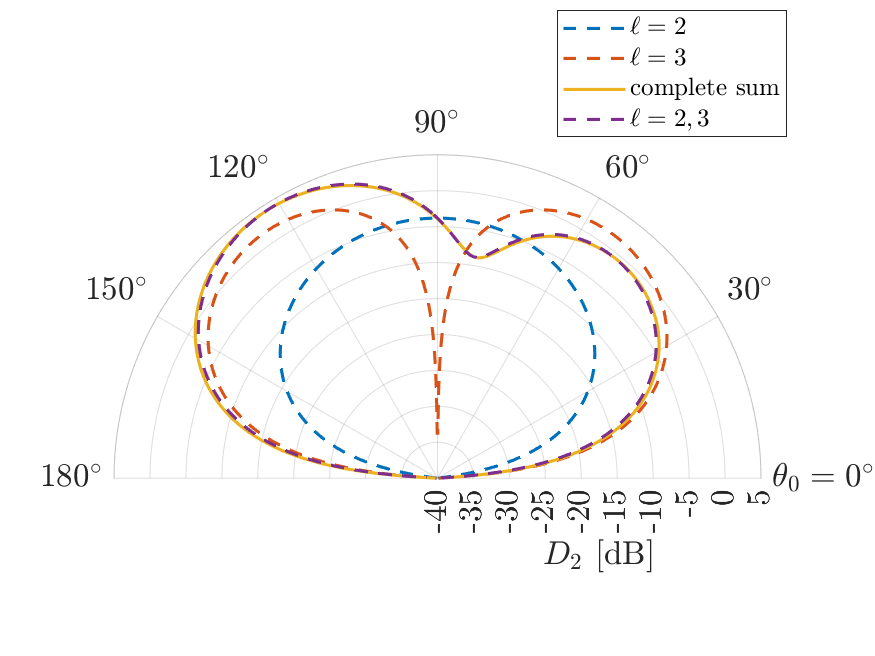}\label{fig:m2_decomp_B08}}\hfill
\caption{Decomposition of the harmonic directivity $D_m$ [dB] for the first non-zero harmonic \(m=2\). The curves show the isolated contributions of the first two degrees, \(\ell=2\) and \(\ell=3\), their coherent sum, and the complete harmonic directivity obtained from the full multipole expansion. \inlinecomment{This figure has been updated in the revised manuscript.}}
\label{fig:m2_decomposition}
\end{figure}

To make the previous discussion clearer in angular terms, it is useful to examine the directivity of a single harmonic $m$ while retaining only the first $\ell$ contributions. For a fixed harmonic index \(m\), we define the harmonic directivity (see Eq. \eqref{m directivity} in Appendix \ref{appB}) as
\begin{equation}
D_m(\theta_0,\phi_0)
=
\frac{4\pi
\left|\displaystyle\sum_{\ell=|m|}^{\infty} (-i)^{\ell+1}\,A_{\ell m}\,Y_{\ell m}(\theta_0,\phi_0)\right|^2}
{\displaystyle\sum_{\ell=|m|}^{\infty} \left|A_{\ell m}\right|^2}
\nonumber
\end{equation}
It represent\added{s} the acoustic power radiated per unit solid angle in the direction \((\theta_0,\phi_0)\), normalized by the \deleted[id=FF]{total}\added{average} power radiated by that harmonic. For the first non-zero rotor harmonic of the present two-bladed propellers, namely \(m=2\), the first two admissible degrees are \(\ell=2\) and \(\ell=3\). Figure \ref{fig:m2_decomposition} shows the corresponding isolated contributions to $D_m$, their coherent sum, and the complete harmonic directivity reconstructed from the full multipole expansion. From this figure emerges that in all four cases the curve obtained by retaining only \(\ell=2\) and \(\ell=3\) is almost indistinguishable from the complete harmonic directivity. This provides direct angular evidence that the first two multipoles already contain essentially all the information needed to reconstruct the \deleted[id=FF]{exact} far-field pattern. Higher $\ell$-orders therefore play only a marginal role, even though their coefficients are not identically zero. The two leading multipoles have clearly different angular signatures. The actual directivity is obtained by coherently summing the underlying complex modal amplitudes. The difference between the isolated \(\ell=2\) and \(\ell=3\) contributions and their coherent combination is therefore caused by phase interference. In particular, the main lobe of the harmonic directivity is generally displaced toward \(\theta_0>90^\circ\), and this displacement is correctly reproduced by the coherent sum of the first two multipoles. Both the direction and magnitude of the shift are thus controlled primarily by the relative amplitude and phase of \(A_{22}\) and \(A_{32}\). The relative importance of the two leading multipoles depends on the acoustic compactness, \deleted[id=FF]{and} on the blade geometry and on the operating conditions. The difference between the two propellers is even more evident in this decomposition. At \(M_t=0.2\), Propeller A, shown in panel \ref{fig:m2_decomp_A02}, is clearly dominated by the first admissible multipole, \(\ell=2\). By contrast, Propeller B at the same Mach number, panel \ref{fig:m2_decomp_B02}, exhibits a markedly different behaviour. In this case, the \(\ell=2\) and \(\ell=3\) contributions are both important, and neither of them alone is sufficient to reproduce the complete directivity. The same distinction persists at \(M_t=0.8\). For Propeller A, panel \ref{fig:m2_decomp_A08}, the \(\ell=2\) term remains the leading contribution, although the \(\ell=3\) correction becomes more visible than at low Mach number. For Propeller B, panel \ref{fig:m2_decomp_B08}, the first two multipoles are comparable in importance, and their interference play\added{s} a major role in shaping the directivity. The agreement between the complete sum and the partial sum retaining only \(\ell=2,3\) confirms that the first two admissible multipoles contain the essential structure of the \(m=2\) radiation pattern, but with a markedly different balance in the two propellers. A physical interpretation of the mechanisms generating these first two multipoles will be given in the following section, through suitable approximations of the surface integral. This will clarify the origin of the different behaviours observed for the two propellers.

\begin{figure}
\centering
\subfloat[Propeller A, $M_t=0.2$.]
{\includegraphics[width=.5\textwidth]{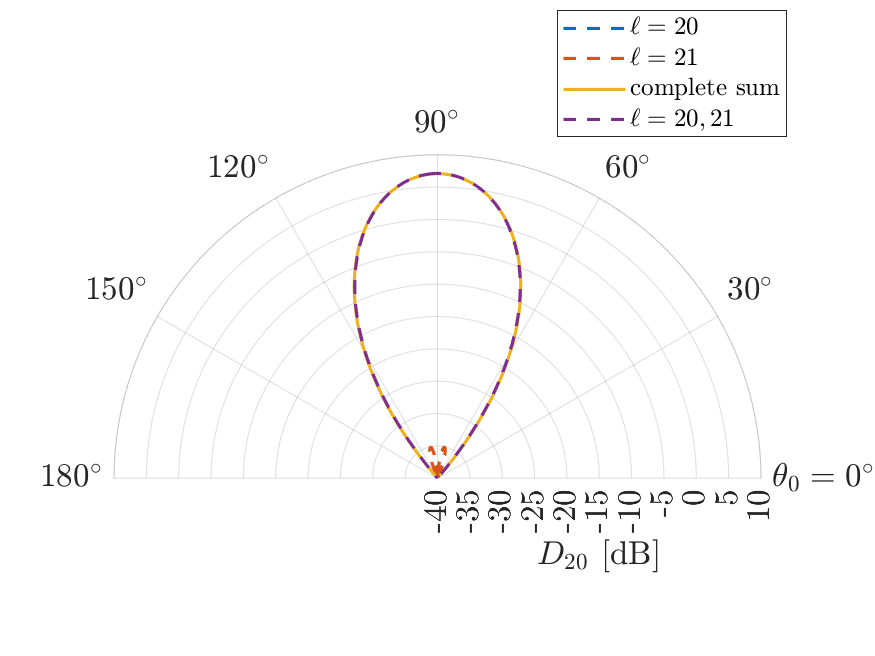}\label{fig:m20_decomp_A02}}\hfill
\subfloat[Propeller B, $M_t=0.2$.]
{\includegraphics[width=.5\textwidth]{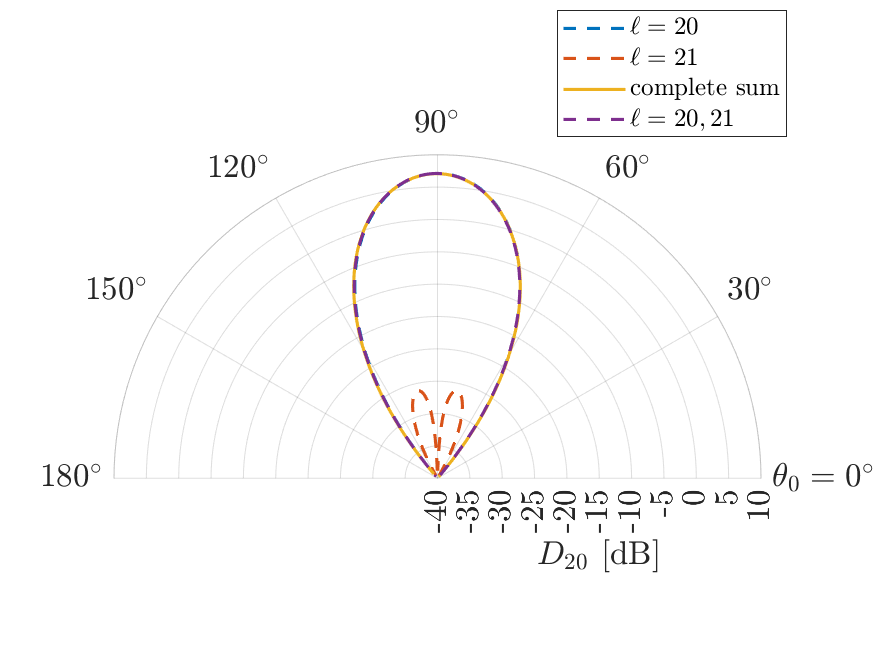}\label{fig:m20_decomp_B02}}\hfill
\subfloat[Propeller A, $M_t=0.8$.]
{\includegraphics[width=.5\textwidth]{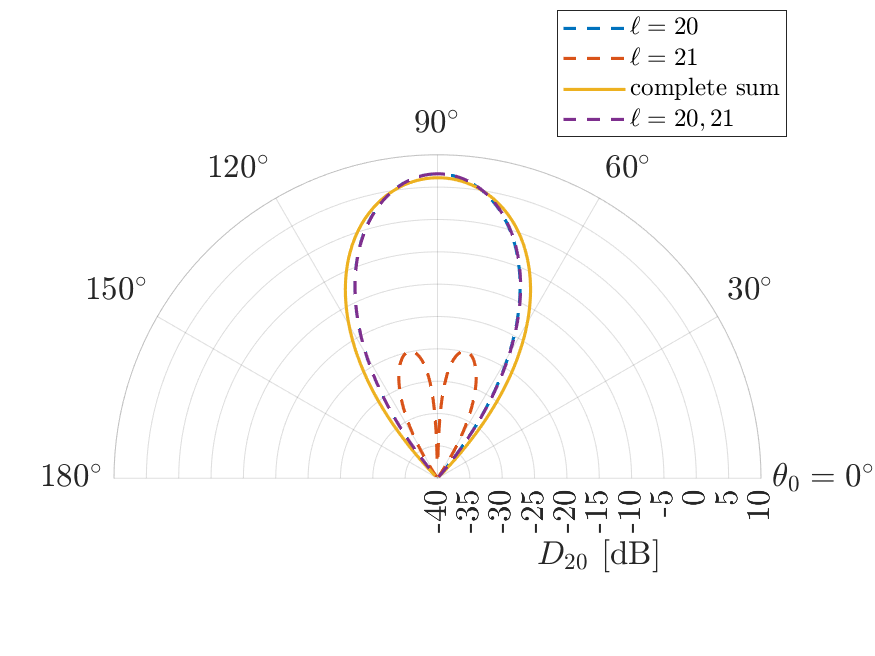}\label{fig:m20_decomp_A08}}\hfill
\subfloat[Propeller B, $M_t=0.8$.]
{\includegraphics[width=.5\textwidth]{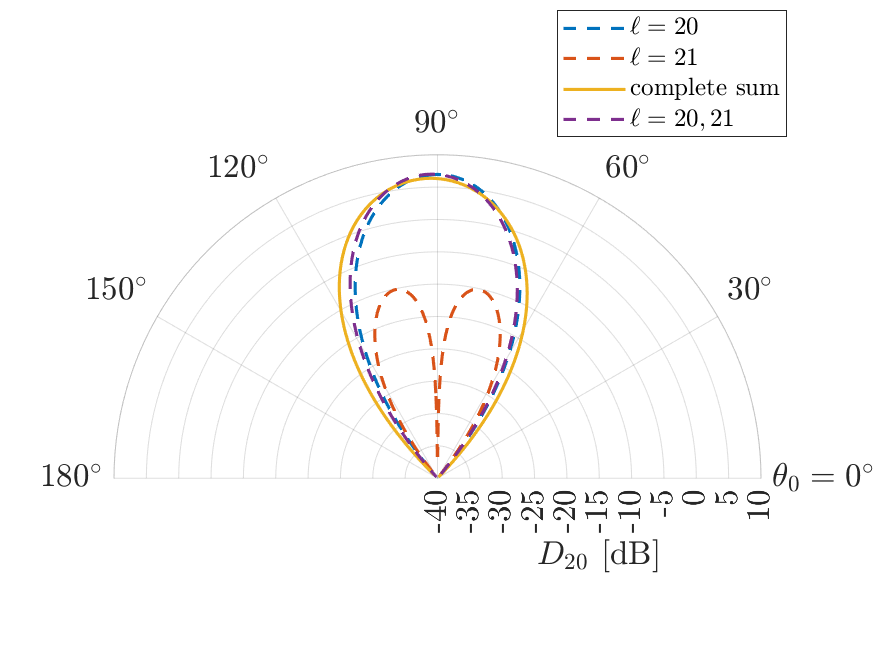}\label{fig:m20_decomp_B08}}\hfill
\caption{Decomposition of the harmonic directivity $D_m$ [dB] for the tenth non-zero harmonic \(m=20\). The curves show the isolated contributions of the first two degrees, \(\ell=20\) and \(\ell=21\), their coherent sum, and the complete harmonic directivity obtained from the full multipole expansion.  \inlinecomment{This figure has been added in the revised manuscript.}}
\label{fig:m20_decomposition}
\end{figure}

\added{As seen from the tables above, the higher harmonics provide a more stringent test of the truncation since the relative contribution of the higher-degree terms tends to increase, especially at \(M_t=0.8\). The same decomposition of the directivity is therefore repeated for the highest retained harmonic, \(m=20\), and the results are reported in figure \ref{fig:m20_decomposition}. Despite this more demanding condition, the coherent sum of the first two admissible degrees, \(\ell=20\) and \(\ell=21\), remains very close to the complete harmonic directivity in all four cases. For Propeller A, the behaviour is again almost entirely controlled by the leading term \(\ell=20\), especially at \(M_t=0.2\), while the \(\ell=21\) contribution remains much weaker. For Propeller B, the \(\ell=21\) contribution is more visible and modifies the shape of the coherent pattern. However, the partial sum retaining only the $\ell=20$ and $\ell=21$ contributions still reproduces the complete result with very good accuracy, particularly near the direction of maximum intensity.}

\added{The angular structure of the two leading multipoles can be characterized more explicitly. For a given harmonic order $m$, the leading symmetric and antisymmetric contributions are associated with the spherical harmonics \citep[p. 157, Eqs. 12-13]{Varshalovich1988}
\begin{equation}
Y_{mm}(\theta_0,\phi_0)
\propto
(\sin\theta_0)^m e^{im\phi_0},
\qquad
Y_{m+1,m}(\theta_0,\phi_0)
\propto
\cos\theta_0\,(\sin\theta_0)^m e^{im\phi_0}
\end{equation}
The corresponding isolated harmonic directivities are therefore proportional to the squared moduli of these functions. The symmetric contribution has a single maximum in the rotor plane, at $\theta_0=\pi/2$. By contrast, the leading antisymmetric contribution vanishes both in the rotor plane and along the rotation axis, and exhibits two maxima symmetrically located on either side of the rotor plane. Differentiation with respect to $\theta_0$ gives the two directions of maximum intensity of the antisymmetric contribution as
\begin{equation}
\theta_{0,\max}^{-}
=
\cos^{-1}\left(\frac{1}{\sqrt{m+1}}\right),
\qquad
\theta_{0,\max}^{+}
=
\pi-
\cos^{-1}\left(\frac{1}{\sqrt{m+1}}\right)
\end{equation}
For the first non-zero harmonic of the present two-bladed propellers, $m=2$, these maxima occur at $\theta_0\approx54.7^\circ$ and $\theta_0\approx125.3^\circ$, in agreement with the isolated $\ell=3$ contribution shown in figure \ref{fig:m2_decomposition}. For $m=20$, the maxima shift to $\theta_0\approx77.4^\circ$ and $\theta_0\approx102.6^\circ$, consistently with the narrower angular distribution of the isolated $\ell=21$ contribution shown in figure \ref{fig:m20_decomposition}. Thus, as the harmonic order increases, the antisymmetric lobes become progressively concentrated around the rotor plane. The complete harmonic directivity, however, does not generally attain its maxima at these angles, since it results from the coherent superposition of the symmetric and antisymmetric multipoles.}

\subsection{Compactness}

\inlinecomment{This subsection has been added for clarity.}

The rapid decay of higher-order $\ell$ can be understood from the source-side spherical Bessel function \(j_\ell\), which enters the factor \(g_{\ell m}\) in Eq. \eqref{f and g definitions}. For a given harmonic \(m\), the argument of the spherical Bessel function is \(mM_t\bar r\), where \(\bar r=r/R\) is the non-dimensional spanwise coordinate. When this quantity is small, the small-argument asymptotics gives \citep[Eq. 16.12]{jackson1962}
\begin{equation}
j_\ell(mM_t\bar r)\approx \frac{(mM_t\bar r)^\ell}{(2\ell+1)!!} \added{\left(
1-\frac{(mM_t\bar r)^2}{2(2\ell+3)}
\right)},
\qquad
mM_t\bar r\ll \ell 
\label{bessel approx}
\end{equation}
This expression explains why large values of \(\ell\) are difficult to excite. Since \(\bar r\le 1\), the relevant parameter controlling the excitation of higher multipoles is essentially \(mM_t\). \added{Increasing either the harmonic order or the tip Mach number makes the neglected terms less strongly suppressed, whereas reducing \(M_t\) rapidly concentrates the expansion in the first admissible $\ell$-degrees.} \deleted[id=FF]{In particular, because the first admissible degree is \(\ell=|m|\), the condition \(mM_t\bar r\ll \ell\) is satisfied whenever \(M_t\ll 1\). It follows that, as the tip Mach number decreases, the higher $\ell$-orders are increasingly suppressed and the expansion becomes dominated by the first values of \(\ell\).} \added{For the \(m\)-th harmonic, the condition \(mM_t = k_m R\ll1\) corresponds to acoustic compactness of the entire rotor at that harmonic. In this limit, the condition \(mM_t\bar r\ll\ell\) is always satisfied, so that the source cannot efficiently excite high-order spatial moments. This clarifies why, for an acoustic compact source, the radiated field is dominated by the lowest admissible multipoles. This may be interpreted as the spherical-mode counterpart of the compactness argument commonly used in the literature to simplify radiation integrals. Indeed, it will be shown in the next section that acoustic compactness is closely connected with the possibility of replacing the blade surface by a lifting-line representation. Moreover, the results in the previous subsection show that also marginal compact sources, i.e. in conditions $M_t = k R \lesssim 1$, are predicted accurately by the first multipoles. Indeed, in the directivity results shown in figures \ref{fig:m2_decomposition} and \ref{fig:m20_decomposition}, $k_m R$ ranges from 0.4 to 16, corresponding to conditions that are not acoustically compact.} 

\begin{figure}
\centering
\subfloat[$M_t=0.2$.]
{\includegraphics[width=.5\textwidth]{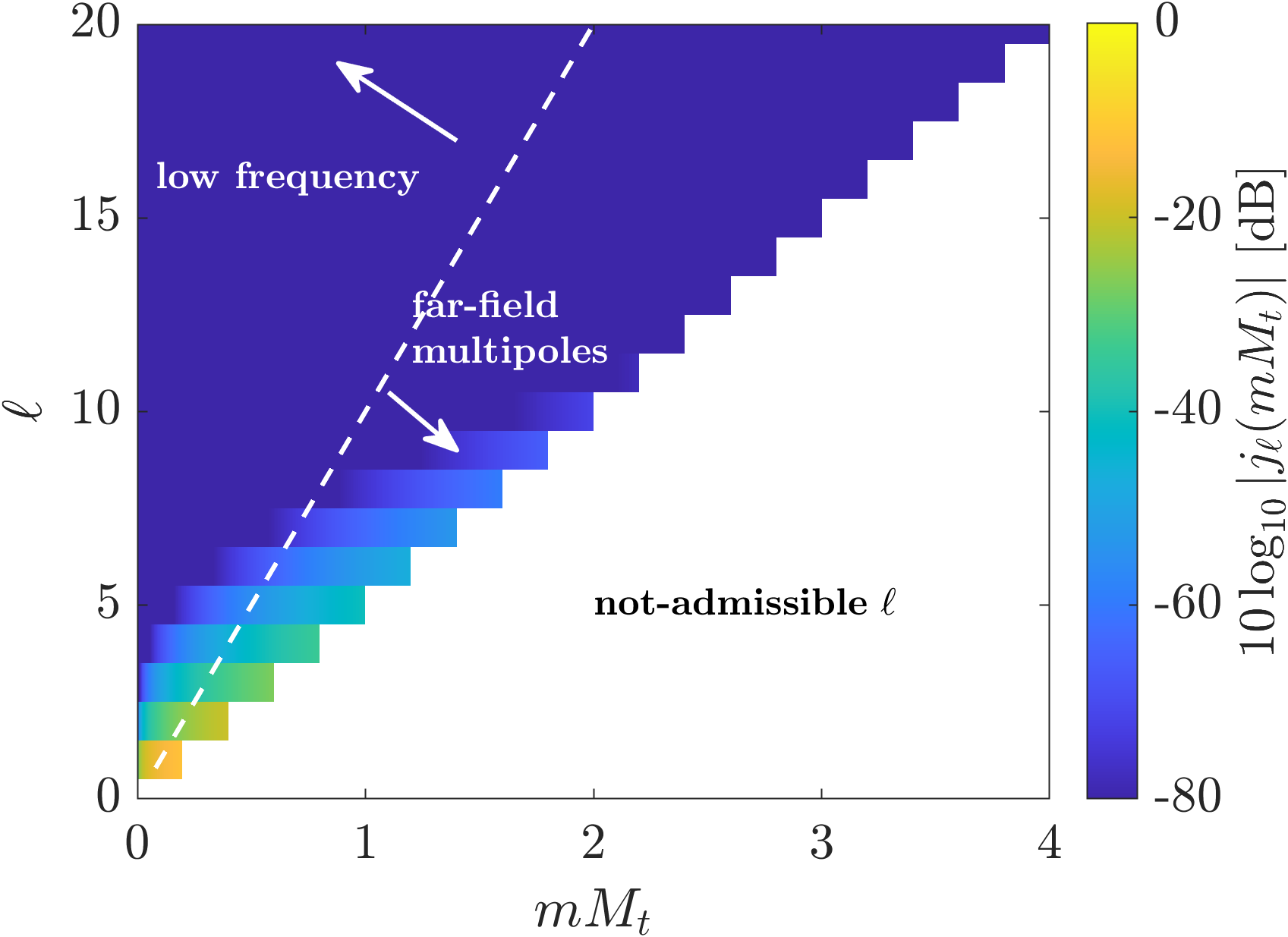}\label{bessel02}}\hfill
\subfloat[$M_t=0.8$.]
{\includegraphics[width=.5\textwidth]{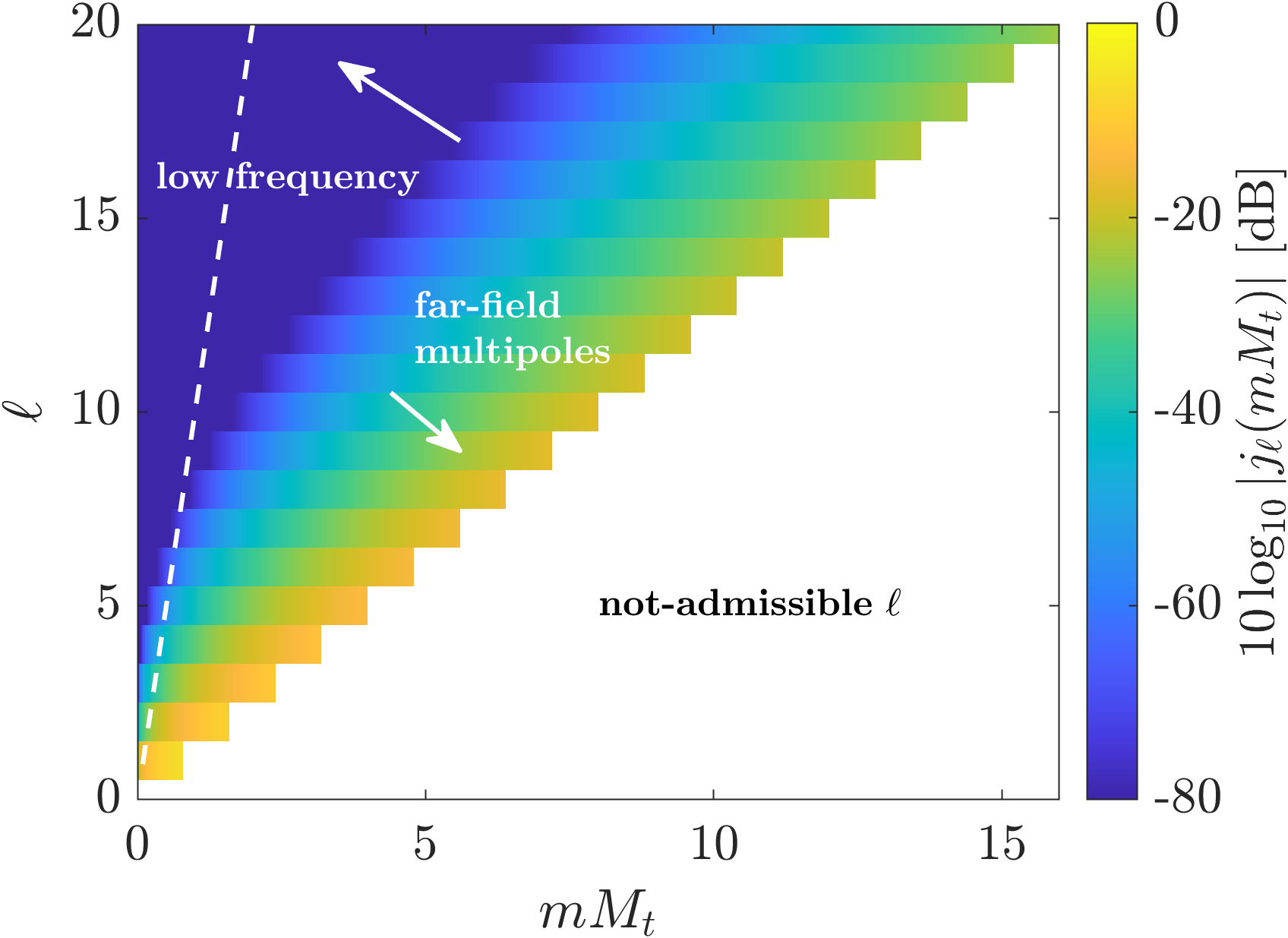}\label{bessel08}}\hfill
\caption{Colour map of \(10\log_{10}|j_\ell(mM_t)|\) [dB] in the \((mM_t,\ell)\)-plane. The region \(\ell<m\) is not admissible. The \deleted[id=FF]{dashed}\added{dotted} line is \(\ell=10\,mM_t\). \inlinecomment{This figure has been updated in the revised manuscript.}}
\label{fig:regime_map_bessel_MtXX}
\end{figure}

Th\deleted[id=FF]{is}\added{e} effect \added{of the rapid decay} on the radiation is especially clear in the far field. Indeed, as far as Eq. \eqref{eq:pressure-formula-FF} is concerned, the \(\ell\)-dependence of the solution is governed mainly by the source coefficients \(A_{\ell m}\). The attenuation of the higher multipole degrees, inherited from the source-side Bessel function entering those coefficients, is reflected in the far-field behaviour. The mechanism is illustrated in figure \ref{fig:regime_map_bessel_MtXX}, which shows the magnitude of \(j_\ell(mM_t)\) in the \((mM_t,\ell)\)-plane. The region \(\ell<m\) is excluded because it is not admissible for spherical harmonics of order \(m\). The dotted line corresponds to \(\ell=10\,mM_t\). Above this line, one is safely in the low-frequency regime \(mM_t\ll \ell\), where the Bessel function is already strongly attenuated. This explains why convergence is much faster at \(M_t=0.2\) (panel \ref{bessel02}) than at \(M_t=0.8\) (panel \ref{bessel08}): at lower tip Mach number the admissible multipoles lie deeper inside the attenuated region of the map, whereas at higher tip Mach number the first few orders remain closer to the active region and therefore play a more important role. The same diagram also helps interpret the far-field condition. Since the far-field approximation requires $\ell \ll mM_t\bar r_0$, the multipoles lying below the dotted line automatically satisfy this requirement for a distant observer, for example \(\bar r_0\approx 100\). In this sense, the region below the dotted curve may be viewed as containing the multipoles that can both be efficiently generated at the source and propagate into the far field for the observer distances considered here.

The rapid decay of the multipole series has an important computational implication. Once the coefficients \(A_{\ell m}\) have been computed, the acoustic field at any observer position is obtained by evaluating only the observer-side functions \(f_{\ell m}(\mathbf{x}_0)\). The cost of these computations therefore depends on the number of retained multipoles rather than on repeated blade integrations. \deleted[id=FF]{As shown previously in Fruncillo, this source/observer separation can yield substantial savings with respect to formulations in which the observer dependence remains embedded inside the source integrals (Hanson).}\added{This source-observer separation was examined quantitatively in a previous work \citep{fruncillo2025}, where computations over increasing numbers of microphone locations showed savings of up to approximately two orders of magnitude relative to Hanson’s formulation \citep{Hanson1980}. The same observer-side scaling applies to the present framework, since the spherical expansion of the Green’s function and the subsequent field reconstruction are unchanged, while the present work differs in the source representation and in the derivation of the corresponding multipole coefficients.}

\section{Approximations of the Surface Integral}
\label{sec:Approximations}

The proposed formulation provides the computation of the multipole coefficients\added{, by integrating over} \deleted[id=FF]{on} the \added{blade} surface \deleted[id=FF]{geometry} through Eq. \eqref{eq:Alm-start}. However, in that form the physical content of the source remains embedded in a surface integral. Suitable approximations of the geometry are therefore valuable because they make the structure of the acoustic source more explicit. In particular, they can reveal how the different geometric and aerodynamic information enter the leading multipole coefficients, clarifying their symmetry and phase properties, and providing orderings that explain which mechanisms dominate in different operating regimes. \added{As discussed above, the present framework can take advantage of any suitable approximation applied to the representation and integration of the source terms. The following analysis focuses on two commonly used source-side approximations that are particularly useful for interpreting the physical nature of the resulting multipoles.}

\subsection{Lifting surface}
\label{sec:Lifting Surface}
\begin{figure}
\centering
\def\svgwidth{0.8\linewidth} 
\begingroup%
  \makeatletter%
  \providecommand\color[2][]{%
    \errmessage{(Inkscape) Color is used for the text in Inkscape, but the package 'color.sty' is not loaded}%
    \renewcommand\color[2][]{}%
  }%
  \providecommand\transparent[1]{%
    \errmessage{(Inkscape) Transparency is used (non-zero) for the text in Inkscape, but the package 'transparent.sty' is not loaded}%
    \renewcommand\transparent[1]{}%
  }%
  \providecommand\rotatebox[2]{#2}%
  \newcommand*\fsize{\dimexpr\f@size pt\relax}%
  \newcommand*\lineheight[1]{\fontsize{\fsize}{#1\fsize}\selectfont}%
  \ifx\svgwidth\undefined%
    \setlength{\unitlength}{503.98677315bp}%
    \ifx\svgscale\undefined%
      \relax%
    \else%
      \setlength{\unitlength}{\unitlength * \real{\svgscale}}%
    \fi%
  \else%
    \setlength{\unitlength}{\svgwidth}%
  \fi%
  \global\let\svgwidth\undefined%
  \global\let\svgscale\undefined%
  \makeatother%
  \begin{picture}(1,0.31851468)%
    \lineheight{1}%
    \setlength\tabcolsep{0pt}%
    \put(0,0){\includegraphics[width=\unitlength,page=1]{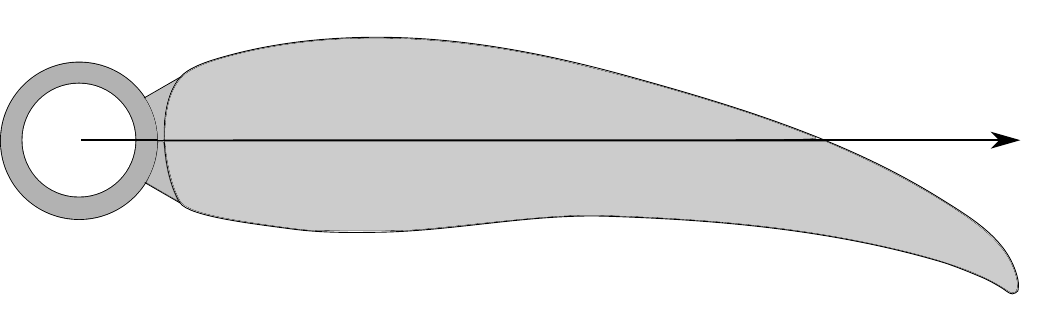}}%
    \put(0.0459974,0.20673261){\color[rgb]{0,0,0}\makebox(0,0)[lt]{\lineheight{1.25}\smash{\begin{tabular}[t]{l}$\Omega$   \\\end{tabular}}}}%
    \put(0.71615857,0.02077157){\color[rgb]{0,0,0}\makebox(0,0)[lt]{\lineheight{1.25}\smash{\begin{tabular}[t]{l}trailing edge\end{tabular}}}}%
    \put(0.69603453,0.30341574){\color[rgb]{0,0,0}\makebox(0,0)[lt]{\lineheight{1.25}\smash{\begin{tabular}[t]{l}leading edge\end{tabular}}}}%
    \put(0,0){\includegraphics[width=\unitlength,page=2]{liftingsurface.pdf}}%
    \put(0.70118154,0.14329587){\color[rgb]{0,0,0}\makebox(0,0)[lt]{\lineheight{1.25}\smash{\begin{tabular}[t]{l}$\phi'_{TE}$\\\end{tabular}}}}%
    \put(0.96322305,0.20096761){\color[rgb]{0,0,0}\makebox(0,0)[lt]{\lineheight{1.25}\smash{\begin{tabular}[t]{l}$x'$\end{tabular}}}}%
    \put(0,0){\includegraphics[width=\unitlength,page=3]{liftingsurface.pdf}}%
  \end{picture}%
\endgroup%

\caption{Lifting-surface geometry. The \(x'\)-axis is attached to the rotor, passing through the mid-chord of a reference blade section. The trailing-edge position of each section is then described by the azimuthal offset \(\phi'_{TE}\) measured from this line, with counter-clockwise angles taken as positive (negative value in the present example).}
\label{liftingsurface geometry}
\end{figure}

In this section we specialise the general expression in Eq. \eqref{eq:Alm-start} to the lifting-surface case. Since the earliest studies of propeller noise \citep{Gutin1936,Garrick1954}, this has been among the most widely adopted modelling approaches. The main benefit is to represent the source field on a geometry that is simpler than the physical one. Here we consider the case in which the reference surface is collapsed onto the equatorial plane, described in rotating coordinates $(r,\phi')$ at $\theta=\pi/2$ \added{(the primes on $r$ and $\theta$ are omitted hereafter)}. This assumption is commonly used as a starting point, whereby the propeller surface is approximated by its planform \citep{Carley1999,glegg2024}. Under this approximation, Eq. \eqref{eq:Alm-start} reduces to
\begin{equation}
A^{(LS)}_{\ell m}
 = 
Nik_m\int_{R_{hub}}^{R}\int_0^{2\pi}
\left[
\mathbf f(r,\phi')\cdot(\nabla_{\mathbf{x}'}g_{\ell m})_0
- i \rho_0 a_s k_m V_{n}(r,\phi')    g_{\ell m,0}
\right] 
r  dr  d\phi'
\label{eq:Alm1}
\end{equation}
where the subscript $0$ denotes evaluation on the plane at $(r,\phi')$. This approximation is consistent with the classical two-dimensional thin-airfoil theory \citep{Ashley1965}. Geometrically, the reference surface lies in the plane of the relative-velocity vector; a small deviation from this plane therefore corresponds to a small aerodynamic incidence too. At a fixed radial station $r$, the blade section is modelled as a two-dimensional airfoil, whose upper and lower surfaces are described using a dimensionless chordwise coordinate. Although the $\phi'$-integration in Eq. \eqref{eq:Alm1} is written over $[0,2\pi]$, the integrand is non-zero only over the chordwise extent of the blade at each radius. Denoting by $\phi'_{TE}(r)$ the (co-rotating) azimuth\added{al curve describing} \deleted[id=FF]{of} the trailing edge, as in figure \ref{liftingsurface geometry}, we parametrize the chordwise direction by the coordinate $\xi\in[0,1]$ as
\begin{equation}
    \phi'=\phi'_{\mathrm{TE}}(r)+\frac{c(r)}{r}\,\xi,
    \qquad
    d\phi'=\frac{c(r)}{r}\,d\xi
    \label{eq:xi-def}
\end{equation}
so that $\xi=0$ and $\xi=1$ correspond to the trailing and leading edges, respectively. 

Let $z^{(\pm)}(r,\xi)$ denote the normal displacement of the physical blade surface from the equatorial plane, on the upper ($+$) and lower ($-$) sides, and define the camber (mean) line and thickness functions as
\begin{equation}
    z_{cl}=\frac{z^{(+)}+z^{(-)}}{2},
    \qquad
    z_{th}=z^{(+)}-z^{(-)}
\end{equation}
respectively. For a section at small twist angle $\beta(r)$ with respect to the plane of rotation, the upper and lower surfaces can be written as
\begin{align}
    z^{(\pm)}(r,\xi) &= z_{cl}(r,\xi) \pm \frac{1}{2}z_{th}(r,\xi) + c(r)\,\beta(r)\,\xi
\end{align}
It is often convenient to normalise by the local chord, writing $z = c\,h$. This gives the outward (non-unit) normals on the upper and lower surfaces as
\begin{align}
    \mathbf n^{(\pm)} &= \left(\mp \frac{\partial h_{cl}}{\partial \xi}-\frac{1}{2}\frac{\partial h_{th}}{\partial \xi}\mp\beta\right)\,\boldsymbol{\hat\phi}' \mp \boldsymbol{\hat\theta}
    \label{normals}
\end{align}
where $h_{cl}=z_{cl}/c$ and $h_{th}=z_{th}/c$. From the surface geometry we can now obtain the expressions for the force per unit area and for the normal velocity. Retaining only the pressure perturbation contribution to the loading, \(\mathbf{f}^{(\pm)} \approx p^{(\pm)}\,\hat{\mathbf{n}}^{(\pm)}\), that is, neglecting the viscous term which is typically negligible in the Reynolds-number range relevant to propellers \citep{glegg2024}, the net force (sum of the contributions from the two sides) can be decomposed in its components along $\boldsymbol{\hat\theta}$ and $\boldsymbol{\hat\phi}'$ directions as
\begin{equation}
    \mathbf f_{\theta} = \Delta p\,\boldsymbol{\hat\theta}, \qquad 
    \mathbf f_{\phi'} = \left[\left(\beta+\frac{\partial h_{cl}}{\partial \xi}\right)\,\Delta p \;-\;\frac{1}{2}\frac{\partial h_{th}}{\partial \xi}\,\Sigma p\right]\boldsymbol{\hat\phi}'
\end{equation}
where $\Delta p = p^{(-)}-p^{(+)}$ is the pressure jump from upper to lower surface and $\Sigma p = p^{(+)}+p^{(-)}$ is their sum.

A similar reduction applies to the normal velocity. From Eq. \eqref{normals} it is easy to see that the only geometric contribution to $V_n$ arises from the thickness distribution, since it has the same sign on both sides, so that
\begin{equation}
    V_n = -\Omega\,\frac{\partial z_{th}}{\partial \phi'}
    = -\Omega\,r\,\frac{\partial h_{th}}{\partial \xi}
\end{equation}

This clarifies the well-known physical mechanisms which contribute to propeller noise: a normal-loading term $\mathbf f_{\theta}$ associated with the pressure jump acting on the plane and whose primary effect is lift when the blade twist is small; a tangential-loading term $\mathbf f_{\phi'}$, which involves both the pressure jump and sum and is mainly related to drag; and a thickness term $h_{th}$, arising from the normal velocity induced by the upper and lower surfaces. Accordingly, each multipole coefficient $A_{\ell\deleted[id=FF]{,} m}$ can be expressed in terms of these quantities.

Making explicit the azimuthal dependence of spherical harmonics, $Y^*_{\ell m}(\theta,\phi') = Y^*_{\ell m}(\theta,0)\,e^{-im\phi'}$, the function $g_{\ell m}$ and its gradient components on the rotor plane are
\begin{equation}
\begin{aligned} 
g_{\ell m,0} &= j_\ell(k_m r)\, Y^*_{\ell m}\!\left(\tfrac{\pi}{2},\phi'\right) = j_\ell(k_m r)\, Y^*_{\ell m}\!\left(\tfrac{\pi}{2},0\right)\,e^{-im\phi'}  \\ 
\frac{1}{r}(\partial_\theta g_{\ell m})_0 &= \frac{1}{r}j_\ell(k_m r)\, \left.\frac{\partial Y^*_{\ell m}}{\partial\theta}\right|_{\theta=\pi/2} = \frac{1}{r}j_\ell(k_m r)\, \left.\frac{\partial Y^*_{\ell m}(\theta,0)}{\partial\theta}\right|_{\theta=\pi/2}\,e^{-im\phi'}  \\ 
\frac{1}{r}(\partial_{\phi'} g_{\ell m})_0 &= \frac{1}{r}j_\ell(k_m r)\, \left.\frac{\partial Y^*_{\ell m}}{\partial\phi'}\right|_{\theta=\pi/2} = -im\,\frac{1}{r}j_\ell(k_m r)\, Y^*_{\ell m}\!\left(\tfrac{\pi}{2},0\right)\,e^{-im\phi'} 
\end{aligned}
\end{equation}
Substituting into Eq. \eqref{eq:Alm1} yields \inlinecomment{The formatting of the equations has been refined.}
\begin{equation}
\begin{aligned}
A^{(LS)}_{\ell m}
&=
N i k_m  \int_{R_{hub}}^{R} 
j_\ell(k_m r) \frac{c(r)}{r} e^{-im\phi'_{\mathrm{TE}}(r)}\int_0^{\deleted[id=FF]{2\pi}\added{1}} 
\Bigg[
 \left|\mathbf f_{\theta} (r,\xi)\right| \left.\frac{\partial Y^*_{\ell m}(\theta,0)}{\partial\theta}\right|_{\theta=\pi/2} + \\
&
- i m\, \left|\mathbf f_{\phi'}(r,\xi) \right|\, Y^*_{\ell m}\!\left(\tfrac{\pi}{2},0\right)
+\, i \rho_0\,a_s k_m  \Omega r^2\, \frac{\partial h_{th}(r,\xi)}{\partial \xi}\, Y^*_{\ell m}\!\left(\tfrac{\pi}{2},0\right)
\Bigg] e^{-im\frac{c(r)}{r}\,\xi}\, d\xi\,dr
\end{aligned}
\label{eq:Alm1.2a}
\end{equation}

The thickness term in Eq. \eqref{eq:Alm1.2a} is conveniently rewritten by integrating by parts in $\xi$\added{, noting that $ h_{th}(r,0)= h_{th}(r,1)=0$}. Introducing $f_L$ and $f_D$ as dimensionless pressure coefficients, such that $\left|\mathbf f_{\theta} \right| = 1/2\rho_0 (\Omega r)^2 f_L$ and $\left|\mathbf f_{\phi'} \right| = 1/2\rho_0 (\Omega r)^2 f_D$, the following expression is obtained
\begin{equation}
\begin{aligned}
A^{(LS)}_{\ell m}
&=
N i \rho_0 k_m  \int_{R_{hub}}^{R} 
j_\ell(k_m r) \frac{c(r)}{r} e^{-im\phi'_{\mathrm{TE}}(r)}\int_0^{\deleted[id=FF]{2\pi}\added{1}} 
\Bigg[
 \frac{1}{2} (\Omega r)^2f_L(r,\xi) \left.\frac{\partial Y^*_{\ell m}(\theta,0)}{\partial\theta}\right|_{\theta=\pi/2} + \\
&
- \frac{1}{2} i m\, (\Omega r)^2f_D(r,\xi)\, Y^*_{\ell m}\!\left(\tfrac{\pi}{2},0\right)
-\, \,a^2_s k^2_m c(r) r h_{th}(r,\xi)\, Y^*_{\ell m}\!\left(\tfrac{\pi}{2},0\right)
\Bigg] e^{-im\frac{c(r)}{r}\,\xi}\, d\xi\,dr
\end{aligned}
\label{eq:Alm1.2}
\end{equation}

Rewriting using the non-dimensional spanwise coordinate $\bar r=r/R$, the hub ratio $\bar R_{hub}=R_{hub}/R$, the chord-to-diameter ratio $\gamma(\bar r)=c(\bar r)/(2R)$, and the tip Mach number $M_t=\Omega R/a_s$, we can collect the terms and obtain the lifting-surface multipole coefficients as
\begin{equation} 
\begin{aligned} 
A^{(LS)}_{\ell m} =
& i \rho_0 a_s^2 M_t^3 N m B_{L_{\ell m}} \int_{\bar{R}_{hub}}^{1} j_\ell \left(m M_t \bar{r}\right) \gamma(\bar r) \bar{r} e^{-im\phi'_{TE}(\bar r)} \left[ \int_0^{1} f_L\bigl(\bar{r},\xi\bigr) e^{-im\frac{2\gamma(\bar r)}{\bar{r}}\xi} d\xi \right] d\bar{r} +\\[2mm] 
- 4 & i \rho_0 a_s^2 M_t^3 N m^3 B_{T_{\ell m}} \int_{\bar{R}_{hub}}^{1} j_\ell \left(m M_t \bar{r}\right) \gamma(\bar r)^2 e^{-im\phi'_{TE}(\bar r)} \left[\int_0^{1} h_{th}\bigl(\bar{r},\xi\bigr) e^{-im\frac{2\gamma(\bar r)}{\bar{r}}\xi}d\xi\right] d\bar{r}  +\\[2mm] 
&  \rho_0 a_s^2 M_t^3 N m^2 B_{D_{\ell m}} \int_{\bar{R}_{hub}}^{1} j_\ell \left(m M_t \bar{r}\right) \gamma(\bar r) \bar{r} e^{-im\phi'_{TE}(\bar r)} \left[ \int_0^{1} f_D\bigl(\bar{r},\xi\bigr) e^{-im\frac{2\gamma(\bar r)}{\bar{r}}\xi} d\xi \right] d\bar{r} 
\end{aligned} 
\label{eq:A2lm} 
\end{equation}
where $B_{L_{\ell m}}=\left.\frac{\partial Y^*_{\ell m}(\theta,0)}{\partial\theta}\right|_{\theta=\pi/2}$ and $B_{T_{\ell m}} =B_{D_{\ell m}} = Y^*_{\ell m}\!\left(\tfrac{\pi}{2},0\right)$, i.e. they are the values of the spherical harmonics and their \(\theta\)-derivative evaluated on the equatorial plane $\theta=\pi/2$. These factors encode the angular structure of the radiation and can be written explicitly in terms of $\ell$ and $m$, as reported in the Appendix \ref{appA}. Contributions proportional to $B_{T_{\ell m}}$ are symmetric with respect to $\theta_0=\pi/2$ (equal modulus above and below the plane) and include the thickness\added{, represented by the second term in Eq. \eqref{eq:A2lm},} and \deleted[id=FF]{drag terms} \added{torque-loading, represented by the third term}. Owing to the parity properties of the spherical harmonics, only terms with \(\ell-\added{|}m\added{|}\) even contribute to this part. By contrast, contributions proportional to $B_{L_{\ell m}}$ are antisymmetric with respect to $\theta_0=\pi/2$, and therefore produce an antisymmetric directivity pattern. These terms are associated with \deleted[id=FF]{lift} \added{thrust-loading, represented by the first term in Eq. \eqref{eq:A2lm},} and involve only values of \(\ell\) such that \(\ell-|m|\) is odd. 

This behaviour clarifies which physical mechanisms enter the dominant multipole coefficients and, in turn, explains the directivity patterns discussed in Section \ref{sec:dominant-multipoles}. In the lifting-surface framework, the symmetric family of multipoles collects the contribution of thickness and tangential-loading \added{(torque)}, whereas the antisymmetric family is primarily associated with the normal-loading \added{(thrust)} contribution, whose main aerodynamic effect is lift when the blade twist and incidence are small. \added{The different symmetry properties of the thrust- and torque-noise contributions were already recognized in the classical propeller-noise literature, in particular by \citet{Kurtz1970}. The present formulation recasts this distinction in terms of spherical multipoles, showing how they populate different families of coefficients.} This distinction is particularly useful for interpreting the behaviour of Propeller A, for which the lifting-surface approximation is especially appropriate because it figures blade operating at small angles. Under these conditions, the pressure jump across the section remains comparatively weak, so that the lift-related contribution is smaller than the symmetric contribution associated mainly with thickness and, to a lesser extent, drag-like loading. As a consequence, the first antisymmetric multipole is much weaker than the first symmetric one. This explains why the antisymmetric component in figures \ref{fig:m2_decomp_A02} and \ref{fig:m2_decomp_A08} is so small. 

\subsubsection{Far-field pressure}
Once closed-form expressions for the multipole coefficients have been derived, both the acoustic pressure and the radiated sound power can be obtained directly from the general relations Eqs. \eqref{eq:pressure_formula} and \eqref{eq:power formula}. In particular, substituting Eq. \eqref{eq:A2lm} into Eq. \eqref{eq:pressure-formula-FF} yields the complex far-field pressure for the lifting surface
\begin{equation}
\begin{aligned}
    \hat{p}^{(FF,LS)}_{m}\left(\mathbf{x}_0\right)
    \approx & \rho_0 a_s^2 M_t^2 N \frac{e^{im M_t \bar{r}_0}}{\bar{r}_0} \sum^{\infty}_{\ell=|m|} \left(-i\right)^{\ell} Y_{\ell m}\!\left(\theta_0,\phi_0\right) \\[2mm] 
    \Bigg\{ &B_{L_{\ell m}}
    \int_{\bar{R}_{hub}}^{1}  j_\ell\!\left(m M_t \bar{r}\right) 
    \gamma(\bar r)  \bar{r} 
    e^{-im\phi'_{TE}(\bar r)}
    \left[ \int_0^{1} f_L\bigl(\bar{r},\xi\bigr) 
    e^{-im\frac{2\gamma(\bar r)}{\bar{r}}\xi}  d\xi \right] d\bar{r} +\\[2mm]
     - 4  m^2 & B_{T_{\ell m}}
    \int_{\bar{R}_{hub}}^{1}  j_\ell\!\left(m M_t \bar{r}\right) 
    \gamma(\bar r)^2 
    e^{-im\phi'_{TE}(\bar r)}
    \left[\int_0^{1} h_{th}\bigl(\bar{r},\xi\bigr) 
    e^{-im\frac{2\gamma(\bar r)}{\bar{r}}\xi}d\xi\right] d\bar{r} +\\[2mm]
     - i  m & B_{D_{\ell m}}
    \int_{\bar{R}_{hub}}^{1}  j_\ell\!\left(m M_t \bar{r}\right) 
    \gamma(\bar r) \bar{r} 
    e^{-im\phi'_{TE}(\bar r)}
    \left[ \int_0^{1} f_D \bigl(\bar{r},\xi\bigr) 
    e^{-im\frac{2\gamma(\bar r)}{\bar{r}}\xi}  d\xi \right] d\bar{r}\Bigg\}
\end{aligned}
\label{}
\end{equation}
These expressions \deleted[id=FF]{are equivalent}\added{can be related} to those obtained by \citet[Eq. 36]{Hanson1980} in his far-field helicoidal-surface theory, when restricted to zero forward velocity. \added{In hovering conditions, Hanson's helicoidal coordinates reduce to a cylindrical description. His far-field approximation is based on the geometric condition $r_0\gg R$, which can be compared with the present criterion $k_m r_0\gg\ell$. Indeed, the dominant spherical degrees satisfy $\ell\simeq m$. Hence, for $kR=M_t\lesssim 1$, the present far-field condition is at least as restrictive as Hanson's condition, and therefore implies it.} A straightforward comparison can be made and two differences are worth emphasizing. First, in the present formulation each harmonic $m$ is represented through an infinite sum over the index $\ell$, whereas Hanson's far-field results are written in closed form. Second, the present multipole expansion introduces spherical Bessel functions whose argument depends only on the source coordinates and is therefore independent of the observer position. In Hanson's formulation, by contrast, the corresponding Bessel functions encode the observer location; consequently, the source integrals must be recomputed when the observer position changes. This separation of source and observer is a key practical advantage of the present approach when evaluating pressure at many microphone locations. Moreover, the present framework retains the same formal structure whether or not a far-field assumption is invoked: the pressure is expressed in the same way (but including spherical Hankel functions in the observer factor, instead of the asymptotic limit), and no separate near-field derivation is required. The distinction between near and far field is reflected primarily in the rate of convergence of the truncated series, rather than in a change of the analytical form of the solution. A further advantage concerns the evaluation of radiated power. In Hanson's formulation the power is typically obtained by an additional integration of the far-field pressure over a control sphere \citep{Zhong2020}. In the present multipole approach, instead, the radiated power can be expressed directly in terms of the multipole coefficients by Eq. \eqref{eq:power formula}, so that its computation reduces to a simple summation.

To understand the phase shift among the three terms, their different symmetry must be taken into account. The \deleted[id=FF]{lift}\added{thrust} term is proportional to \((-i)^\ell\), but for fixed \(m\) only terms with \(\ell-|m|\) odd contribute; the first non-zero term is therefore \(\ell=|m|+1\). For the thickness and \deleted[id=FF]{drag}\added{torque} terms, instead, the first contribution is \(\ell=|m|\). Hence, $\text{\deleted[id=FF]{lift}\added{thrust}} \propto (-i)^{|m|+1}, \text{thickness} \propto -(-i)^{|m|}, \text{\deleted[id=FF]{drag}\added{torque}} \propto (-i)^{|m|+1}$: if the source integrals are real (or if they have approximately the same argument), one obtains that \deleted[id=FF]{lift}\added{thrust} and \deleted[id=FF]{drag}\added{torque} are in phase, whereas thickness is in quadrature with them. In the lifting-surface case, however, the phase also contains the contribution of the complex source integrals, and this effect may become significant at higher frequency. 

As shown in \deleted[id=FF]{s}\added{S}ection \ref{sec:dominant-multipoles}, the first two multipoles are sufficient to obtain an accurate description of the far-field, implying that all three source contributions (\deleted[id=FF]{lift}\added{thrust}, \deleted[id=FF]{drag}\added{torque}, and thickness) must in general be retained. Under this condition, the asymptotic form of the spherical harmonics for the lowest degrees at fixed order $m$ is particularly informative. The contributions that are symmetric with respect to the plane of rotation (e.g., thickness and \deleted[id=FF]{drag}\added{torque}) are associated with the first non-zero sectoral term $Y_{mm}(\theta_0,\phi_0)\ \propto\ (\sin\theta_0)^{m}\,e^{im\phi_0}$ \citep[p. 157, Eq. 12]{Varshalovich1988}. In contrast, an antisymmetric contribution (e.g., \deleted[id=FF]{lift}\added{thrust}) requires a factor that changes sign across the plane, and is captured by $Y_{m+1,m}(\theta_0,\phi_0)\ \propto\ \cos\theta_0\,(\sin\theta_0)^{m}\,e^{im\phi_0}$ \citep[p. 157, Eq. 13]{Varshalovich1988}. These expressions allow straightforward considerations about the directivity of a single $m$-harmonic, as will be discussed later.

\subsection{Lifting line}
\label{sec:Lifting Line}

The approach developed in the previous \deleted[id=FF]{s}\added{S}ection \ref{sec:Lifting Surface} is well suited to propellers whose blades are moderately twisted and operate at small incidence. Many practical propellers, however, feature large inflow angles and high-aspect-ratio blades. In this section we therefore specialise \deleted[id=FF]{Goldstein's integral} \added{Eq. \eqref{eq:Alm-start}} to the case of slender blades operating at arbitrary incidence. The blade is approximated by a spanwise lifting curve, along which compact airfoil sections are distributed. \added{An early treatment based on this approximation was given by \citet{clark1974}, who observed that the lifting-line idealisation is acoustically equivalent to assuming chordwise compactness, \(k c\ll 1\). This condition is naturally satisfied in the regime considered here. Indeed, if the rotor is at most marginally compact, $M_t = k R \lesssim 1$, and the blade has a high aspect ratio, \(c/(2R)\ll1\), then \(k c\ll 1\). A related analysis was presented by \citet{Brouwer1992}, who applied matched asymptotic expansions to the aerodynamic and acoustic fields of high-aspect-ratio propellers. Chordwise compactness therefore provides a clear asymptotic basis for reducing the surface integral to a lifting-line representation while retaining the spanwise variation of the acoustic source.}

\begin{figure}
\centering
\def\svgwidth{0.8\linewidth} 
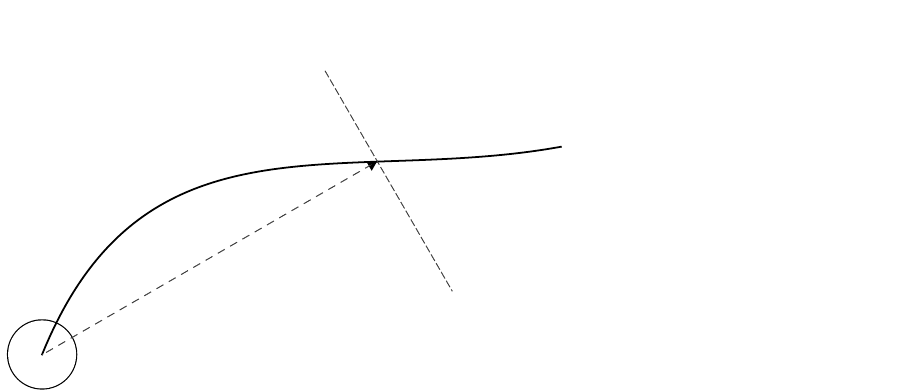
\caption{Lifting-line geometry.}
\label{liftingline geometry}
\end{figure}

Fix a station point $\mathbf{x}'_s(\lambda)$ in the frame co-rotating with the lifting curve, with radius $r_s=|\mathbf{x}'_s|\in[R_{\mathrm{hub}},R]$. \inlinecomment{The notation for $\mathbf{x}'_s$ has been checked.} Let $2R$ be the outer scale and $c$ the inner scale, and consider the chord-to-diameter ratio as a small parameter $\varepsilon_\ell=c/(2R)=\gamma\ll1$. The sectional patch around $\mathbf{x}'_s$ is parametrised by a local coordinate $\boldsymbol\xi$ in the airfoil section as
\begin{equation}
    \mathbf{x}'(\lambda,\mu;\varepsilon_\ell)
    =\mathbf{x}'_s(\lambda)+\varepsilon_\ell \boldsymbol\xi(\lambda,\mu)
\end{equation}
where $\lambda$ is a spanwise coordinate and $\mu$ parametrises the closed section contour. The full blade surface is swept out by moving these closed sectional curves along the span. From an aerodynamic viewpoint, see figure \ref{liftingline geometry}, it is natural to assume that the local sections remain aligned with the incoming relative flow, so that the radial coordinate provides the most convenient spanwise parametrisation. Accordingly, in what follows we take \(\lambda=r\). If the airfoil section is viewed as the surface generated by intersecting the blade with spheres centred at the hub, and if \(dl = \bigl|\partial_\mu \boldsymbol\xi(\lambda,\mu)\bigr|\,d\mu\) denotes the arc-length element of the section contour, then the surface element is simply written as $dS =\varepsilon_\ell dl\,dr$. With this choice, the original surface integral can be reduced to a contour integral over each compact section followed by a spanwise integration along the lifting line.

The acoustic function $g_{\ell m}$ can be expanded around the line $\mathbf{x}'_s$ as
\begin{equation}
\begin{aligned}
g_{\ell m}(\mathbf{x}')
&= g_{\ell m,s}
 + \varepsilon_\ell \boldsymbol\xi\cdot\nabla_s g_{\ell m,s}
 + O(\varepsilon_\ell^2)\\[2mm]
\nabla_{\mathbf{x}}g_{\ell m}(\mathbf{x}')
&= \nabla_s g_{\ell m,s}
 + \varepsilon_\ell \left(\nabla_s\nabla_s g_{\ell m,s}\right)\cdot\boldsymbol\xi
 + O(\varepsilon_\ell^2)
\end{aligned}
\label{eq:taylor-g}
\end{equation}
where the subscript $s$ denotes evaluation at $\mathbf{x}'_s$ and $\nabla_s$ acts on the source coordinates. The Taylor expansion is applied only to $g_{\ell m}$ and its gradient whereas the sources $\mathbf f$ and $V_n$ are \deleted[id=FF]{kept in their exact form}\added{retained without approximation}.

Substituting Eq. \eqref{eq:taylor-g} into Eq. \eqref{eq:Alm-start}, the multipole coefficient $A^{(LL)}_{\ell m}$ for the lifting line becomes
\begin{equation}
\begin{aligned}
A^{(LL)}_{\ell m}
\approx N  i k_m \int^R_{R_{hub}}\Bigg\{
\varepsilon_\ell & \oint_{\partial S} 
 \Big[\underbrace{ \mathbf f\cdot\nabla_s g_{\ell m,s}}_{\text{force term}}
- \underbrace{i \rho_0 a_s k_m  V_n g_{\ell m,s}}_{\text{velocity term}}\Big]
dl +\\
\qquad\qquad
+ \varepsilon^2_\ell &\oint_{\partial S} 
 \Big[ \underbrace{\mathbf f\cdot\bigl(\nabla_s\nabla_s g_{\ell m,s}\bigr)\boldsymbol\xi}_{\text{force moment}}
- \underbrace{i  \rho_0 a_s k_m  V_n \boldsymbol\xi\cdot\nabla_s g_{\ell m,s}}_{\text{velocity moment}}\Big]
dl
\Bigg\} dr
\end{aligned}
\label{eq:Alm-station-raw}
\end{equation}
We then define the compact force and velocity moments over $\partial S$ as
\begin{equation}
\begin{aligned}
\mathbf F_s &= \added{\varepsilon_\ell} \oint_{\partial S} \mathbf f \,  dl,
& Q_{V,s} &= \added{\varepsilon_\ell} \oint_{\partial S} V_{n}  \,  dl, \\
\mathsf M_s^{(f)} &= \added{\varepsilon^2_\ell} \oint_{\partial S} \mathbf f\otimes\boldsymbol\xi  \,  dl,
& \mathbf M_s^{(V)} &= \added{\varepsilon^2_\ell} \oint_{\partial S} V_{n}\,\boldsymbol\xi   \,  dl
\end{aligned}
\label{eq:LL-moments-nonorth}
\end{equation}
Here $\mathbf F_s$ is the resultant sectional force (per unit length, acting on the fluid), $Q_{V,s}$ is the net volume flux through the sectional patch, $\mathsf M_s^{(f)}$ is the first moment of the force, and $\mathbf M_s^{(V)}$ is the first moment of the normal velocity. Using these definitions we obtain the compact expression
\begin{equation}
\begin{aligned}
A^{(LL)}_{\ell m}
\approx N  i k_m \int^R_{R_{hub}}\Bigg[
& \deleted[id=FF]{\varepsilon_\ell\Big(}
  \mathbf F_s\cdot\nabla_s g_{\ell m,s}
 - i  \rho_0 a_s k_m  Q_{V,s} g_{\ell m,s}
\deleted[id=FF]{\Big)} + \\[2mm]
+& \deleted[id=FF]{\varepsilon^2_\ell\Big(}
  \mathsf M_s^{(f)} : \nabla_s\nabla_s g_{\ell m,s}
 - i  \rho_0 a_s k_m  \mathbf M_s^{(V)}\cdot\nabla_s g_{\ell m,s}
\deleted[id=FF]{\Big)}
\Bigg] dr
\end{aligned}
\label{eq:LL-Alm-moment-expansion}
\end{equation}
At first order, the expansion yields two terms. For an impermeable blade section, the velocity term $Q_{V,s}$ vanishes by local mass conservation, so that the first non-zero contribution is force-related $\mathbf F_s$, in agreement with the analysis of \citet{brouwer1989}. To better understand the sectional force, it is convenient to decompose \(\mathbf f\) into a component normal to the airfoil section ($(\mathbf f\cdot\hat{\mathbf{x}}'_s) \hat{\mathbf{x}}'_s$) and a component lying within the section plane ($\mathbf f_\perp$)
\begin{equation} 
\mathbf f = (\mathbf f\cdot\hat{\mathbf{x}}'_s) \hat{\mathbf{x}}'_s+\mathbf f_\perp, \qquad \mathbf f_\perp = \mathbf f-(\mathbf f\cdot\hat{\mathbf{x}}'_s)\hat{\mathbf{x}}'_s 
\end{equation} 
The corresponding scalar and vector moments over the sectional patch $\partial S$ are 
\begin{equation} 
\begin{aligned} 
F_r &=\added{\varepsilon_\ell}\oint_{\partial S}(\mathbf f\cdot\hat{\mathbf{x}}'_s)  dl, & \mathbf F_\perp &=\added{\varepsilon_\ell}\oint_{\partial S}\mathbf f_{\perp}  dl, \\[3pt] 
\mathbf I_\perp^{(f)} &=\added{\varepsilon^2_\ell}\oint_{\partial S}\boldsymbol\xi  (\mathbf f\cdot\hat{\mathbf{x}}'_s)  dl, & \mathsf M_\perp^{(f)} &=\added{\varepsilon^2_\ell}\oint_{\partial S}\mathbf f_{\perp}\otimes\boldsymbol\xi  dl
\end{aligned} 
\end{equation} 
Here $F_r$ and $\mathbf F_\perp$ are the radial and in-plane components of the resultant sectional force. The quantities $\mathbf I_\perp^{(f)}$ and $\mathsf M_\perp^{(f)}$ are the first moments of the radial and in-plane force components. In terms of these sectional moments, the compact force moments in Eq. \eqref{eq:LL-Alm-moment-expansion} can be written as 
\begin{equation} \mathbf F_s = F_r \hat{\mathbf{x}}'_s + \mathbf F_\perp, \qquad \mathsf M_s^{(f)} = \hat{\mathbf{x}}'_s\otimes\mathbf I_\perp^{(f)} + \mathsf M_\perp^{(f)} 
\end{equation} 
The in-plane tensor $\mathsf M_\perp^{(f)}$ contains the first moments of the in-plane force about axes in the section plane. It is useful to distinguish the symmetric and skew-symmetric parts of $\mathsf M_\perp^{(f)}$. The skew-symmetric part $\mathsf M_\perp^{(f),\text{skew}}$ is directly related to the \deleted[id=FF]{classical} aerodynamic moment of the airfoil section about the radial axis. Indeed, the scalar pitching moment about $\hat{\mathbf{x}}'_s$ is
\begin{equation}
M_{\text{pitch}} = \added{\varepsilon^2_\ell} \oint_{\partial S} (\boldsymbol\xi\times\mathbf f_{\perp})\cdot\hat{\mathbf{x}}'_s dl
\end{equation} 
which can be expressed in terms of $\mathsf M_\perp^{(f),\text{skew}}$. However, because the Hessian $\nabla_s\nabla_s g_{\ell m,s}$ is symmetric, only the symmetric part $\mathsf M_\perp^{(f),\text{sym}}$ contributes to the acoustic term $\mathsf M_s^{(f)} : \nabla_s\nabla_s g_{\ell m,s}$ in Eq. \eqref{eq:LL-Alm-moment-expansion}. The skew-symmetric part, and hence the classical sectional pitching moment, does not enter the solution at the order considered here.

As for the velocity moment $\mathbf M_s^{(V)}$, if the surface velocity may be regarded as approximately uniform over the section, then its first moment reduces to
\begin{equation}
\mathbf M_s^{(V)}
= \added{\varepsilon^2_\ell}
\oint_{\partial S}
(\mathbf V\cdot \mathbf n)\,\boldsymbol\xi\,dl
\approx \added{\varepsilon^2_\ell}
\mathbf u_s\cdot
\oint_{\partial S}
\mathbf n\otimes\boldsymbol\xi\,dl
= 
A_s\,\mathbf u_s
\end{equation}
where \(\mathbf u_s\) denotes the local sectional velocity and \(A_s\) the sectional area.

\subsubsection{Straight line in the equatorial plane}
The general expressions found above can be simplified further for geometries representative of many practical propellers. Away from the hub and tip regions where three-dimensional effects are strongest, the local aerodynamic loading is approximately two-dimensional \citep{johnson2013}. The resultant sectional force is then dominated by the lift and drag components acting in the local airfoil plane, denoted by \(\mathbf L_s\) and \(\mathbf D_s\) (forces per unit span exerted on the fluid), whereas the radial component may be neglected. Accordingly, all moments associated with the radial force become negligible; in particular $F_r \approx 0$ and $\mathbf I_\perp^{(f)} \approx \mathbf 0$.

As discussed by \citet{Brouwer1992}, a thin-airfoil assumption is not strictly required in a lifting-line formulation, but it is convenient when seeking a simpler description. In particular, for a generic airfoil the symmetric force-moment tensor \(\mathsf M_\perp^{(f),\mathrm{sym}}\) should be regarded as being of the same order as the first moment of the velocity \(\mathbf M_s^{(V)}\), according to Eq. \eqref{eq:LL-Alm-moment-expansion}. Under a thin-airfoil approximation, however, the leading symmetric part depends only on the first moment of the pressure jump \(\Delta p\) about the chosen sectional reference point through which the lifting line passes. If that point is taken sufficiently close to the center of pressure, this leading-order symmetric contribution vanishes. Under these assumptions, the dominant sectional moments reduce to the resultant lift-drag force and to the area-related velocity moment
\begin{equation}
\mathbf F_s \approx \mathbf F_\perp \approx \mathbf L_s+\mathbf D_s,
\qquad
Q_{V,s}\approx 0,
\qquad
\mathsf M_s^{(f)} \approx \mathsf M_\perp^{(f),\mathrm{sym}} \approx 0,
\qquad
\mathbf M_s^{(V)} \approx  A_s \mathbf u_s
\label{simplified moments}
\end{equation}

Substituting Eqs. \eqref{simplified moments} in Eq. \eqref{eq:LL-Alm-moment-expansion} then gives
\begin{equation}
A^{(LL)}_{\ell m}
= N i k_m \int_{R_{hub}}^{R}\frac{j_\ell(m M_t \bar r)}{r}
\nabla_{\Omega_s}Y_{\ell m}^*(\hat{\mathbf{x}}'_s)\cdot
\left[  \deleted[id=FF]{\varepsilon_\ell\left(\right.}\mathbf L_\perp(r)+\mathbf D_\perp(r)\deleted[id=FF]{\left.\right)}
- \deleted[id=FF]{\varepsilon^2_\ell} i\rho_0 a_s k_m  A_s(r) \mathbf u_s(r)\right] dr
\label{eq:Blm-master-new}
\end{equation}
where \(\nabla_{\Omega_s}\) denotes the angular gradient with respect to the source coordinates.

To facilitate comparison with the lifting-surface results, we now restrict attention to the case in which the lifting line lies in the equatorial plane, \(\theta=\pi/2\). In this case
\begin{equation}
\nabla_{\Omega_s}Y_{\ell m}^*(\hat{\mathbf{x}}'_s)
=
\left.\frac{\partial Y_{\ell m}^*(\theta,0)}{\partial\theta}\right|_{\theta=\pi/2}
e^{-im\phi'_s}\,\hat{\boldsymbol\theta}
-im\,Y_{\ell m}^*\!\left(\tfrac{\pi}{2},0\right)e^{-im\phi'_s}\,\hat{\boldsymbol\phi}'
\label{eq:LL-angular-gradient}
\end{equation}
with \(\phi'_s\) denoting the azimuthal position of the lifting line.

In this approach, the local induced velocity \citep{Bramwell2001} can be taken into account. Since $\varphi_i$ is the inflow angle associated with the axial induced velocity, the magnitude of the local relative velocity is given by
\begin{equation}
|\mathbf u_s|=\frac{\Omega r}{\cos\varphi_i}
\end{equation}
The sectional lift and drag are then written as $\mathbf L_s+\mathbf D_s = q_s c\left(c_l\,\hat{\mathbf{L}}+c_d\,\hat{\mathbf{D}}\right)$, where
\begin{equation}
\hat{\mathbf L}
=
\cos\varphi_i\,\hat{\boldsymbol\theta}
+\sin\varphi_i\,\hat{\boldsymbol\phi}',
\qquad
\hat{\mathbf D}
=
\hat{\mathbf u}_s
=
-\sin\varphi_i\,\hat{\boldsymbol\theta}
+\cos\varphi_i\,\hat{\boldsymbol\phi}'
\end{equation}
and $q_s=\frac{1}{2}\rho_0\Omega^2 r^2/\cos^2\varphi_i$ is the local dynamic pressure.

Substituting these expressions into Eq. \eqref{eq:Blm-master-new} yields the first-order multipole coefficients
\begin{equation}
\begin{aligned}
A^{(LL,1)}_{\ell m}
=&\,
i \rho_0 a_s^2 m M_t^3 N
B_{L_{\ell m}}
\int_{\bar R_{hub}}^{1} j_\ell(m M_t \bar r)
\frac{\gamma(\bar r)}{\cos\varphi_i(\bar r)}
\left[c_l(\bar r)-c_d(\bar r)\tan\varphi_i(\bar r)\right]
\bar r\,e^{-im\phi'_s(\bar r)}\,d\bar r +
\\[2pt]
+&
\rho_0 a_s^2 m^2 M_t^3 N
B_{D_{\ell m}}
\int_{\bar R_{hub}}^{1} j_\ell(m M_t \bar r)
\frac{\gamma(\bar r)}{\cos\varphi_i(\bar r)}
\left[c_l(\bar r)\tan\varphi_i(\bar r)+c_d(\bar r)\right]
\bar r\,e^{-im\phi'_s(\bar r)}\,d\bar r
\end{aligned}
\label{eq:Blm-rot-harm-E-first}
\end{equation}
and the second-order multipole coefficients
\begin{equation}
\begin{aligned}
A^{(LL,2)}_{\ell m}
=&\,
-4 i \rho_0 a_s^2 m^3 M_t^3 N B_{T_{\ell m}}
\int_{\bar R_{hub}}^{1} j_\ell(m M_t \bar r)\,
\gamma^2(\bar r) \bar A_s(\bar r)\,
e^{-im\phi'_s(\bar r)}\,d\bar r +
\\[2pt]
&-
4 \rho_0 a_s^2 m^2 M_t^3 N B_{T'_{\ell m}}
\int_{\bar R_{hub}}^{1} j_\ell(m M_t \bar r)\,
\gamma^2(\bar r) \bar A_s(\bar r)\tan\varphi_i(\bar r)\,
e^{-im\phi'_s(\bar r)}\,d\bar r
\end{aligned}
\label{eq:Blm-rot-harm-E-second}
\end{equation}
where $\bar A_s = A_s/c^2$ is the non-dimensional section area. As in the lifting-surface case, symmetry with respect to the plane of rotation remains embedded in the factors $B_{L_{\ell m}}=B_{T'_{\ell m}}$ and $B_{D_{\ell m}}=B_{T_{\ell m}}$. In the present formulation, however, this property follows from the specific choice of placing the lifting line in the plane of rotation. For a more general lifting-line geometry, the corresponding symmetry information would no longer be contained solely in these equatorial factors, but would also enter through the spanwise integral via the spatial description of the line itself. In that case, an explicit parametrization of the lifting-line geometry would be required.

The leading-order term, Eq. \eqref{eq:Blm-rot-harm-E-first}, is governed by sectional lift and drag, both of which scale with the slenderness parameter \(\varepsilon_\ell=\gamma\), whereas the thickness-related corrections, Eq. \eqref{eq:Blm-rot-harm-E-second}, enter only at order \(\gamma^2\). To facilitate comparison with the lifting-surface formulation, the two terms in Eq. \eqref{eq:Blm-rot-harm-E-first} are referred to here as the ``\deleted[id=FF]{lift}\added{thrust}'' and ``\deleted[id=FF]{drag}\added{torque}'' contributions. \deleted[id=FF]{In the lifting-line framework, however, these terms are not associated exactly with directions normal and tangential to the plane of rotation. As a result, the first term may also contain a non-negligible drag-related contribution, just as the second may include a lift-related component.} The first term in Eq. \eqref{eq:Blm-rot-harm-E-second} is clearly related to the sectional area and is therefore again denoted as the ``thickness'' contribution. The last term is referred to as \added{the} ``\deleted[id=FF]{induced}\added{antisymmetric} thickness'' \added{contribution}, \deleted[id=FF]{since it}\added{and} vanishes in the absence of \deleted[id=FF]{induced}\added{axial} velocity. \added{It arises from resolving the velocity source into components normal and tangential to the rotor plane.} The lifting-line approximation therefore retains the aerodynamic effects associated with incidence and induced velocity, while the thickness contribution appears only as a higher-order correction in the slender-blade parameter. This distinction is important because large incidence can have a substantial impact by strengthening the loading-related terms and, in particular, the antisymmetric part of the radiation. In the present conditions, however, this effect is expected to be less pronounced. Indeed, in hovering flight the inflow angle is \deleted[id=FF]{usually} \added{generally} small \citep{Bramwell2001}, and blade element momentum theory, Eq. \eqref{bemt relation}, \deleted[id=FF]{suggests}\added{gives} the scaling \(\varphi_i=O(\varepsilon_\ell^{1/2})\). Within the lifting-line framework\deleted[id=FF]{ this implies that, under hovering conditions}, the corrections introduced by induced velocity play \added{therefore} a secondary role \added{in hover. In particular, the antisymmetric-thickness contribution remains correspondingly small. In forward flight, by contrast, the axial velocity would also include the contribution associated with the translational motion, so that the relative importance of this term could be appreciably different.}

This interpretation also helps explain the directivity patterns discussed in \deleted[id=FF]{s}\added{S}ection \ref{sec:dominant-multipoles}. The dominant antisymmetric contribution is associated with the \deleted[id=FF]{lift}\added{thrust} term, whereas the leading symmetric contribution is associated \deleted[id=FF]{mainly} with the \deleted[id=FF]{drag}\added{torque} term, and with thickness entering at higher order. This structure is particularly relevant for Propeller B, for which the lifting-line approximation is more appropriate because the blade operates at higher incidence. This explains why the antisymmetric multipoles in figures \ref{fig:m2_decomp_B02} and \ref{fig:m2_decomp_B08} play a much more visible role in shaping the directivity. Rather than providing only a weak correction to the dominant symmetric radiation, as in Propeller A, they become large enough to modify the lobe orientation and contribute significantly to the observed fore-aft asymmetry.

\subsubsection{Far-field pressure}
Far-field pressure for the lifting line can be found by \deleted[id=FF]{using} substituting Eqs. \eqref{eq:Blm-rot-harm-E-first} and \eqref{eq:Blm-rot-harm-E-second} in Eq. \eqref{eq:pressure-formula-FF}, yielding
\begin{equation}
\begin{aligned}
    \hat{p}^{(FF,LL)}_{m}\left(\mathbf{x}_0\right)
    \approx & \rho_0 a_s^2 M_t^2 N \frac{e^{im M_t \bar{r}_0}}{\bar{r}_0} \sum^{\infty}_{\ell=|m|} \left(-i\right)^{\ell} Y_{\ell m}\!\left(\theta_0,\phi_0\right) \\[2mm]
    \Bigg\{ &B_{L_{\ell m}}
    \int_{\bar{R}_{hub}}^{1}  j_\ell\!\left(m M_t \bar{r}\right) 
    \frac{\gamma(\bar r) }{\cos\varphi_i(\bar r)}\left[c_l(\bar r)-c_d(\bar r)\tan\varphi_i(\bar r)\right] \bar r e^{- i m \phi'_s(\bar r)} d\bar{r} +\\[2mm]
     -i m &B_{D_{\ell m}}
    \int_{\bar{R}_{hub}}^{1}  j_\ell\!\left(m M_t \bar{r}\right) 
    \frac{\gamma(\bar r)}{\cos\varphi_i(\bar r)}\left[ c_l(\bar r) \tan\varphi_i(\bar r)+c_d(\bar r)\right] \bar r e^{- i m \phi'_s(\bar r)} d\bar{r} +\\[2mm]
     - 4  m^2 &B_{T_{\ell m}}
    \int_{\bar{R}_{hub}}^{1}  j_\ell\!\left(m M_t \bar{r}\right) 
    \gamma(\bar r)^2 A_s(\bar r)
    e^{-im\phi'_{s}(\bar r)} d\bar{r}  +\\[2mm]
     + 4i  m &B_{T'_{\ell m}}
    \int_{\bar{R}_{hub}}^{1}  j_\ell\!\left(m M_t \bar{r}\right) 
    \gamma(\bar r)^2  A_s(\bar r) \tan\varphi_i(\bar r)
    e^{-im\phi'_{s}(\bar r)} d\bar{r} \Bigg\}
\end{aligned}
\end{equation}
The phase structure follows the same symmetry argument discussed for the lifting-surface formulation, but is somewhat simpler here because no additional chordwise phase factor is present. The \deleted[id=FF]{lift}\added{thrust} and \deleted[id=FF]{drag}\added{torque} terms are associated with the leading antisymmetric and symmetric multipoles, respectively, and, when the spanwise integrals are real or have similar arguments, the two contributions are approximately in phase. The thickness-related terms, by contrast, are in quadrature with \deleted[id=FF]{lift}\added{thrust} and \deleted[id=FF]{drag}\added{torque}. More generally, the overall phase of each contribution also contains the azimuthal factor \(e^{-im\phi'_s(\bar r)}\), so that the above phase relations should be regarded as the baseline structure, upon which the spanwise source distribution may give additional corrections.

As in the lifting-surface case, the lowest admissible multipoles are the most informative. Symmetric contributions (\(\ell=|m|\)), such as the \deleted[id=FF]{drag}\added{torque}-related and thickness terms, are associated with the first non-zero sectoral harmonic \(Y_{mm}\), whereas antisymmetric contributions (\(\ell=|m|+1\)), such as \deleted[id=FF]{lift}\added{thrust} and \deleted[id=FF]{induced}\added{antisymmetric} thickness, first appear through \(Y_{m+1,m}\).

\subsection{Comparison with \deleted[id=FF]{exact} \added{full surface} integration}
\label{sec:Comparison}

In this section we discuss the results obtained from the presented approximations. First, the predictions obtained with the lifting-surface and lifting-line formulations are compared against the \deleted[id=FF]{``exact''} prediction of Eq. \eqref{eq:Alm-start}, i.e. based on the surface integral evaluated on the true blade geometry. The purpose of this comparison is to assess when the two approximations are applicable and to quantify their accuracy over representative operating conditions. The same propellers A and B introduced in \deleted[id=FF]{s}\added{S}ection \ref{sec:dominant-multipoles} are considered here. In general, the lifting-surface (LS) formulation should perform better for Propeller A, whereas the lifting-line (LL) formulation is anticipated to be more suitable for Propeller B. Since both propellers are unswept, the trailing-edge azimuthal phase is taken as $\phi'_{\mathrm{TE}} = \added{-}c/(2r)$ for the LS model, while $\phi'_s = c/(4r)$ is adopted for the LL model, implying that the lifting line passes through the quarter-chord line of each airfoil section.

For clarity, the main acoustic quantities reported in the figures are defined in the Appendix \ref{appB}. A quantity of particular interest in the present framework is the total directivity of the radiated pressure field
\begin{equation}
    D(\theta_0,\phi_0)
=
\frac{4\pi\displaystyle\sum_{m=-\infty}^{\infty}\frac{1}{m^2}
\left|\sum_{\ell=|m|}^{\infty} (-i)^{\ell+1}\,A_{\ell m}\,Y_{\ell m}(\theta_0,\phi_0)\right|^2}
{\displaystyle\sum_{m=-\infty}^{\infty}\frac{1}{m^2}\sum_{\ell=|m|}^{\infty}\left|A_{\ell m}\right|^2}
\nonumber
\end{equation}
It measures the acoustic radiation in a given direction relative to its spherical average. It is frequently expressed in dB so that directions with \(D(\theta_0,\phi_0)\ge 0\) dB correspond to above-average radiation, whereas directions with \(D<0\) dB identify weakly radiating regions. 

\begin{figure}
\centering
\subfloat[Propeller A, $M_t=0.2$.]
{\includegraphics[width=.5\textwidth]{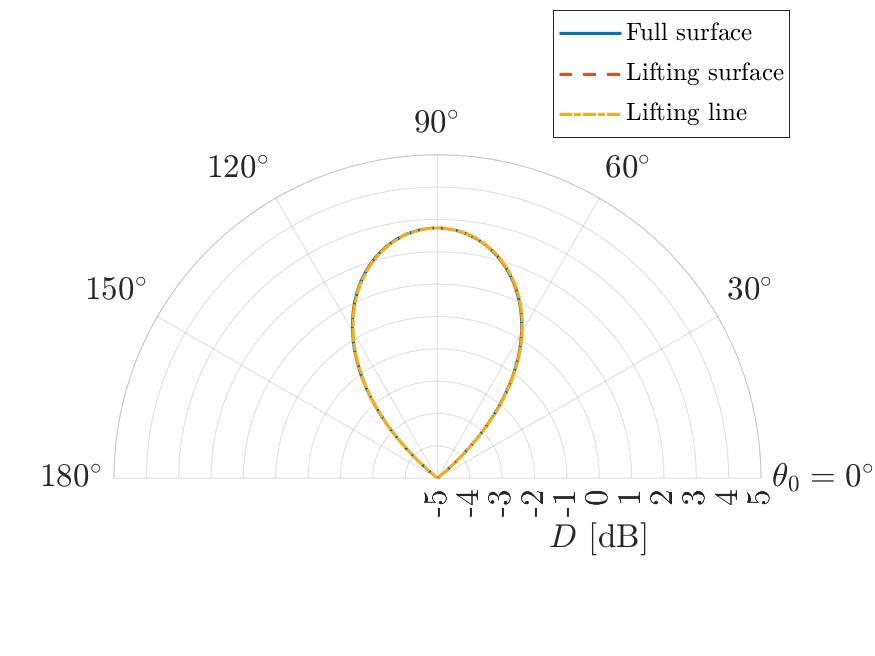}\label{Dir_PropAM02}}\hfill
\subfloat[Propeller B, $M_t=0.2$.]
{\includegraphics[width=.5\textwidth]{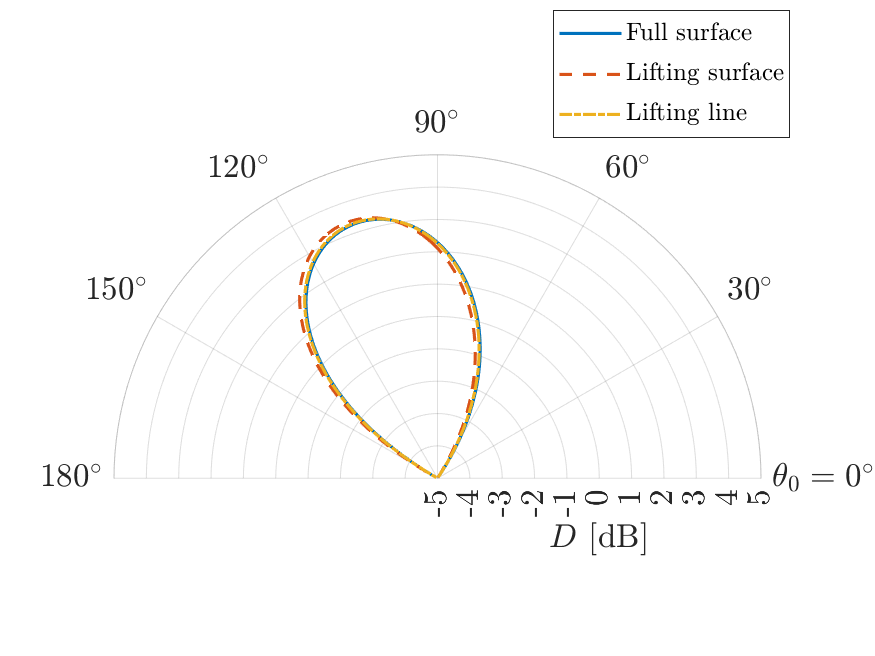}\label{Dir_PropBM02}}\hfill
\subfloat[Propeller A, $M_t=0.8$.]
{\includegraphics[width=.5\textwidth]{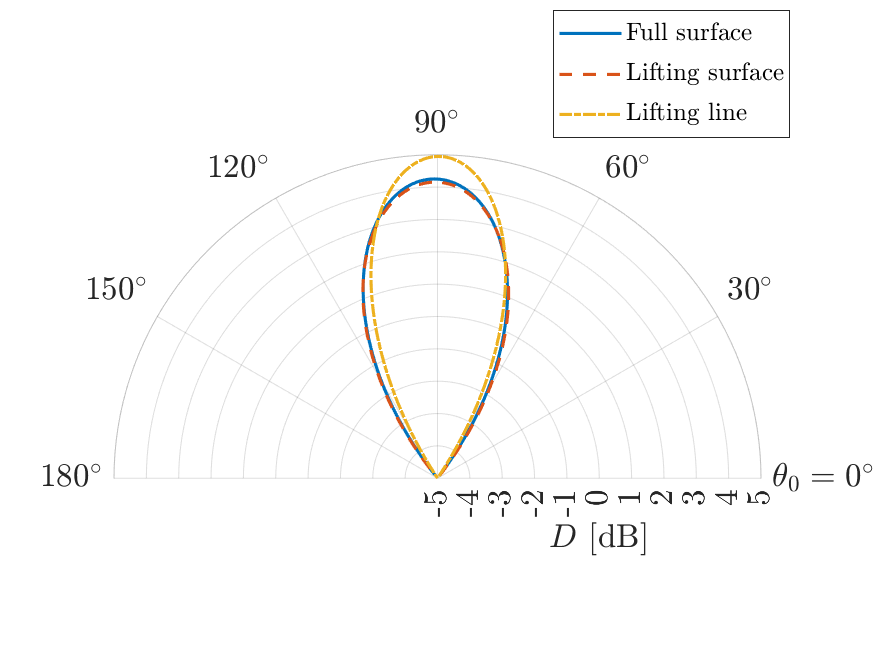}\label{Dir_PropAM08}}\hfill
\subfloat[Propeller B, $M_t=0.8$.]
{\includegraphics[width=.5\textwidth]{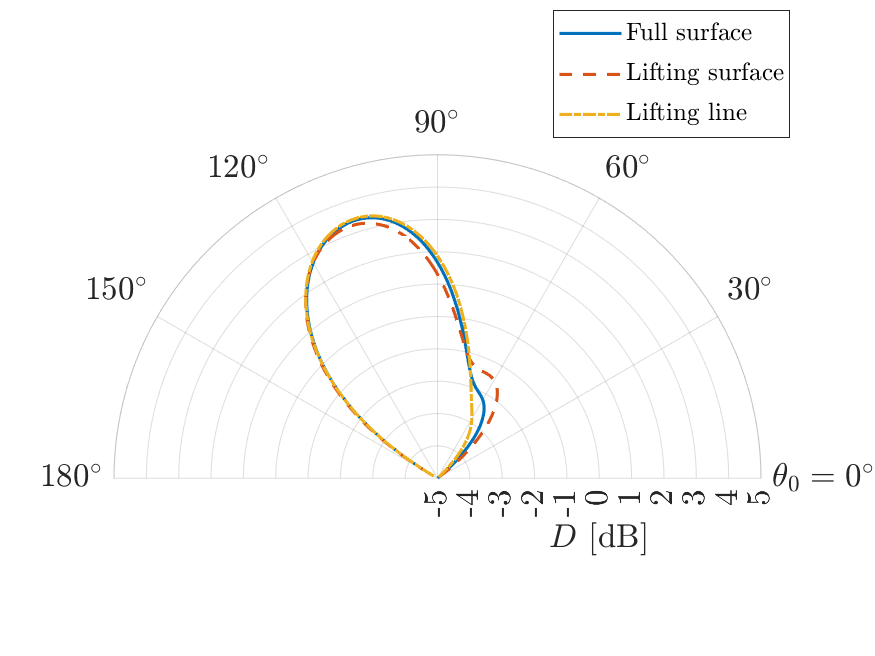}\label{Dir_PropBM08}}\hfill
\caption{Directivity $D$ [dB] in polar axes, comparing exact surface integration with the lifting-surface and lifting-line approximations. \inlinecomment{This figure has been updated in the revised manuscript.}}
\label{directivity_comparisons}
\end{figure}

Figure \ref{directivity_comparisons} shows that both reduced models reproduce the main features of the radiation pattern with generally good accuracy, while also highlighting the expected differences in their respective ranges of validity. Overall, the lifting-surface (LS) approximation provides a\deleted[id=FF]{n impressive} \added{very} close \deleted[id=FF]{agreement} \added{correspondence} with the \deleted[id=FF]{exact solution}\added{full surface integration}. In particular, for Propeller A at \(M_t=0.2\) (panel \ref{Dir_PropAM02}), both approximations are in \deleted[id=FF]{very close}\added{excellent} agreement\deleted[id=FF]{ with the exact solution}. The main lobe is nearly symmetric about the rotor plane, and both reduced models accurately reproduce its orientation, width, and peak level. In this low-Mach-number condition, the first few harmonics dominate and large-chord phase effects remain weak, so that both simplified descriptions provide an adequate representation of the source. At the higher tip Mach number \(M_t=0.8\) (panel \ref{Dir_PropAM08}), the difference between the two approximations becomes more evident. The LS solution remains close to the exact one, correctly reproducing both the overall lobe shape and the slight displacement of the maximum away from \(\theta_0=90^\circ\). By contrast, the LL model predicts a stronger peak located too close to the rotor plane. This behaviour is consistent with the fact that, as \(M_t\) increases, higher harmonics become more relevant and chordwise phase effects play a more visible role. For Propeller B at \(M_t=0.2\) (panel \ref{Dir_PropBM02}), both reduced models again provide a good approximation of the exact directivity, including the tilted main lobe and the overall fore-aft asymmetry. A similar conclusion holds for Propeller B at \(M_t=0.8\) (panel \ref{Dir_PropBM08}): the \deleted[id=FF]{LS}\added{LL} formulation remains almost indistinguishable from the \deleted[id=FF]{exact}\added{reference} solution\added{, particularly in the directions of strongest radiation}, while the \deleted[id=FF]{LL}\added{LS} model also performs well but predicts \deleted[id=FF]{a}\added{slightly} weaker \deleted[id=FF]{peak}\added{values}. LL prediction is closer to the exact solution than it was for Propeller A at the same Mach number, since the relatively small chord of Propeller B makes the compact LL representation more reliable. Overall, \deleted[id=FF]{Fig.}\added{figure} \ref{directivity_comparisons} confirms that the approximations provide \added{an} accurate description of the directivity in their respective regime\added{s} of validity. \deleted[id=FF]{At the same time, the}\added{The} results show that LS is \deleted[id=FF]{the more}\added{a} robust approximation over the range of cases considered here, whereas LL becomes increasingly competitive as the blade geometry and loading approach the slender, higher-incidence regime for which it is intended. It is, however, important to emphasize that under the hovering conditions considered here the blade twist is generally not very large, since stall must be avoided. In particular, the twist remains small near the tip (figure \ref{AeroSpan}), which is also the region that radiates most strongly in subsonic conditions \citep{Parry1989}. This aspect tends to favour the lifting-surface approximation. \deleted[id=FF]{In other operating conditions, such as those associated with high forward speed, the blade twist is significantly larger, and one may therefore expect the lifting-line approximation to improve relative to the lifting-surface one.}

The rapid convergence of the multipole series discussed in \deleted[id=FF]{s}\added{S}ection \ref{sec:dominant-multipoles} allows, at least in the low-frequency regime, to understand the location of the dominant directivity lobe directly from the first two admissible multipoles. If the acoustic field is dominated by the first non-zero rotor harmonic \(m=N\), and only the leading terms \(\ell=N\) and \(\ell=N+1\) are retained, then the directivity reduces to
\begin{equation}
D(\theta_0,\phi_0)
\propto
\left|
(-i)^{N+1}A_{NN}Y_{NN}(\theta_0,\phi_0)
+
(-i)^{N+2}A_{N+1,N}Y_{N+1,N}(\theta_0,\phi_0)
\right|^2
\end{equation}
At low frequency, using the small-argument approximation of the spherical Bessel functions, Eq. \eqref{bessel approx}, one finds that
\begin{equation}
A_{NN}=O(M_t^{N+3}),
\qquad
A_{N+1,N}=O(M_t^{N+4})
\nonumber
\end{equation}
Therefore, the antisymmetric contribution associated with $(\ell,m)=(N+1,N)$ is smaller by one additional power of $M_t$ than the symmetric one associated with $(\ell,m)=(N,N)$. This implies that, in the low-frequency limit, the directivity is primarily governed by the symmetric component, while the antisymmetric correction becomes progressively important only as $M_t$ increases, or when the \deleted[id=FF]{lift}\added{thrust} contribution (that is captured by antisymmetric \added{multi}poles) is sufficiently strong. It is also worth noting that, since both retained \deleted[id=FF]{harmonics}\added{multipoles} share the same azimuthal factor $e^{iN\phi_0}$, this dependence cancels out upon taking the modulus squared. As a result, the directivity depends only on $\theta_0$.

The harmonic power spectra in \deleted[id=FF]{Fig.}\added{figure} \ref{power_comparisons} \added{(see Eq. \eqref{eq:power_calculation} in Appendix \ref{appB})} provide a complementary assessment of the accuracy of the two approximate models, since they reveal the rate at which the acoustic energy decays in frequency. At the lower tip Mach number, shown in panels \ref{Pow_PropAM02} and \ref{Pow_PropBM02}, both approximations reproduce the low-pass character of the spectrum well. In agreement with experimental evidence \citep{Trebble1987a,Trebble1987b}, the radiated power decreases rapidly with increasing \(m\), so that the acoustic output is dominated by the first few blade-passing harmonics. In this regime, the LS and LL predictions are both very close to the \deleted[id=FF]{exact solution}\added{full surface integration} over the dominant \deleted[id=FF]{low-order} harmonics. For Propeller B at \(M_t=0.2\), the agreement remains excellent over the whole range shown, whereas for Propeller A at \(M_t=0.2\) the LL approximation begins to overpredict the highest harmonics. However, these discrepancies occur only where the spectrum is already many tens of decibels below the first harmonic and therefore have little practical influence on the overall tonal content. At the higher tip Mach number, shown in panels \ref{Pow_PropAM08} and \ref{Pow_PropBM08}, the spectral decay becomes markedly weaker and a substantially larger number of harmonics contributes to the radiated power\deleted[id=FF]{, as already anticipated in section \ref{sec:dominant-multipoles}}. Under these conditions, the differences between the two reduced formulations become more informative. For Propeller A at \(M_t=0.8\), the LS approximation follows the \deleted[id=FF]{exact}\added{correct} trend reasonably well, although it tends to underestimate the level of the higher harmonics. By contrast, the LL approximation produces a spectrum that is too flat, with a systematic overprediction that becomes increasingly pronounced as \(m\) increases. This behaviour is consistent with the fact that, for the larger-chord propeller, \deleted[id=FF]{large-chord and} distributed-loading effects play an important role in shaping the high-harmonic content. This is in agreement with the larger directivity error observed in panel \ref{Dir_PropAM08}. For Propeller B at \(M_t=0.8\), the discrepancies are much smaller. The LS approximation is again almost indistinguishable from the exact solution, while the LL model shows only a small discrepancy across the spectrum. This improved behaviour is consistent with the smaller chord of Propeller B, for which compact sectional loading remains an accurate description even when the spectrum broadens. Overall, the spectral comparison complements the directivity results and confirms the expected validity trends of the two simplified models.
\begin{figure}
\centering
\subfloat[Propeller A, $M_t=0.2$.]
{\includegraphics[width=.5\textwidth]{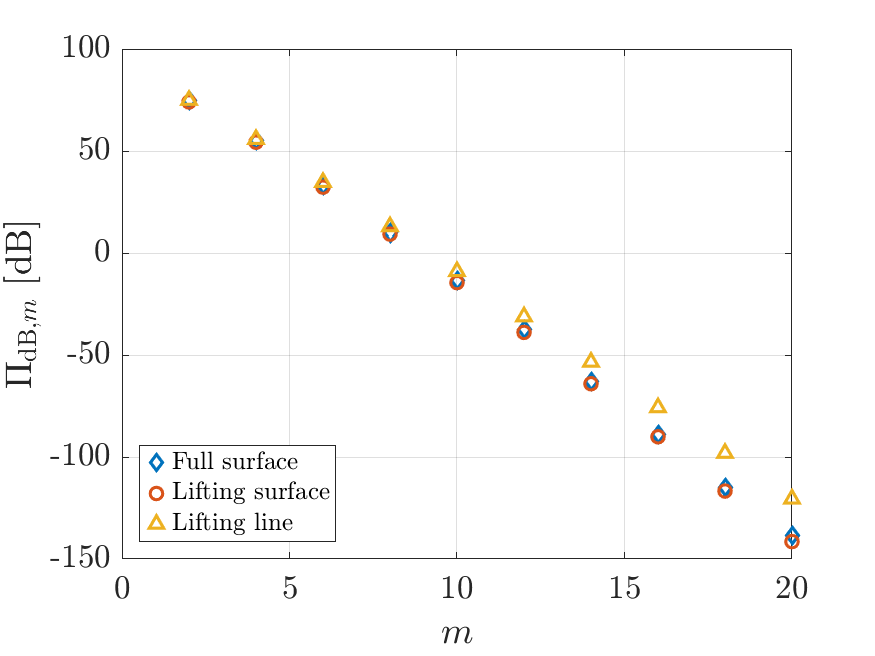}\label{Pow_PropAM02}}\hfill
\subfloat[Propeller B, $M_t=0.2$.]
{\includegraphics[width=.5\textwidth]{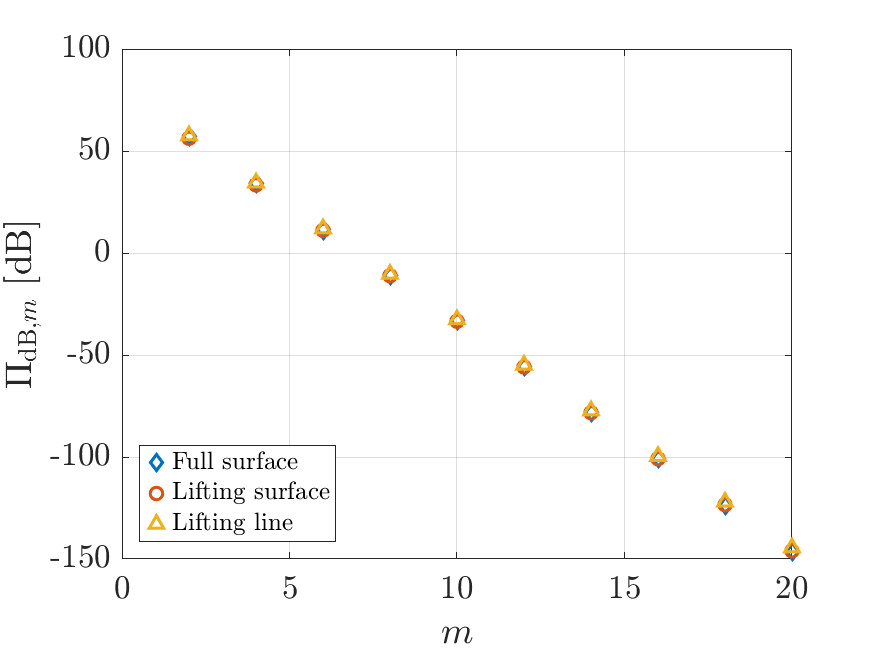}\label{Pow_PropBM02}}\hfill
\subfloat[Propeller A, $M_t=0.8$.]
{\includegraphics[width=.5\textwidth]{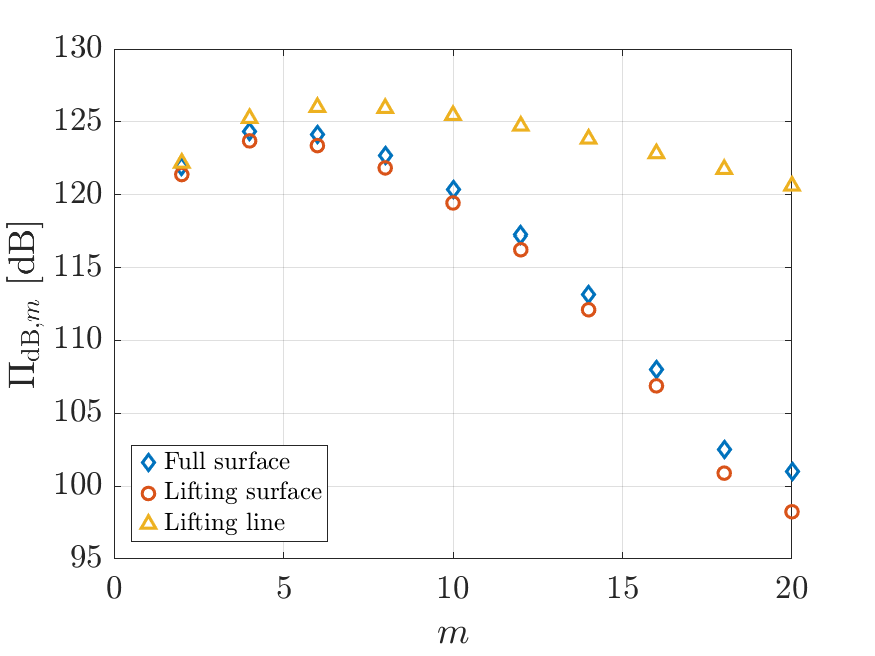}\label{Pow_PropAM08}}\hfill
\subfloat[Propeller B, $M_t=0.8$.]
{\includegraphics[width=.5\textwidth]{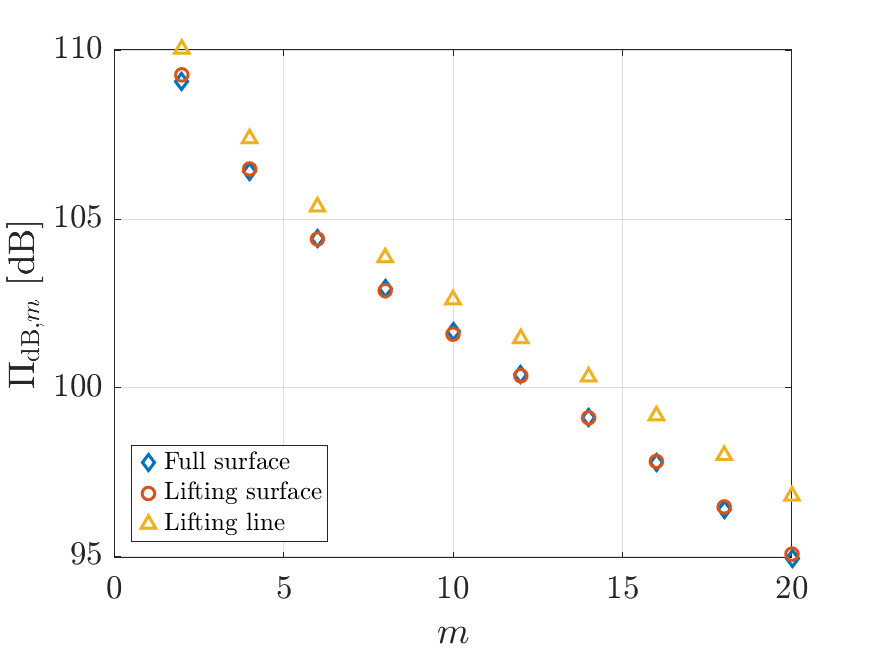}\label{Pow_PropBM08}}\hfill
\caption{\deleted[id=FF]{Acoustic}\added{Sound} power \added{level} per harmonic as a function of the azimuthal index $m$, comparing exact surface integration with the lifting-surface and lifting-line approximations. \inlinecomment{This figure has been updated in the revised manuscript.}}
\label{power_comparisons}
\end{figure}

\begin{figure}
\centering
\subfloat[Propeller A, $M_t=0.2$, $\theta_0\approx 90^\circ$.]
{\includegraphics[width=.5\textwidth]{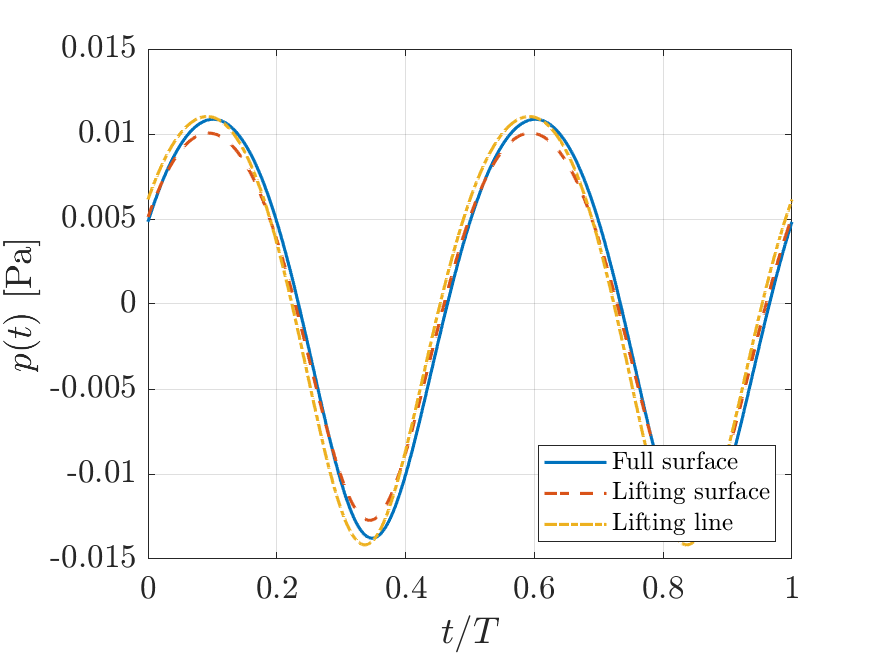}\label{Time_PropAM02}}\hfill
\subfloat[Propeller B, $M_t=0.2$, $\theta_0\approx 107^\circ$.]
{\includegraphics[width=.5\textwidth]{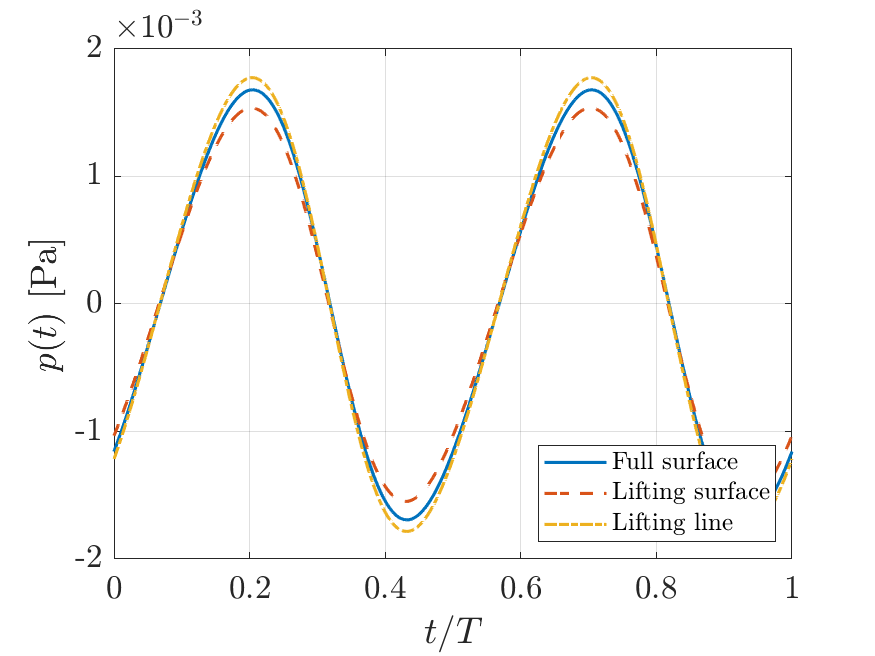}\label{Time_PropBM02}}\hfill
\subfloat[Propeller A, $M_t=0.8$, $\theta_0\approx 91^\circ$.]
{\includegraphics[width=.5\textwidth]{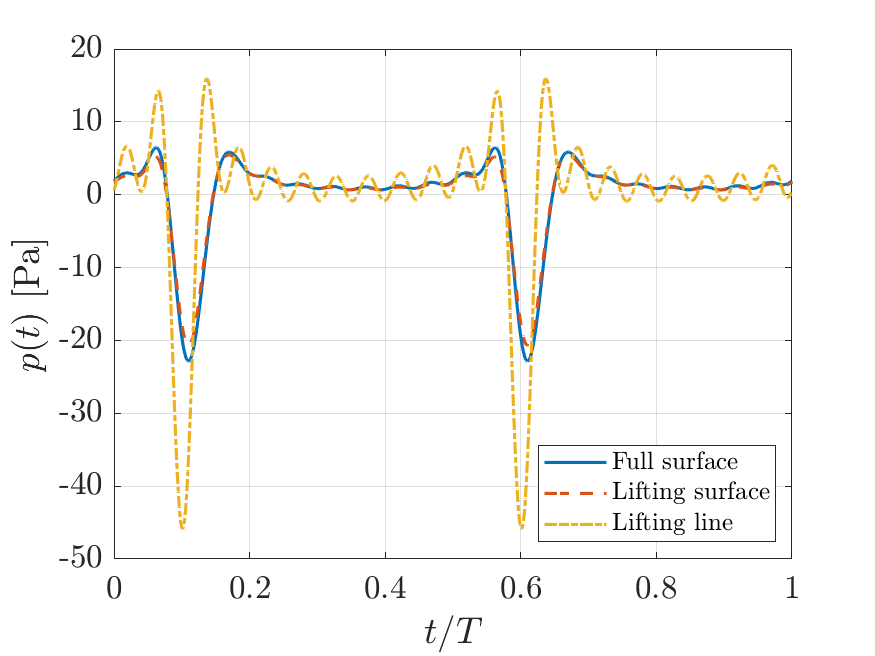}\label{Time_PropAM08}}\hfill
\subfloat[Propeller B, $M_t=0.8$, $\theta_0\approx 108^\circ$.]
{\includegraphics[width=.5\textwidth]{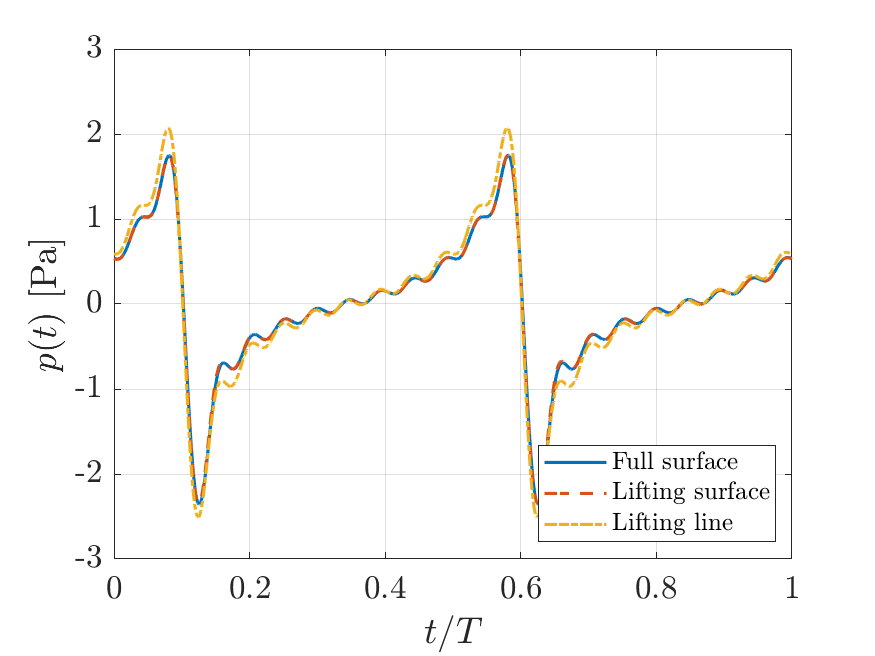}\label{Time_PropBM08}}\hfill
\caption{Far-field acoustic pressure time history over one rotor period, \(T=2\pi/\Omega\), at a fixed observer location in the meridional plane \(\phi_0=0^\circ\), with \(r_0=5\) m and \(\theta_0\) corresponding to the direction of maximum directivity. The results obtained from exact surface integration are compared with the lifting-surface and lifting-line approximations. \inlinecomment{This figure has been updated in the revised manuscript.}}
\label{time_comparisons}
\end{figure}

The time histories shown in \deleted[id=FF]{Fig.}\added{figure} \ref{time_comparisons} provide a \deleted[id=FF]{complementary} time-domain view of the trends already observed in the directivity and harmonic-power comparisons. In each case, the pressure signal is evaluated in the direction of maximum radiation, at a distance \(r_0=50R=5\) m from the propeller centre. At the lower tip Mach number, shown in panels \ref{Time_PropAM02} and \ref{Time_PropBM02}, the waveform is nearly sinusoidal and is dominated by the first blade-passing harmonic, \(m=N\). In agreement with the spectral results, both reduced models reproduce the overall periodic behaviour well in this regime. For Propeller A, the two approximations remain very close to the exact solution, with only minor differences near the extrema. For Propeller B, the agreement is good, confirming that when the spectrum is strongly concentrated in the first few harmonics, both simplified descriptions retain the dominant acoustic content. At the higher tip Mach number, shown in panels \ref{Time_PropAM08} and \ref{Time_PropBM08}, the waveform becomes significantly more structured, with steeper gradients, sharper extrema, and visible small-scale oscillations over the main periodic variation. These features directly reflect the stronger contribution of the higher harmonics, already evident in the corresponding power spectra. In this regime, the differences between the two approximations become more pronounced. For Propeller A (panel \ref{Time_PropAM08}), the LS formulation still follows the \deleted[id=FF]{exact}\added{correct} waveform reasonably well, including the timing and overall shape of the main peaks, although some smoothing of the extrema is visible. By contrast, the LL model exhibits a marked deterioration and the pressure peaks are substantially overpredicted. For Propeller B at \(M_t=0.8\) (panel \ref{Time_PropBM08}), the discrepancies remain much smaller and both reduced models preserve the correct phase and shape of the waveform. Overall, the time-domain comparison is fully consistent with the directivity and power results.

Overall, the LL approximation is an efficient and accurate tool when the acoustic response is dominated by the first few harmonics and the main interest lies in the global tonal content. The LS approximation becomes preferable when the higher-frequency part of the spectrum has a visible impact on the waveform, since it provides a more accurate representation of the frequency structure associated with large-chord and distributed-loading effects. In addition, both approximations are advantageous in terms of computational efficiency, as their use generally leads to a reduction in computational cost when compared with \deleted[id=FF]{exact}\added{full surface} integration. A timing benchmark was designed to isolate the cost of calculating the multipole coefficients \(A_{\ell m}\), since the evaluation of pressure and radiated power involves the same steps in all formulations, namely Eqs. \eqref{eq:pressure_formula} and \eqref{eq:power formula}. The computational cost was assessed for Propeller B at \(M_t=0.8\), retaining harmonics up to \(m=20\) and only the first two multipoles (\(\ell_{\max}-|m|=1\)). The median execution times measured over 100 runs are \(\deleted[id=FF]{8.752}\added{6.895}\times10^{-2}\,\mathrm{s}\) for the \deleted[id=FF]{exact}\added{full} surface integration, \(\deleted[id=FF]{2.015}\added{1.046}\times10^{-3}\,\mathrm{s}\) for the lifting-surface formulation, and \(\deleted[id=FF]{6.985}\added{5.778}\times10^{-4}\,\mathrm{s}\) for the lifting-line formulation. Thus, in the present implementation, both reduced models are more than one order of magnitude faster\deleted[id=FF]{ than the exact one}, with speed-up factors of about \(\deleted[id=FF]{43}\added{66}\) for LS and \(\deleted[id=FF]{125}\added{119}\) for LL. These results are consistent with the structure of the three solutions. The LL formulation is the least expensive because, once the sectional quantities are known, the cost reduces to spanwise operations only. The LS formulation is slightly more expensive because, in addition to the spanwise integration, it also evaluates chordwise Fourier-type integrals. However, with the present truncation at \(m=20\), this additional cost remains modest, since the chordwise quantities are computed once per harmonic and then reused for both retained multipoles. This explains why the difference between LS and LL is limited. The much larger cost of the \deleted[id=FF]{exact method}\added{full surface computation} reflects not only the evaluation of the surface integral itself, but also all the operations required to construct and process the three-dimensional blade geometry. \deleted[id=FF]{The exact formulation}\added{It} is therefore penalized both by the size of the surface discretization and by the geometric preprocessing, which is absent, or considerably lighter, in the simplified formulations. As a result, the latter retain essentially the same acoustic accuracy while achieving a markedly lower computational cost.

\subsubsection{\deleted[id=FF]{Lift, drag}\added{Loading} and thickness contributions}

\begin{figure}
\centering
\subfloat[Lifting surface, Propeller A.]
{\includegraphics[width=.5\textwidth]{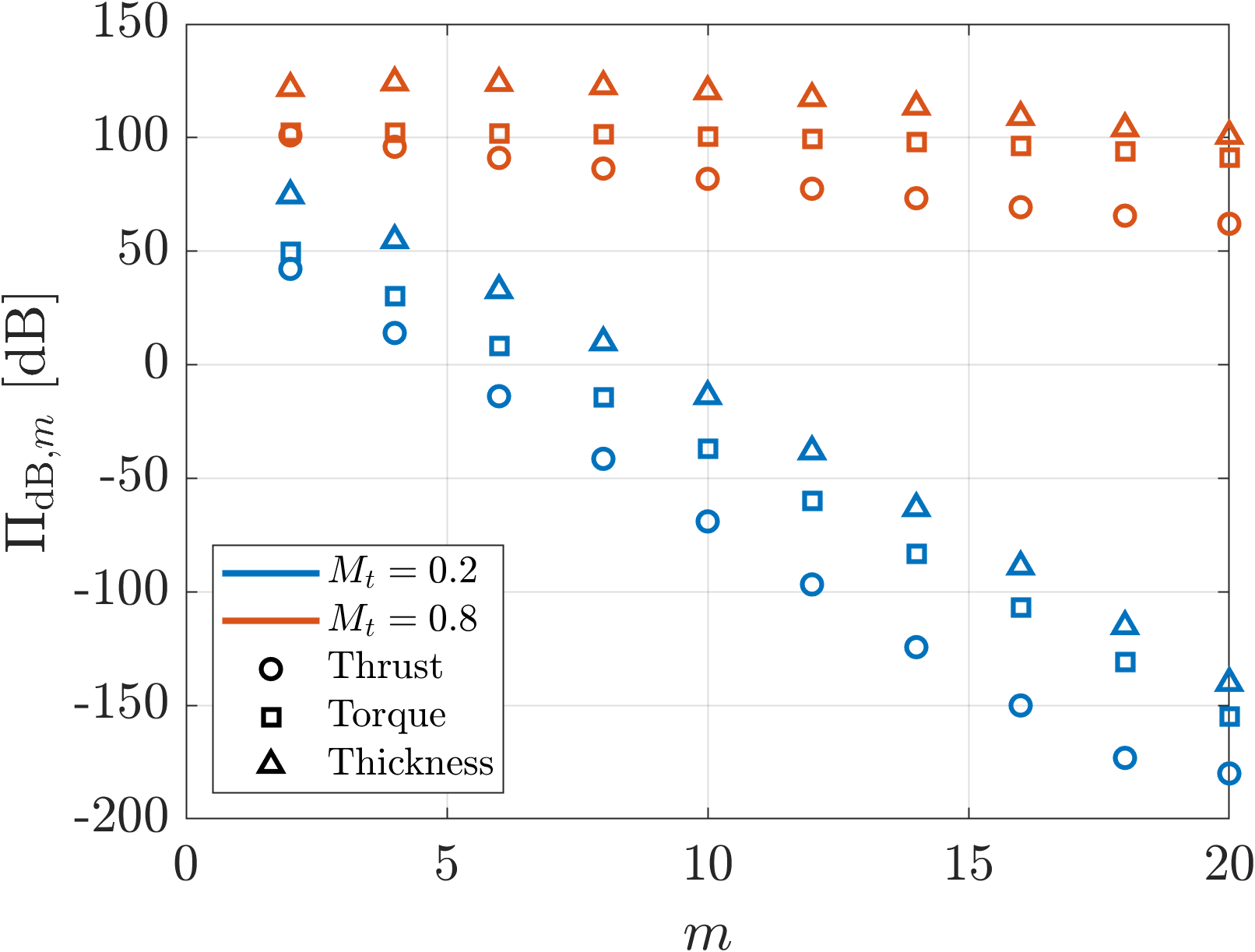}\label{LS_power_contrib_Mt02_Mt08}}\hfill
\subfloat[Lifting line, Propeller B.]
{\includegraphics[width=.5\textwidth]{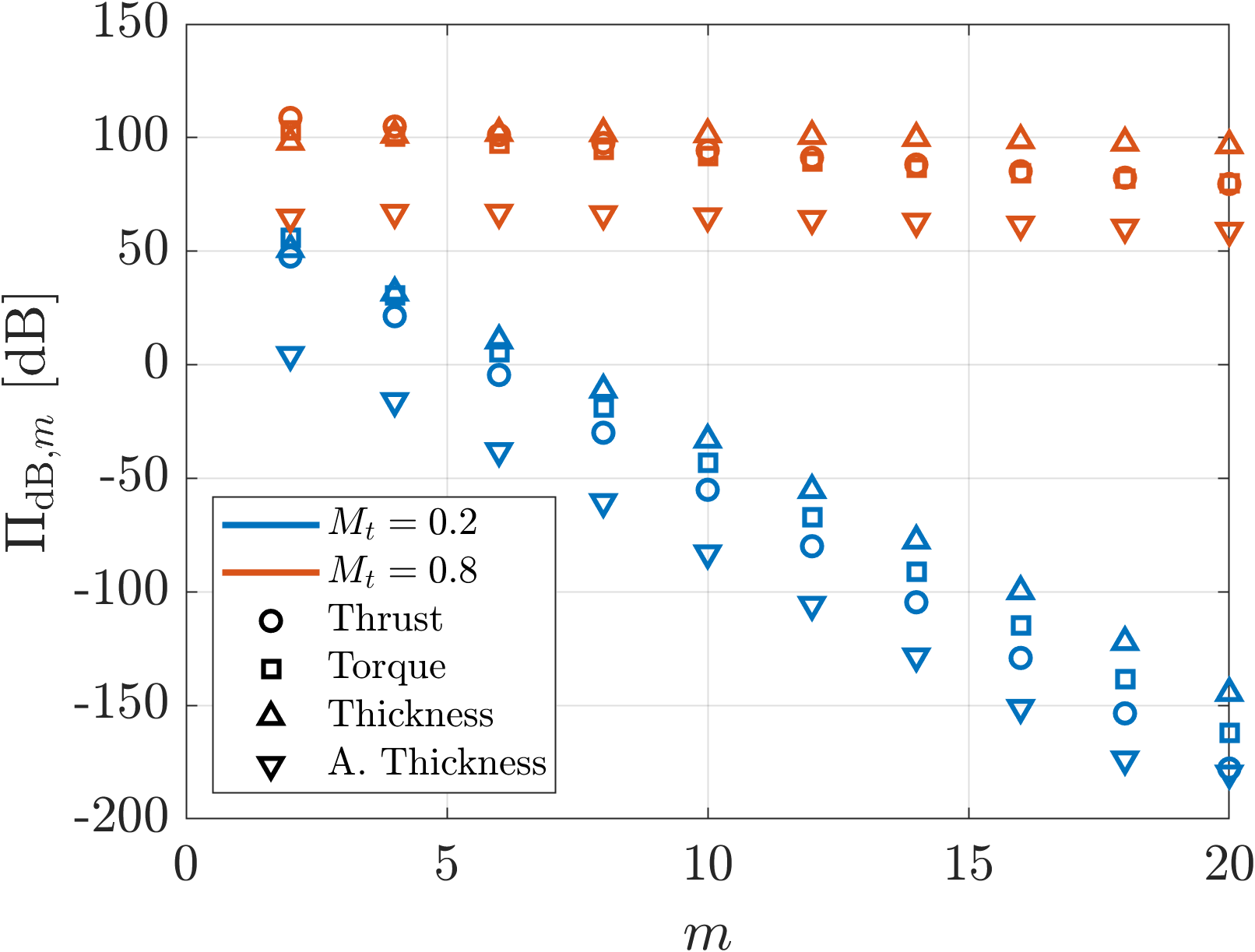}\label{LL_power_contrib_Mt02_Mt08}}
\caption{Sound power level \(\Pi_{dB,m}\) as a function of harmonic index \(m\), comparing the different contributions in the two approximate formulations: (a) lifting-surface (LS), with \deleted[id=FF]{lift}\added{thrust}, \deleted[id=FF]{drag}\added{torque}, and thickness contributions; (b) lifting-line (LL), with \deleted[id=FF]{lift}\added{thrust},  \deleted[id=FF]{drag}\added{torque}, thickness, and \deleted[id=FF]{induced-} \added{antisymmetric} thickness contributions. \inlinecomment{This figure has been updated in the revised manuscript.}}
\label{fig:power_contrib_LS_LL}
\end{figure}

To quantify how the different physical mechanisms play a role in the multipole expansion, we decompose the radiated acoustic power to understand the contribution of the first admissible multipoles. Figure \ref{fig:power_contrib_LS_LL} reports the sound power level associated with each rotor harmonic, decomposed into the individual model contributions (different markers) and shown for the two tip Mach numbers, \(M_t=0.2\) and \(M_t=0.8\) (different colours). \deleted[id=FF]{It should be recalled that the labels ``lift'' and ``drag'' are used here only in an approximate interpretative sense. More precisely, the former corresponds to the antisymmetric contribution, associated primarily with loading normal to the reference plane, whereas the latter corresponds to the symmetric contribution, associated mainly with tangential loading.} \added{The power levels are obtained by evaluating each contribution from its corresponding separated multipole coefficients $A_{\ell m}$. Consequently, the cross-interference terms between different contributions, which are present in the total radiated power, are not included.}

For Propeller A, panel \ref{LS_power_contrib_Mt02_Mt08} shows that the three lifting-surface contributions remain clearly separated in level \added{at $M_t=0.2$}. Over most of the harmonics displayed, the thickness term is the dominant one, while the two loading-related terms remain smaller and exhibit a relative importance that varies with harmonic index. The same ordering is evident in \deleted[id=FF]{T}\added{t}able \ref{tab:LS_power_contrib_m2_first2ell}, where for the first blade-passing harmonic the thickness contribution exceeds both the \deleted[id=FF]{lift}\added{thrust}-related and \deleted[id=FF]{drag}\added{torque}-related terms at both Mach numbers. This hierarchy is consistent with the assumptions underlying the LS approximation. In the small-incidence condition on which the model is based, thickness and \deleted[id=FF]{loading}\added{thrust} enter at comparable order, whereas the \deleted[id=FF]{drag}\added{torque}-related part appears only as a higher-order correction, in agreement with thin-airfoil arguments \citep{Ashley1965}. For the very small angle of attack considered here (\(\alpha<\deleted[id=FF]{2}\added{3}^\circ\), see panel \ref{fig:aero-span-propA}), the blade loading remains weak enough for the thickness contribution to dominate the radiated power, while the loading terms become more visible as \(M_t\) increases. 

Panel \ref{LL_power_contrib_Mt02_Mt08} reports the same power decomposition for the lifting-line formulation. In this case, \added{the thrust, torque and thickness contributions are of comparable magnitude at the lower harmonics. Thus, although the loadings terms enter at leading order in the slender-blade expansion, the symmetric-thickness contribution remains acoustically significant and becomes increasingly important at higher harmonics. By contrast, the antisymmetric-thickness contribution remains substantially weaker than the other terms at both Mach numbers, confirming its secondary role under the hovering conditions considered here.} \deleted[id=FF]{the hierarchy of the various contributions follows directly from the asymptotic structure of the expansion: lift and drag appear at leading order ($\varepsilon_{\ell}$), whereas thickness enters only at the next order ($\varepsilon^2_{\ell}$). The results reflect this ordering clearly. The dominant acoustic output is associated with lift and drag, while the thickness-related contributions remain weaker throughout the range of harmonics considered. In particular, the induced-thickness term is negligible at both Mach numbers and stays well below the other contributions.}
\begin{table}
\centering
\small
\setlength{\tabcolsep}{6pt}
\begin{tabular}{lcc}
\toprule
 & $M_t=0.2$ & $M_t=0.8$ \\
\midrule
$\ell=m=2$ (Torque) & 49.9 & 102.1 \\
$\ell=m=2$ (Thickness) & 74.5 & 121.8 \\
$\ell=m+1=3$ (Thrust) & 42.2 & 101.0 \\
\bottomrule
\end{tabular}
\caption{Lifting-surface power [dB] contributions for the first blade-passing harmonic $m=2$ restricted to the first two admissible multipoles ($\ell=|m|$ and $\ell=|m|+1$) for Propeller A. \inlinecomment{This table has been updated in the revised manuscript.}}
\label{tab:LS_power_contrib_m2_first2ell}
\end{table}

\begin{table}
\centering
\small
\setlength{\tabcolsep}{6pt}
\begin{tabular}{lcc}
\toprule
 & $M_t=0.2$ & $M_t=0.8$ \\
\midrule
$\ell=m=2$ (Torque) & 55.9 & 103.2 \\
$\ell=m=2$ (Thickness) & 50.9 & 98.1 \\
$\ell=m+1=3$ (Thrust) & 47.6 & 108.7 \\
$\ell=m+1=3$ (A. Thickness) & 4.3 & 64.8 \\
\bottomrule
\end{tabular}
\caption{Lifting-line power [dB] contributions for the first blade-passing harmonic $m=2$ restricted to the first two admissible multipoles ($\ell=|m|$ and $\ell=|m|+1$) for Propeller B. \inlinecomment{This table has been updated in the revised manuscript.}}
\label{tab:LL_power_contrib_m2_first2ell}
\end{table}

The two approximations are based on different hypotheses and therefore capture different physical effects. Despite these differences, there exists an overlap region in which the two models predict the same leading-order multipole content. This common limit is obtained by letting the chordwise compactness parameter, embedded in \(\gamma\), tend to zero in the LS formulation and the inflow angle \(\varphi_i\) tend to zero in the LL formulation. The resulting multipole coefficient can be written as
\begin{equation}
I_{\ell m}
= \deleted[id=FF]{\lim_{\gamma\to 0} A^{(LS)}_{\ell m}
= \lim_{\varphi_i\to 0} A^{(LL,1)}_{\ell m}
=} i \rho_0 a_s^2 m M_t^3 N 
B_{L_{\ell m}}
\int_{\bar R_{\mathrm{hub}}}^{1} j_\ell(m M_t \bar r) 
\gamma(\bar r) c_l(\bar r) \bar r 
e^{-im\phi'_s(\bar r)} d\bar r
\label{eq:intersection_I_lm}
\end{equation}
Equation \eqref{eq:intersection_I_lm} shows that, in the common limit, both descriptions reduce to the same lift-generated multipole coefficient. On the LS side, the limit \(\gamma\ll1\) has two consequences. First, the chordwise phase factor $\exp\!\left[-im\frac{2\gamma(\bar r)}{\bar r}\xi\right]$ tends to unity, so that the large-chord interference retained by the full LS model disappears. Second, the thickness contribution, which is proportional to \(\gamma^2\), becomes smaller than the lift term, which scales as \(\gamma\). The chordwise lift integral then reduces to the sectional lift coefficient, $\int_0^1 f_L(\bar r,\xi)\,d\xi \;\approx\; c_l(\bar r)$, and the only remaining phase information is the spanwise azimuthal factor \(e^{-im\phi'_s(\bar r)}\). In other words, once the blade becomes chordwise compact, the LS model loses sensitivity to the detailed loading distribution along the chord and collapses onto a spanwise lift description. On the LL side, the limit \(\varphi_i\to0\) suppresses the terms that are specific to large inflow angle. In particular, the drag-like contribution associated with the inclined sectional force \deleted[id=FF]{vanishes}\added{becomes negligible}, and the sectional lift direction aligns with \(\deleted[id=FF]{-}\hat{\boldsymbol\theta}\) on the plane \(\theta=\pi/2\). Intuitively, the overlap region corresponds to a blade that is both slender and nearly planar: under these conditions, neither large-chord interference nor inflow-angle corrections are strong enough to affect the acoustic field. Outside this intersection region, the two formulations differ in predictable ways, consistent with the results discussed above. From a practical standpoint, the two methods should therefore be viewed as complementary rather than competing approaches.

\section{Experimental Validation}
\label{sec:Experimental Validation}

This section compares the results obtained with the multipole coefficients derived from the proposed approximations (Eq. \eqref{eq:A2lm} for LS and Eqs. \eqref{eq:Blm-rot-harm-E-first} and \eqref{eq:Blm-rot-harm-E-second} for LL) against experimental acoustic measurements. \added{The validation is divided into two parts.} \deleted[id=FF]{Two different propellers are considered, each chosen to match the assumptions and target regime of one modelling approach: TUC-TUC propeller is used to validate the lifting-surface formulation, while APC propeller is used to validate the lifting-line formulation. Since the experiments refer to different geometries and operating conditions, the purpose of this section is to verify that each approach reproduces the main acoustic features in its intended domain. For both datasets, model and measurements are compared in terms of the sound pressure level as a function of elevation angle at first blade-passing frequency (BPF), as well as the single-microphone pressure spectrum and pressure time history evaluated at the direction of maximum radiation.} 

\added{First, two low tip-Mach number cases are considered using the raw experimental data available for two propellers: the TUC-TUC propeller \citep{bensignor2025,Bensignor2026} and the APC propeller \citep{Grande2022b,Grande2022a}. The raw pressure time histories are available for these datasets, and both the LS and LL approximations are examined. The comparison is carried out in terms of the sound pressure level (SPL, see Eq. \eqref{eq:SPL_calculation} in Appendix \ref{appB}) as a function of elevation angle at the first and second blade-passing frequency (BPF), as well as microphone pressure spectra and pressure time histories measured in directions of strong tonal radiation.}  \deleted[id=FF]{The pressure field at the microphone locations is evaluated using the exact expansion in Eq. \eqref{eq:pressure_formula}, retaining the spherical Hankel functions in the observer factor, since the far-field approximation is not satisfied. For clarity and reproducibility, table \ref{tab:exp-datasets} summarises the geometry and operating conditions of the two experimental datasets.} 

\added{The second part addresses higher-tip-Mach-number conditions, for which the present analysis relies on experimental data available in the literature, in particular the measurements of \citet{Boxwell1978}. This comparison is used to assess the behaviour of the formulation outside the low-tip-Mach-number range covered by the raw TUC-TUC and APC datasets.} 

\added{In all cases, the pressure field at the microphone locations is evaluated using the exact expansion in Eq.~\eqref{eq:pressure_formula}, retaining the spherical Hankel functions in the observer factor, since the far-field approximation is not satisfied. The predicted time-domain signal is reconstructed with Eq. \eqref{pressure fourier} from the retained harmonics. Before discussing the comparisons, it is important to recall that the validation of higher BPF harmonics is particularly delicate. Several mechanisms that are not included in the present isolated-propeller model may affect the measured high-frequency content. These include flow confinement or recirculation in the test environment, small unsteady motions of the rotor, blade-to-blade loading asymmetries, and motor-related acoustic contamination \citep{Stephenson2019,Nardari2019,Zhong2020}. The literature shows that recirculation effects are especially relevant in static or hover conditions in confined facilities, where the recirculated turbulent flow can be re-ingested by the rotor and generate additional unsteady loading \citep{Casalino2023}. The resulting increase is mainly observed at higher BPF harmonics and at observer locations where unsteady-loading noise radiates more efficiently, whereas the first BPF and in-plane radiation may be less affected. For this reason, the first and second BPF directivities are used as the primary quantitative validation of the tonal field predicted by the LS and LL approximations. The higher-frequency spectra and time-domain waveform details are instead interpreted as a more demanding assessment of the model trends.}

\inlinecomment{The two previous subsections devoted to the LS and LL comparisons have been merged into a single subsection.}

\subsection{\added{Low tip-Mach number}}

\begin{figure}
\centering
\subfloat[TUC-TUC propeller.\label{fig:tuctuc_geom}]
{\includegraphics[width=0.45\textwidth]{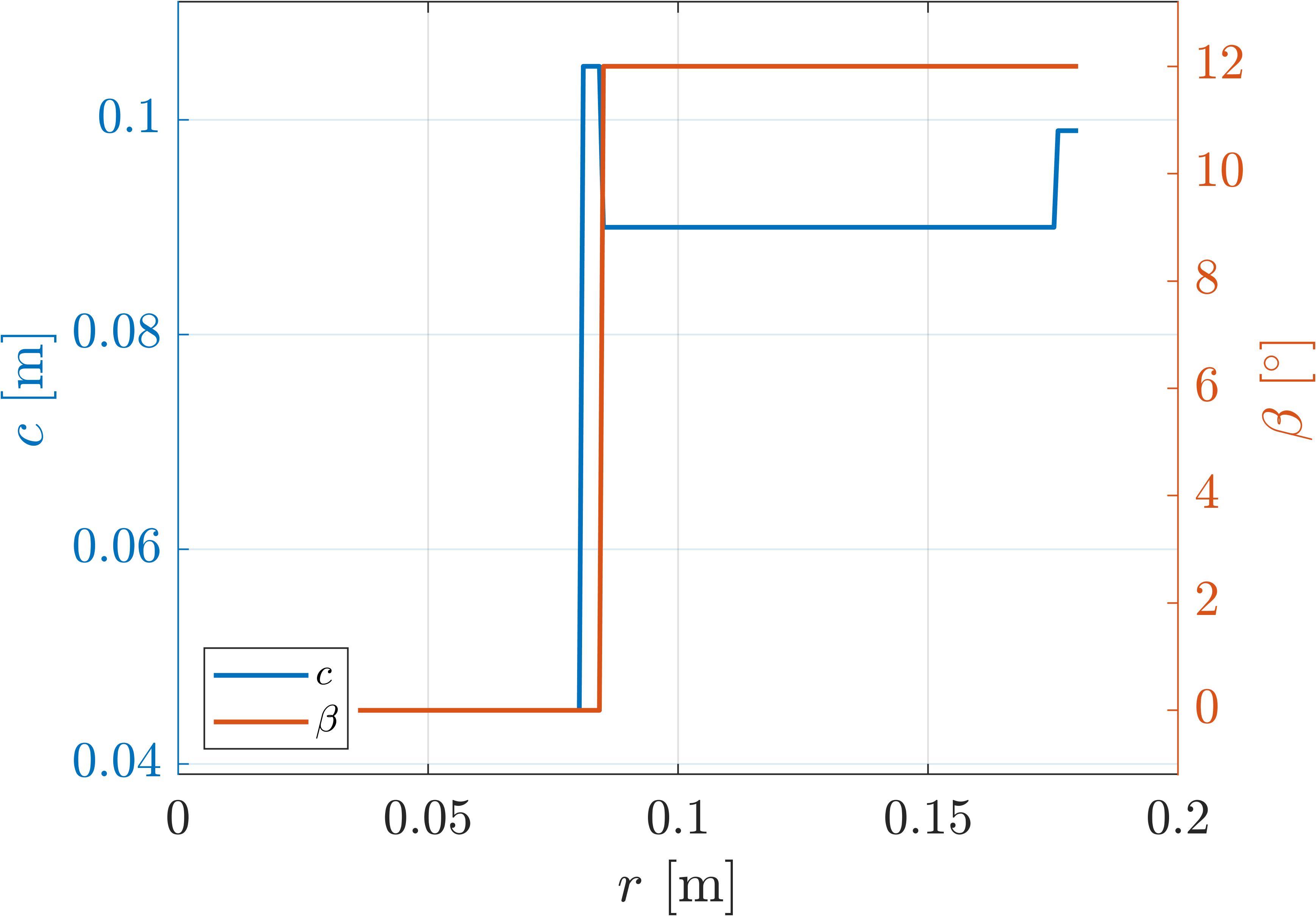}}\hfill
\subfloat[APC propeller.\label{fig:apc_geom}]
{\includegraphics[width=0.45\textwidth]{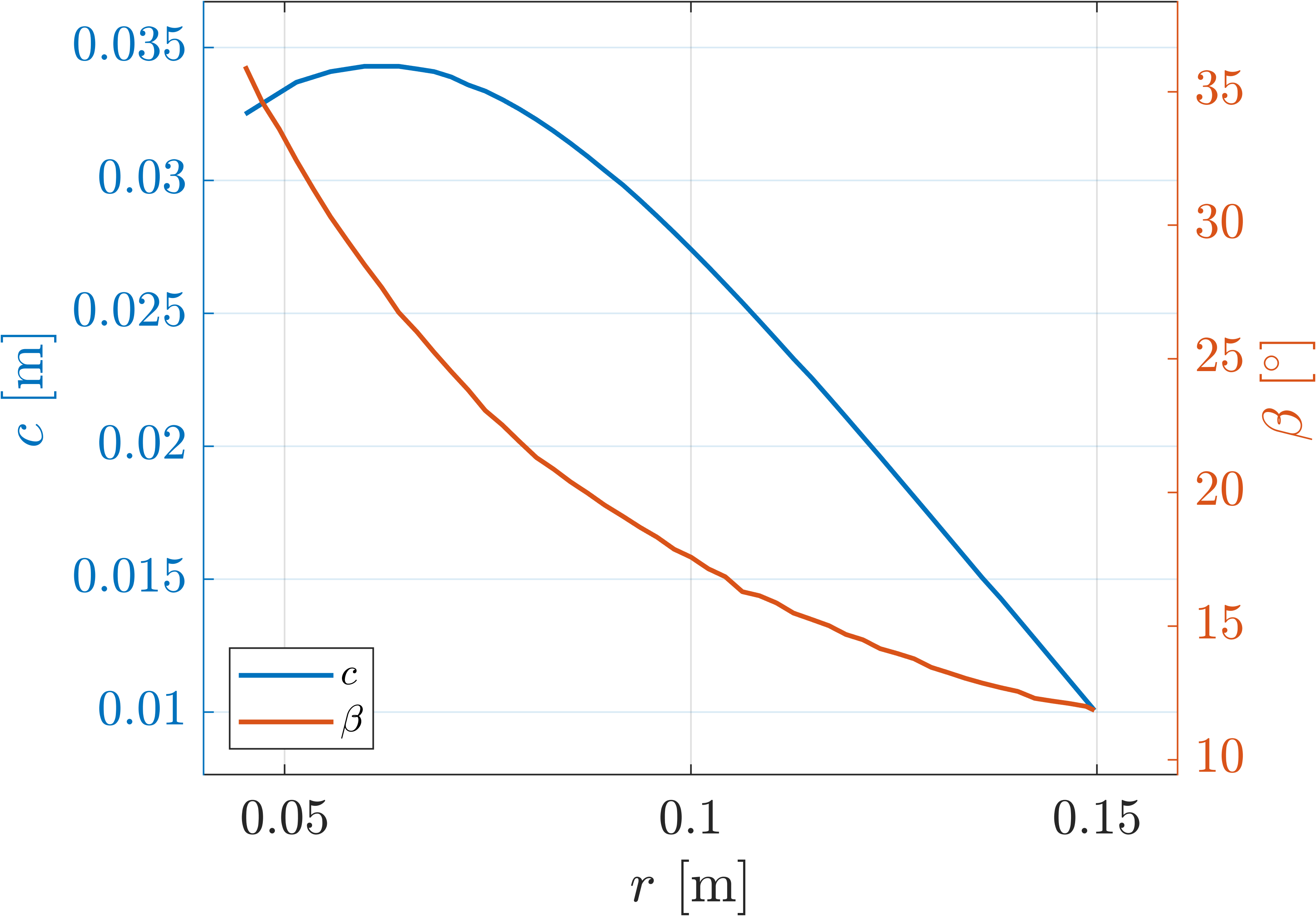}}
\caption{Spanwise chord and twist distributions of the two propellers considered in the low-tip-Mach-number validation. \inlinecomment{This figure has been added in the revised manuscript.}}
\label{fig:lowMt_geom}
\end{figure}

\added{This subsection considers the two low-tip-Mach-number datasets for which the raw pressure time histories are available. For clarity and reproducibility, table \ref{tab:exp-datasets} summarises the geometry and operating conditions of the two experimental datasets, while figure \ref{fig:lowMt_geom} reports the spanwise chord and twist distributions used for the TUC-TUC and APC propellers. The aerodynamic input required by the LS and LL acoustic coefficients is obtained using the same BEMT implementation adopted in Section \ref{sec:dominant-multipoles}. The aerodynamic databases are constructed at the nominal rotational speeds of the two propellers, namely $6000$ and $5000$ rpm for the TUC-TUC and APC configurations, respectively. These values differ slightly from the measured rotational speeds used in the acoustic calculations.}

\begin{table}
\centering
\small
\setlength{\tabcolsep}{7pt}
\begin{tabular}{lcc}
\toprule
 & \textbf{TUC-TUC Propeller} & \textbf{APC Propeller} \\
\midrule
Number of blades $N$ & 2 & 2 \\
Tip radius $R$ [m] & 0.180 & 0.150 \\
Hub radius $R_{\mathrm{hub}}$ [m] & 0.036 & 0.048 \\
Mean chord length $c$ [m] & 0.045/0.090 & 0.024 \\
Twist angle $\beta$ [$^\circ$] & 0/12 (constant) & 33-12 (linear) \\
Airfoil & NACA 0030/0018 & NACA 4412 \\
\midrule
Rotational speed \deleted[id=FF]{$n$}[rpm] & 6121 & 5023 \\
Tip Mach $M_t$ [-] & 0.34 & 0.23 \\
\midrule
Microphone distance $r_0$ [m] & 1.27 & 1.20 \\
Microphone angles $\theta_0$ [$^\circ$] & $\{70,\dots,135\}$ & $\{60,\dots,120\}$ \\
\added{$k r_0$ [-]} & \added{2.39} & \added{1.86} \\
\midrule
Sampling frequency $f_s$ [Hz] & 204800 & 102400 \\
Record length [s] & 30 & 30 \\
Welch window length [samples] & 65536 & 32768 \\
Overlap [\%] & 50 & 50 \\

\bottomrule
\end{tabular}
\caption{Experimental \deleted[id=FF]{datasets}\added{parameters} used for models validation\deleted[id=FF]{.}\added{:} TUC-TUC Propeller \citep{Bensignor2026} \deleted[id=FF]{is used for lifting-surface (LS) validation} and APC Propeller \citep{Grande2022b}\deleted[id=FF]{ for lifting-line (LL) validation}. \inlinecomment{This table has been updated in the revised manuscript.}}
\label{tab:exp-datasets}
\end{table}

\deleted[id=FF]{In this section the lifting-surface model is compared with measurements for TUC-TUC propeller. The propeller} \added{The TUC-TUC propeller} has two blades, each consisting of an inboard and an outboard section: the inboard portion (NACA 0030, chord $c=0.045$ m) has zero twist angle $\beta$, whereas the outboard portion (NACA 0018, chord $c=0.090$ m) is twisted at $\beta=12^\circ$. \added{The midplate is neglected, whereas the tip plate is represented as a NACA 0018 section with an increased chord, following the modelling adopted by \citet{Bensignor2026}.} \deleted[id=FF]{The LS multipole coefficients are computed from the blade geometry and the prescribed aerodynamics obtained with a blade element momentum theory code. The predicted time-domain signal is reconstructed with Eq. \eqref{pressure fourier} from the retained harmonics.} \added{The experimental tonal levels are obtained by integrating the one-sided Welch power spectral density over a 20 Hz frequency band centred on each BPF harmonic.}

\added{Tables~\ref{tab:tuctuc_validation_alm_ls} and~\ref{tab:tuctuc_validation_alm_ll} report the leading multipoles obtained with the LS and LL approximations, for the two propellers. In both cases, the dominant contribution is associated with the lowest admissible order, $\ell=|m|$, with a rapid decay as $\ell-|m|$ increases. The two formulations therefore lead to the same qualitative multipole hierarchy. The LL coefficients are slightly larger, but the relative ordering of the modes is essentially unchanged.}

\begin{table}
\centering
\def~{\hphantom{0}}
\begin{tabular}{ccccc}
$\ell-|m|$ & $m=2$ & $m=4$ & $m=6$ & $m=8$ \\ \hline
0 & $9.274\times 10^{0}$ & $5.372\times 10^{0}$ & $1.659\times 10^{0}$ & $3.850\times 10^{-1}$ \\ 
1 & $1.028\times 10^{0}$ & $2.772\times 10^{-1}$ & $5.527\times 10^{-2}$ & $9.877\times 10^{-3}$ \\ 
2 & $3.811\times 10^{-2}$ & $4.187\times 10^{-2}$ & $1.720\times 10^{-2}$ & $4.701\times 10^{-3}$ \\ 
3 & $5.615\times 10^{-3}$ & $3.011\times 10^{-3}$ & $8.206\times 10^{-4}$ & $1.761\times 10^{-4}$ \\ 
\end{tabular}
\caption{Leading multipoles $|A_{\ell m}^{\mathrm{LS}}| \added{[\mathrm{Pa}]}$ for the TUC-TUC \deleted[id=FF]{lifting-surface} validation case. \inlinecomment{This table has been updated in the revised manuscript.}}
\label{tab:tuctuc_validation_alm_ls}
\end{table}

\begin{table}
\centering
\def~{\hphantom{0}}
\begin{tabular}{ccccc}
$\ell-|m|$ & $m=2$ & $m=4$ & $m=6$ & $m=8$ \\ \hline
0 & $1.006\times 10^{1}$ & $6.532\times 10^{0}$ & $2.445\times 10^{0}$ & $7.490\times 10^{-1}$ \\ 
1 & $1.091\times 10^{0}$ & $3.253\times 10^{-1}$ & $7.595\times 10^{-2}$ & $1.676\times 10^{-2}$ \\ 
2 & $4.118\times 10^{-2}$ & $5.027\times 10^{-2}$ & $2.490\times 10^{-2}$ & $8.956\times 10^{-3}$ \\ 
3 & $5.945\times 10^{-3}$ & $3.519\times 10^{-3}$ & $1.125\times 10^{-3}$ & $2.993\times 10^{-4}$ \\ 
\end{tabular}
\caption{Leading multipoles $|A_{\ell m}^{\mathrm{LL}}| \added{[\mathrm{Pa}]}$ for the TUC-TUC \deleted[id=FF]{lifting-surface} validation case. \inlinecomment{This table has been updated in the revised manuscript.}}
\label{tab:tuctuc_validation_alm_ll}
\end{table}

\deleted[id=FF]{Figure \ref{fig:exp-LS-directivity} compares the measured and predicted first BPF  ($m=2$) levels across the microphone array. The comparison isolates the dominant tonal component and provides a check of the angular accuracy predicted by the LS. Figures \ref{fig:exp-LS-time} and \ref{fig:exp-LS-spectrum} report the time history and the pressure spectrum at $\theta_0=90^\circ$. The spectrum is computed using Welch averaging.}

\begin{figure}
\centering
\subfloat[First \added{and second} blade-passing-frequency sound pressure level in the meridional plane ($\phi_0=0$).]
{\includegraphics[width=0.6\textwidth]{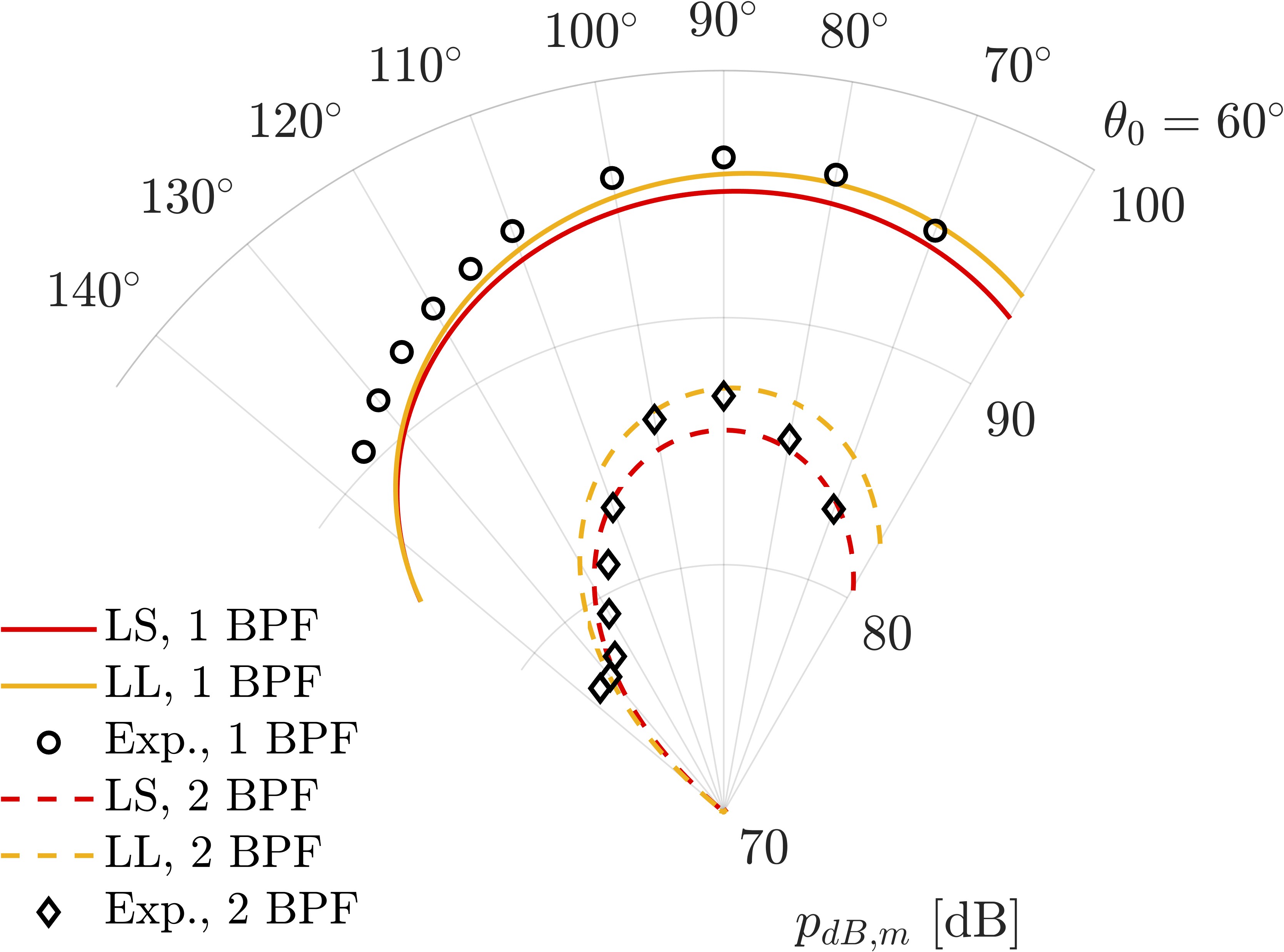}\label{fig:TUCTUC_directivity}}\\
\centering
\subfloat[Time-domain pressure at the observer location $\theta_0=90^\circ$.]
{\includegraphics[width=0.45\textwidth]{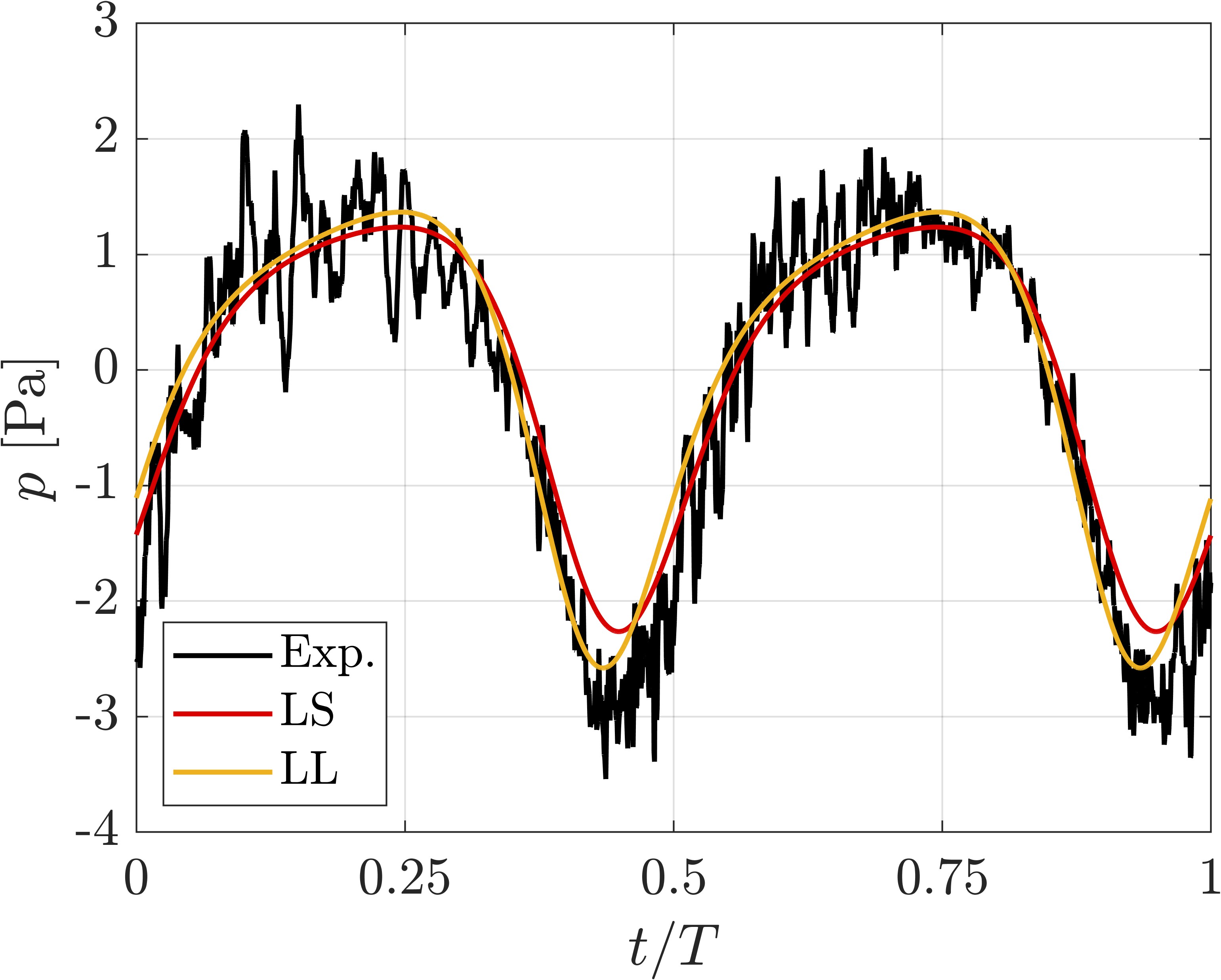} \label{fig:TUCTUC_time90}}\hfill
\subfloat[Narrowband $p_{dB,m}$ spectrum at the observer location $\theta_0=90^\circ$.]
{\includegraphics[width=0.45\textwidth]{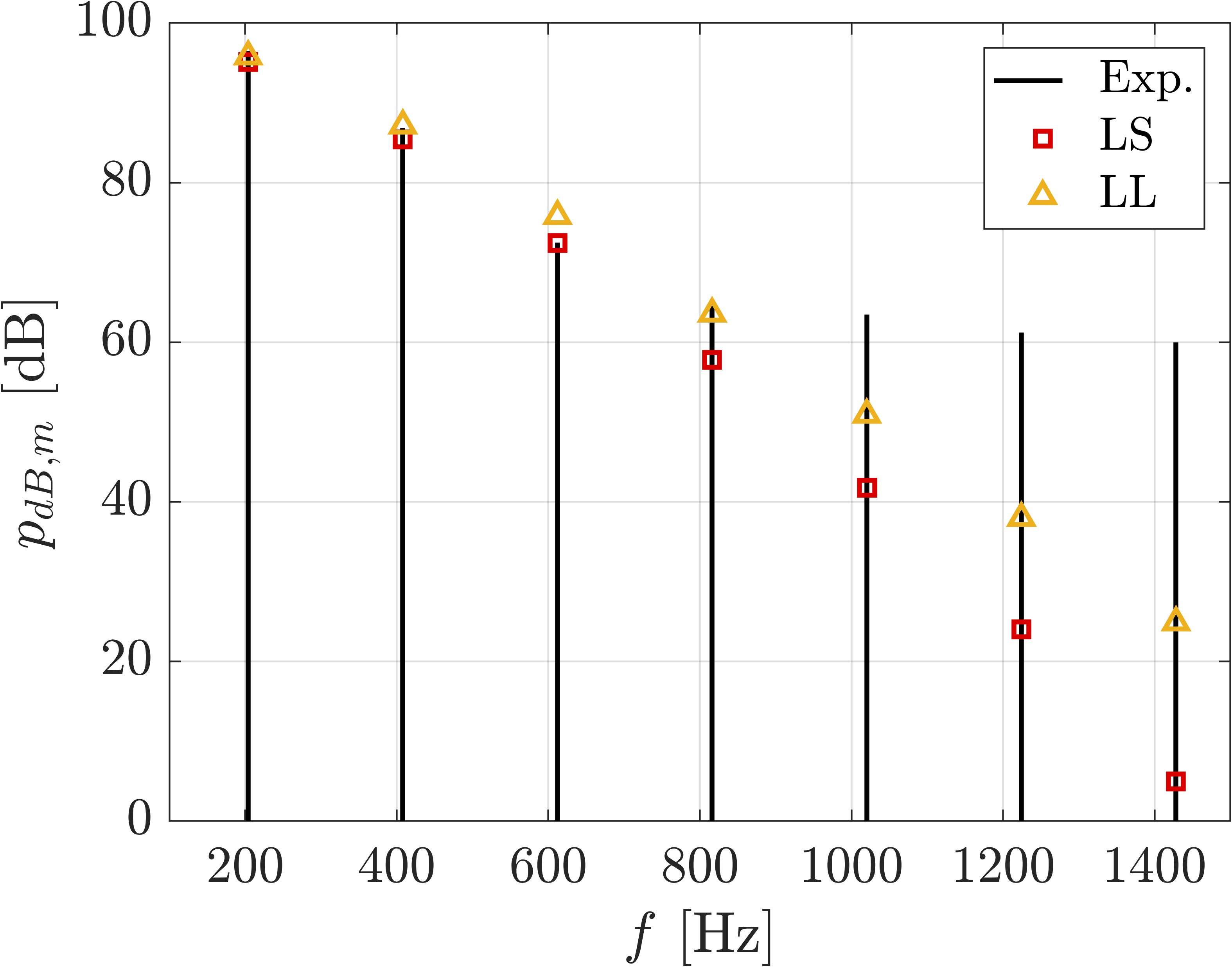}\label{fig:TUCTUC_spectrum90}}\\
\centering
\subfloat[Time-domain pressure at the observer location $\theta_0=100^\circ$.]
{\includegraphics[width=0.45\textwidth]{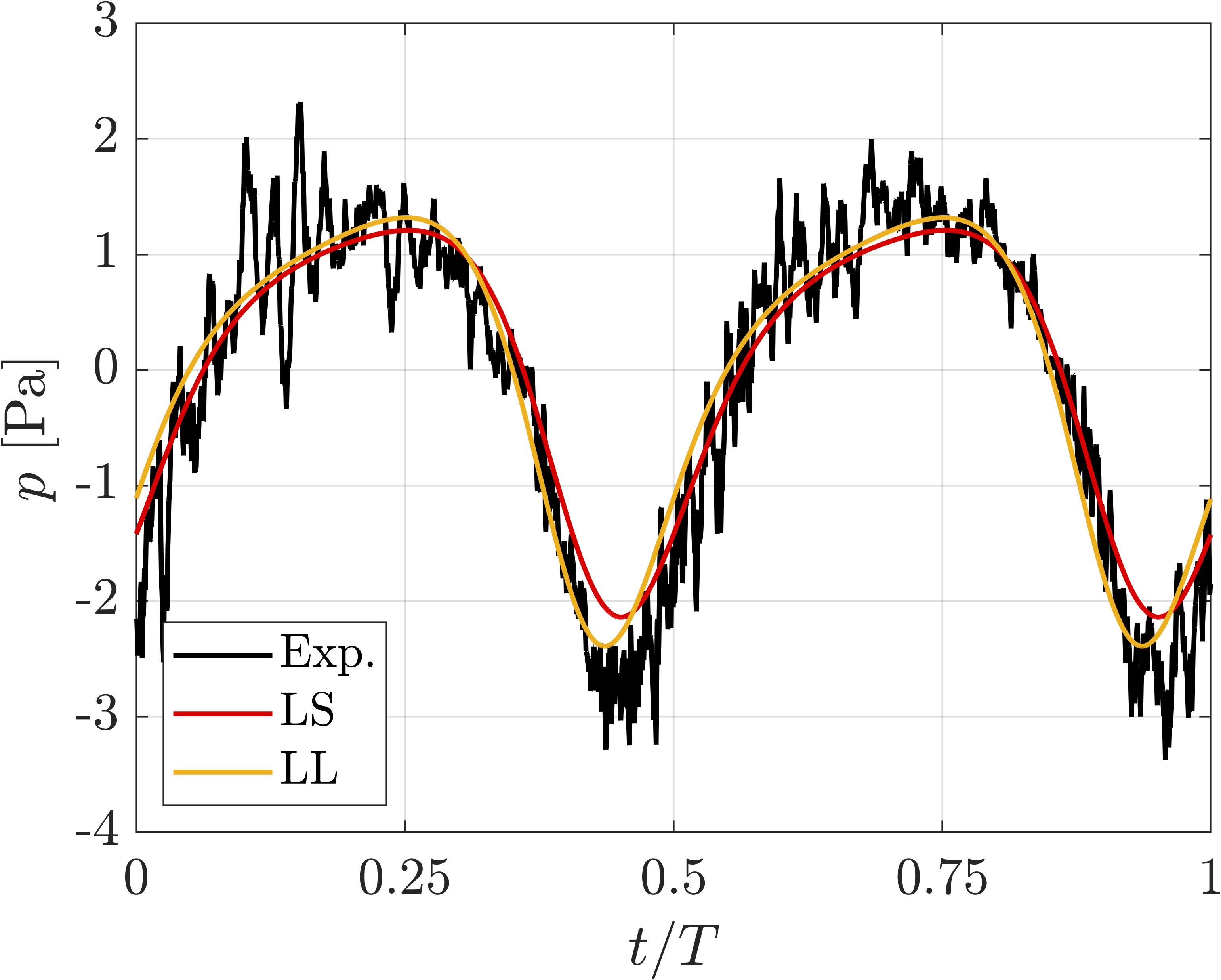} \label{fig:TUCTUC_time100}}\hfill
\subfloat[Narrowband $p_{dB,m}$ spectrum at the observer location $\theta_0=100^\circ$.]
{\includegraphics[width=0.45\textwidth]{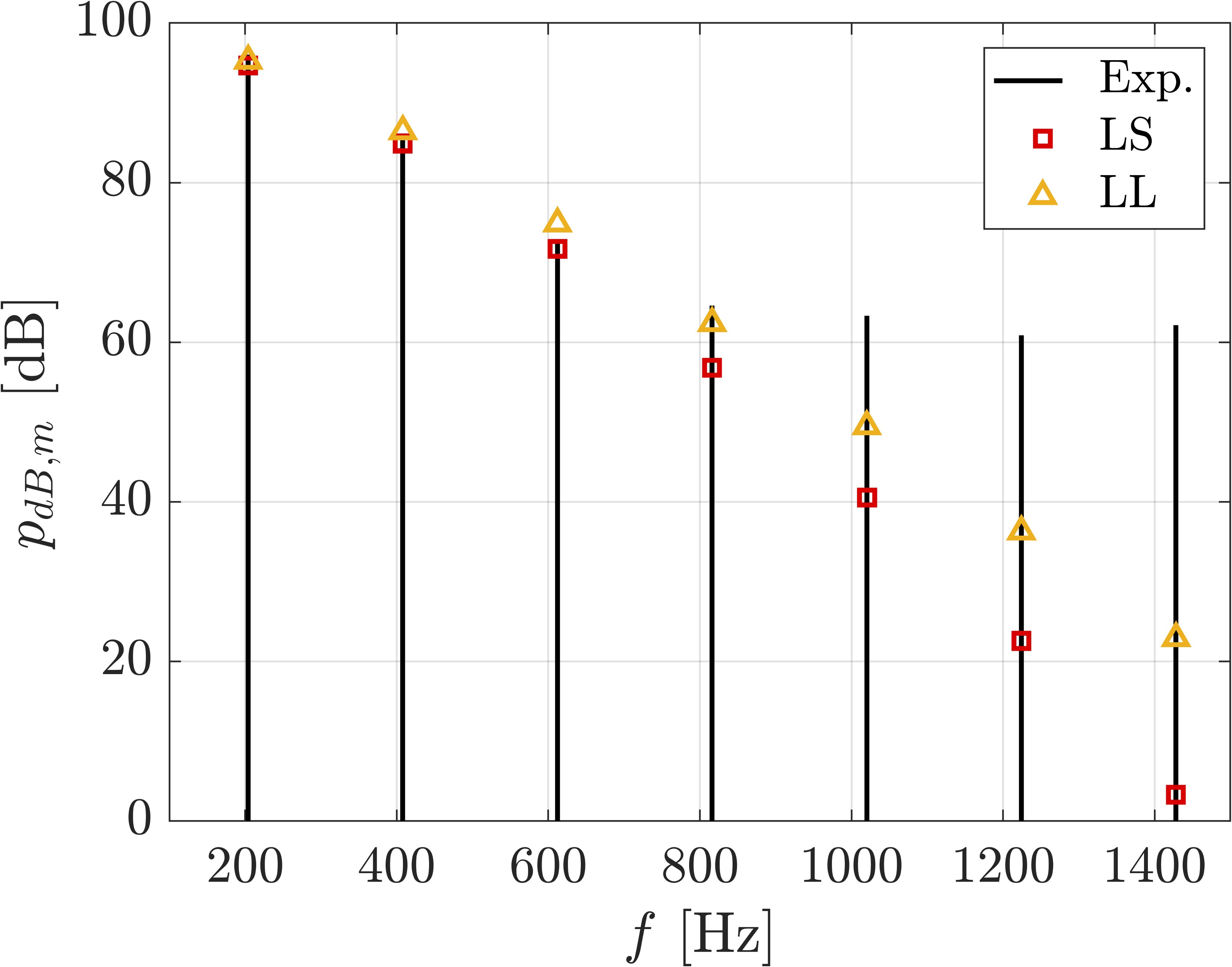}\label{fig:TUCTUC_spectrum100}}\hfill
\caption{Experimental vs \deleted[id=FF]{Lifting Surface}\added{numerical} prediction for TUC-TUC propeller. \inlinecomment{This figure has been updated in the revised manuscript.}}
\label{fig:TUCTUC_validation}
\end{figure}

\added{Figure~\ref{fig:TUCTUC_validation} compares the measured and predicted acoustic field for the TUC-TUC propeller, retaining only the first two non-zero multipole contributions. The first-BPF directivity is captured well by both formulations. The maximum radiation occurs close to the rotor plane, around $\theta_0\approx 90^\circ$, and the level decreases progressively as the observer moves away from this direction. This indicates that the symmetric multipole associated with thickness and drag is dominant, as expected. The signed errors reported in Table~\ref{tab:tuctuc_validation_spl_bpf12} show that, at the first BPF, the LS prediction remains within about 1.9~dB of the measurements over the microphone arc, representing a notably close agreement in view of the $\pm1$ dB accuracy of the microphones. The LL prediction gives a comparable level of agreement, with errors ranging from a slight overprediction at $\theta_0=70^\circ$ to an underprediction of about 1.8~dB at the largest angles. At the second BPF, the agreement remains reasonable, but the errors are larger, consistently with the known sensitivity of higher BPF harmonics to unsteady loading and facility-induced flow recirculation effects \citep{Stephenson2019,Nardari2019,Casalino2023}. The spectra and time histories in figure \ref{fig:TUCTUC_validation} provide a complementary view of the same comparison. The LS and LL reconstructions reproduce the coherent periodic component of the measured pressure signal and place the main tonal peaks at the correct BPF harmonics. The experimental spectra also contain a broadband floor and additional high-frequency content, which are not expected to be captured by the present tonal formulation. These components are visible in the time domain as cycle-to-cycle fluctuations and sharper waveform features, which may be associated with broadband contributions and small unsteady motions \citep{Zhong2020,Yunus2025,Pereira2024}. The comparison therefore indicates that both LS and LL recover the dominant deterministic tonal field.}

\begin{table}
\centering
\begin{tabular}{cccc|ccc}
$\theta_0$ [$^\circ$] & \multicolumn{3}{c|}{$1^{\mathrm{st}}$ BPF} & \multicolumn{3}{c}{$2^{\mathrm{nd}}$ BPF} \\ 
  & Exp. [dB] & Exp.-LS [dB] & Exp.-LL [dB] & Exp. [dB] & Exp.-LS [dB] & Exp.-LL [dB] \\ \hline
70 & 95.03 & +0.67 & -0.25 & 83.04 & -0.27 & -2.29 \\ 
80 & 96.18 & +1.18 & +0.36 & 85.32 & +0.40 & -1.47 \\ 
90 & 96.49 & +1.38 & +0.67 & 86.82 & +1.38 & -0.32 \\ 
100 & 96.05 & +1.34 & +0.77 & 86.12 & +1.21 & -0.31 \\ 
110 & 95.01 & +1.23 & +0.80 & 83.11 & -0.19 & -1.53 \\ 
115 & 94.24 & +1.14 & +0.78 & 81.05 & -0.99 & -2.25 \\ 
120 & 93.52 & +1.24 & +0.95 & 79.26 & -1.21 & -2.38 \\ 
125 & 92.72 & +1.43 & +1.21 & 77.68 & -0.87 & -1.96 \\ 
130 & 91.74 & +1.65 & +1.49 & 77.14 & +0.91 & -0.11 \\ 
135 & 90.60 & +1.91 & +1.82 & 77.08 & +3.62 & +2.68 \\
\end{tabular}
\caption{Experimental SPLs and signed numerical errors at the first and second BPF for the TUC-TUC propeller. \inlinecomment{This table has been added in the revised manuscript.}}
\label{tab:tuctuc_validation_spl_bpf12}
\end{table}

\deleted[id=FF]{The results show that the model captures the dominant tonal behaviour of the propeller with very good accuracy, despite the fact that the aerodynamic inputs are obtained from a low-fidelity blade-element description. The first BPF directivity exhibits the correct overall shape, with a maximum located around $\theta_0\approx 90^\circ$ and a progressive decay as the observer moves away from the rotor plane. This indicates that the symmetric multipole associated with thickness and drag is dominant, as expected. Quantitatively, the discrepancy with the measurements remains small across the microphone array, with a mean error of approximately $-0.97$~dB and a maximum deviation of $-1.53$~dB. The single-microphone spectrum further confirms that the model places the tonal peaks at the correct blade-passing harmonics and reproduces not only the first but also the second and third BPF components with comparable fidelity. The experimental spectrum displays a broadband floor and additional high-frequency content that are not expected to be captured by the present tonal formulation. This behaviour is mirrored in the time history: the predicted waveform matches the coherent periodic component, whereas the experimental trace contains cycle-to-cycle fluctuations and sharper features attributable to broadband noise and higher-order unsteady contributions. Overall, the comparison indicates that the lifting-surface implementation predicts the primary tonal field accurately in terms of directivity and harmonic placement up to at least the third BPF, with residual mismatches that are consistent with the low-order aerodynamic input and with the absence of broadband mechanisms in the current modelling assumptions. Moreover, it was verified that retaining only the first non-zero multipoles in the series (i.e. $\ell_{max}-|m|=1$) yields results that are essentially indistinguishable from the fully converged predictions, even though the present configuration is not strictly in the acoustic far field and does not quite satisfy the low-frequency condition.}


\deleted[id=FF]{This section compares the lifting-line model with the experimental measurements for the APC propeller.} \added{The second low-tip-Mach-number case is the APC propeller.}  Compared with the previous case, this propeller is characterised by a smaller chord length, with a mean spanwise value of $c = 0.024$ m, and by a typical linear twist distribution that decreases towards the tip. \added{Although the experimental study by \citet{Grande2022b} focuses on laminar-separation-bubble noise, the present validation considers the hovering condition, for which this contribution has little or no measurable influence. At $5000$ rpm, the first two BPF tones occur at approximately $167$ and $333$ Hz, respectively, well below the kilohertz frequency range associated with laminar-separation-bubble shedding.} \deleted[id=FF]{The LL multipole coefficients are computed from compact sectional aerodynamics, obtained using the same blade-element code. Figure \ref{fig:exp-LL-directivity} reports measured and predicted first BPF levels. Figures \ref{fig:exp-LL-time} and \ref{fig:exp-LL-spectrum} present the spectrum and time signal at $\theta_0=100^\circ$.}

\added{Tables~\ref{tab:apc_validation_alm_ls} and~\ref{tab:apc_validation_alm_ll} report the leading LS and LL multipoles for the APC case. As for the TUC-TUC propeller, the largest coefficients are associated with $\ell=|m|$, and the modal amplitudes decrease rapidly with increasing $\ell-|m|$. The LS and LL tables are very close in structure, indicating that, for this low-tip-Mach-number operating condition, both approximations represent the same dominant compact loading and thickness-noise content.}

\begin{table}
\centering
\def~{\hphantom{0}}
\begin{tabular}{ccccc}
$\ell-|m|$ & $m=2$ & $m=4$ & $m=6$ & $m=8$ \\ \hline
0 & $3.811\times 10^{-1}$ & $5.020\times 10^{-2}$ & $5.429\times 10^{-3}$ & $5.665\times 10^{-4}$ \\ 
1 & $9.364\times 10^{-2}$ & $1.116\times 10^{-2}$ & $1.080\times 10^{-3}$ & $9.985\times 10^{-5}$ \\ 
2 & $5.695\times 10^{-4}$ & $1.494\times 10^{-4}$ & $2.227\times 10^{-5}$ & $2.802\times 10^{-6}$ \\ 
3 & $2.078\times 10^{-4}$ & $5.162\times 10^{-5}$ & $6.992\times 10^{-6}$ & $7.870\times 10^{-7}$ \\ 
\end{tabular}
\caption{Leading multipoles $|A_{\ell m}^{\mathrm{LS}}| \added{[\mathrm{Pa}]}$ for the APC \deleted[id=FF]{lifting-line} validation case. \inlinecomment{This table has been updated in the revised manuscript.}}
\label{tab:apc_validation_alm_ls}
\end{table}

\begin{table}
\centering
\def~{\hphantom{0}}
\begin{tabular}{ccccc}
$\ell-|m|$ & $m=2$ & $m=4$ & $m=6$ & $m=8$ \\ \hline
0 & $3.042\times 10^{-1}$ & $4.131\times 10^{-2}$ & $4.654\times 10^{-3}$ & $5.056\times 10^{-4}$ \\ 
1 & $9.667\times 10^{-2}$ & $1.158\times 10^{-2}$ & $1.128\times 10^{-3}$ & $1.050\times 10^{-4}$ \\ 
2 & $4.525\times 10^{-4}$ & $1.231\times 10^{-4}$ & $1.913\times 10^{-5}$ & $2.505\times 10^{-6}$ \\ 
3 & $2.131\times 10^{-4}$ & $5.316\times 10^{-5}$ & $7.247\times 10^{-6}$ & $8.217\times 10^{-7}$ \\ 
\end{tabular}
\caption{Leading multipoles $|A_{\ell m}^{\mathrm{LL}}| \added{[\mathrm{Pa}]}$ for the APC \deleted[id=FF]{lifting-line} validation case. \inlinecomment{This table has been updated in the revised manuscript.}}
\label{tab:apc_validation_alm_ll}
\end{table}

\begin{figure}
\centering
\subfloat[First \added{and second} blade-passing-frequency sound pressure level in the meridional plane ($\phi_0=0$).]
{\includegraphics[width=0.6\textwidth]{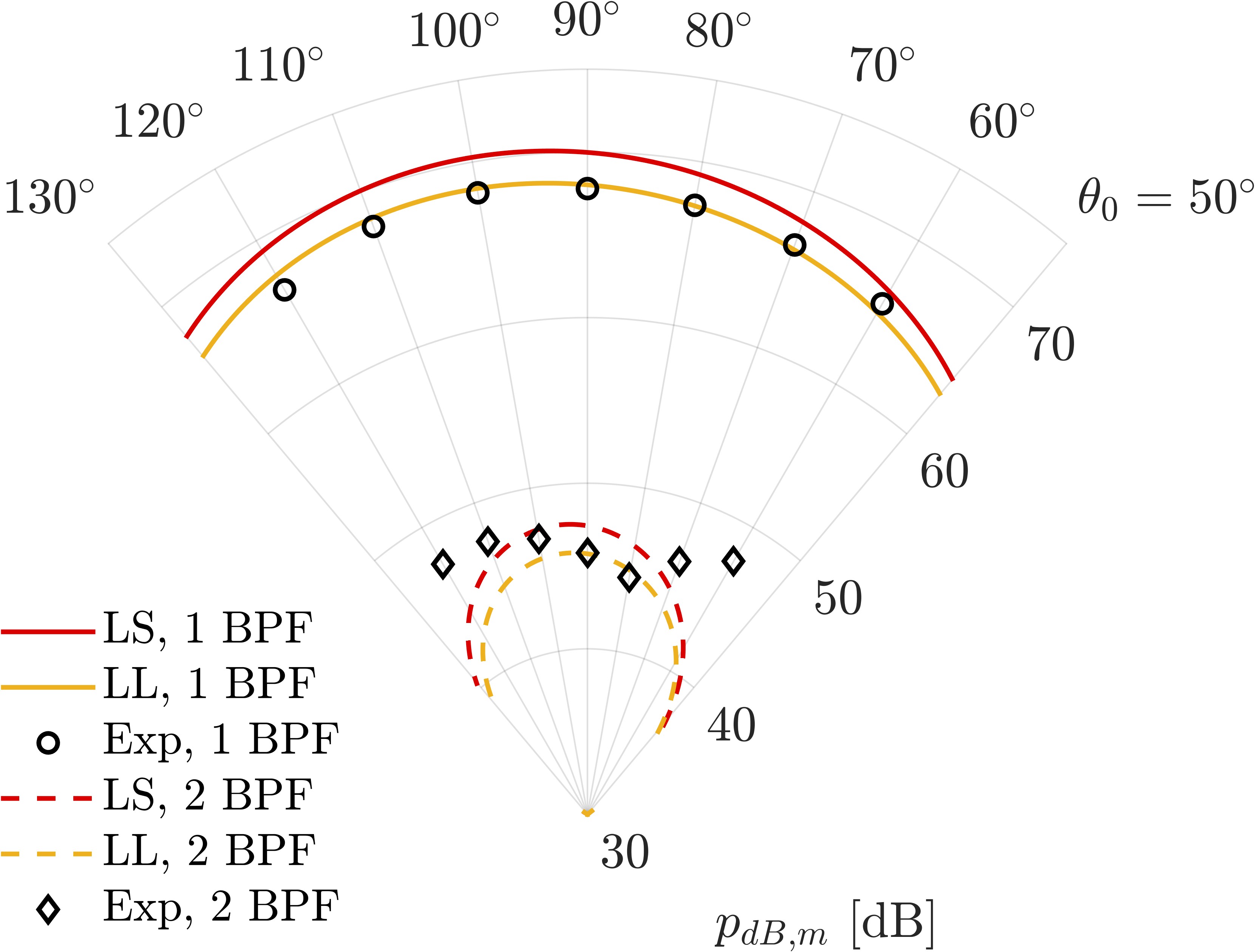}\label{fig:APC_directivity}}\\
\centering
\subfloat[Time-domain pressure at the observer location $\theta_0=90^\circ$.]
{\includegraphics[width=0.45\textwidth]{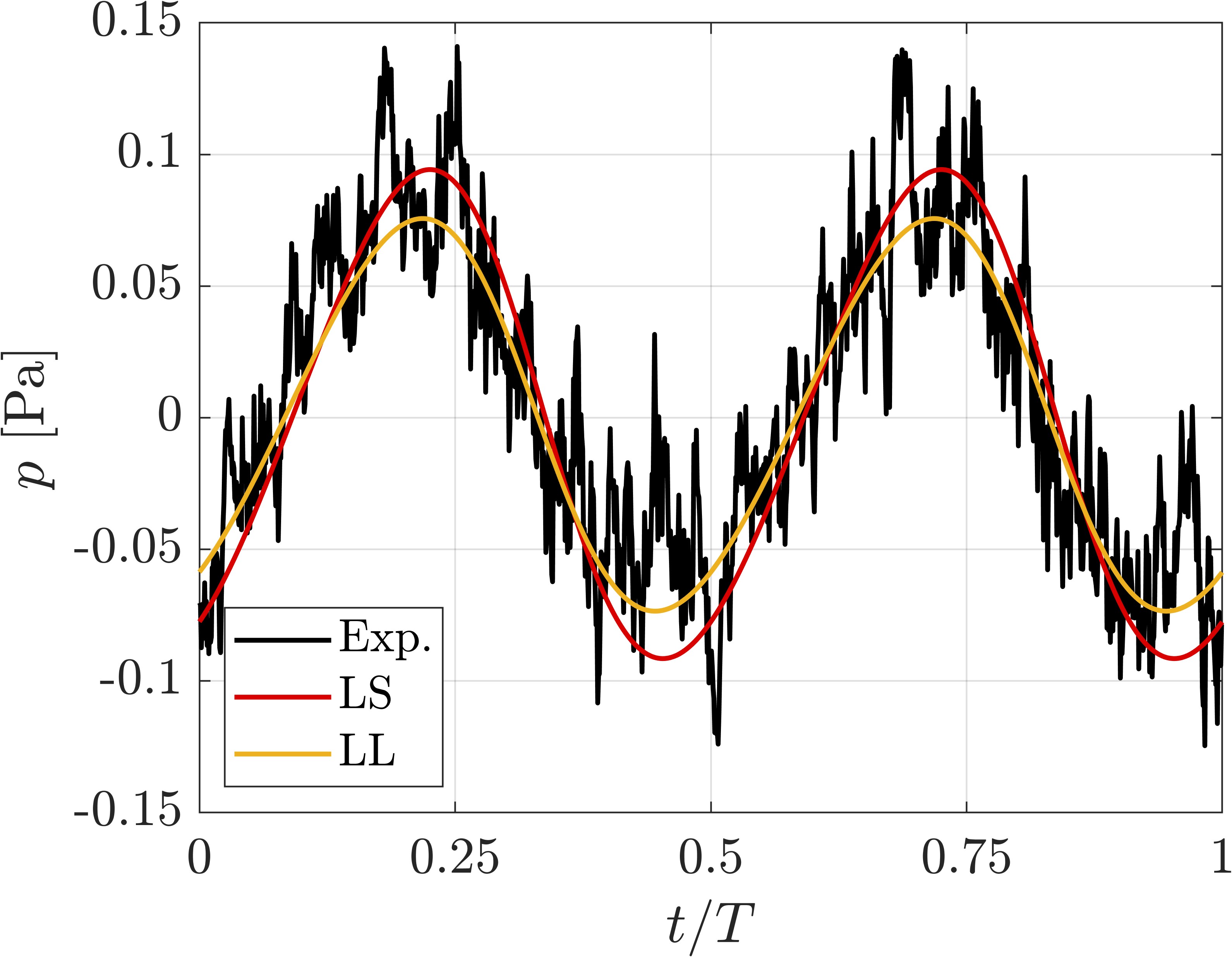} \label{fig:APC_time90}}\hfill
\subfloat[Narrowband $p_{dB,m}$ spectrum at the observer location $\theta_0=90^\circ$.]
{\includegraphics[width=0.45\textwidth]{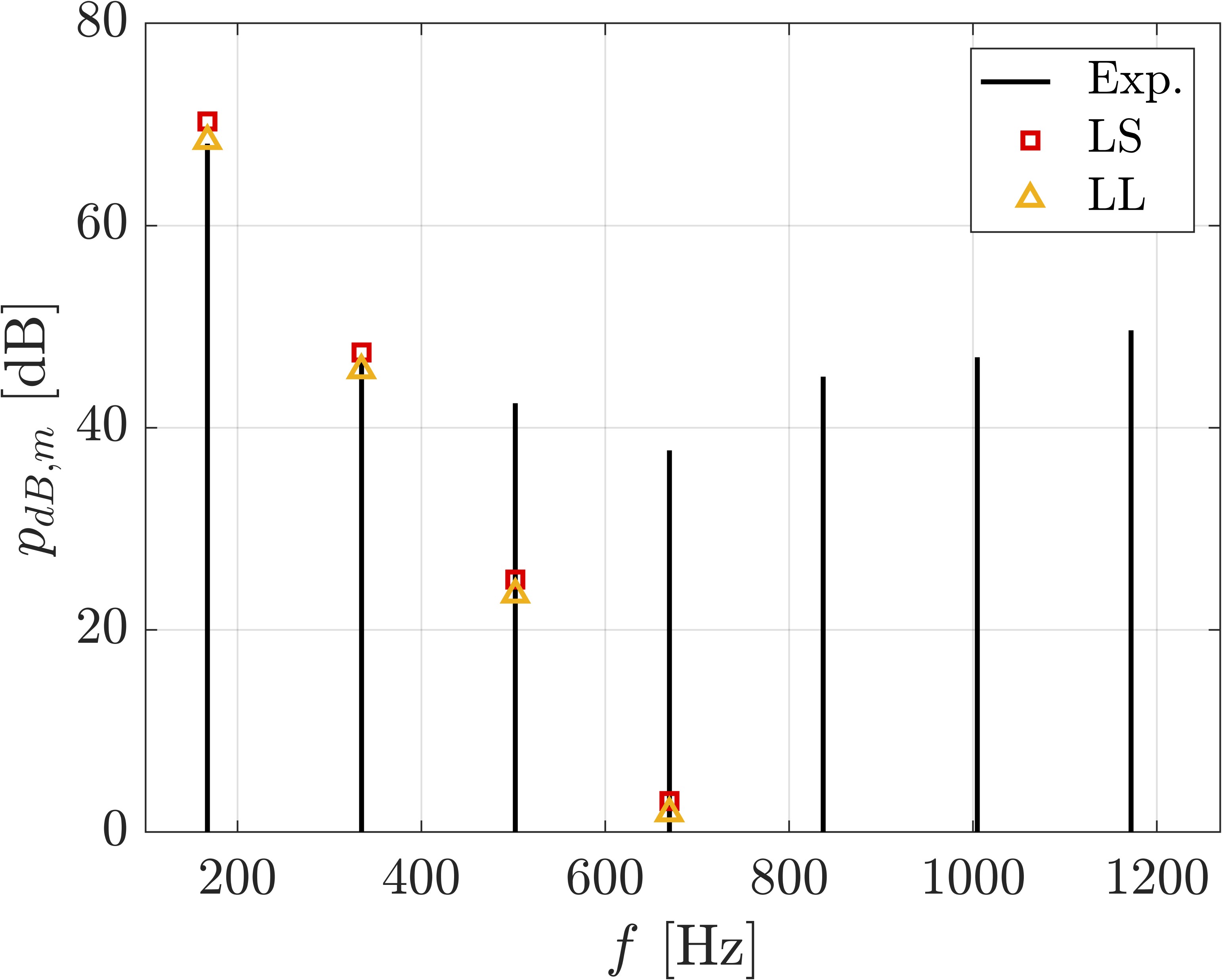}\label{fig:APC_spectrum90}}\\
\centering
\subfloat[Time-domain pressure at the observer location $\theta_0=100^\circ$.]
{\includegraphics[width=0.45\textwidth]{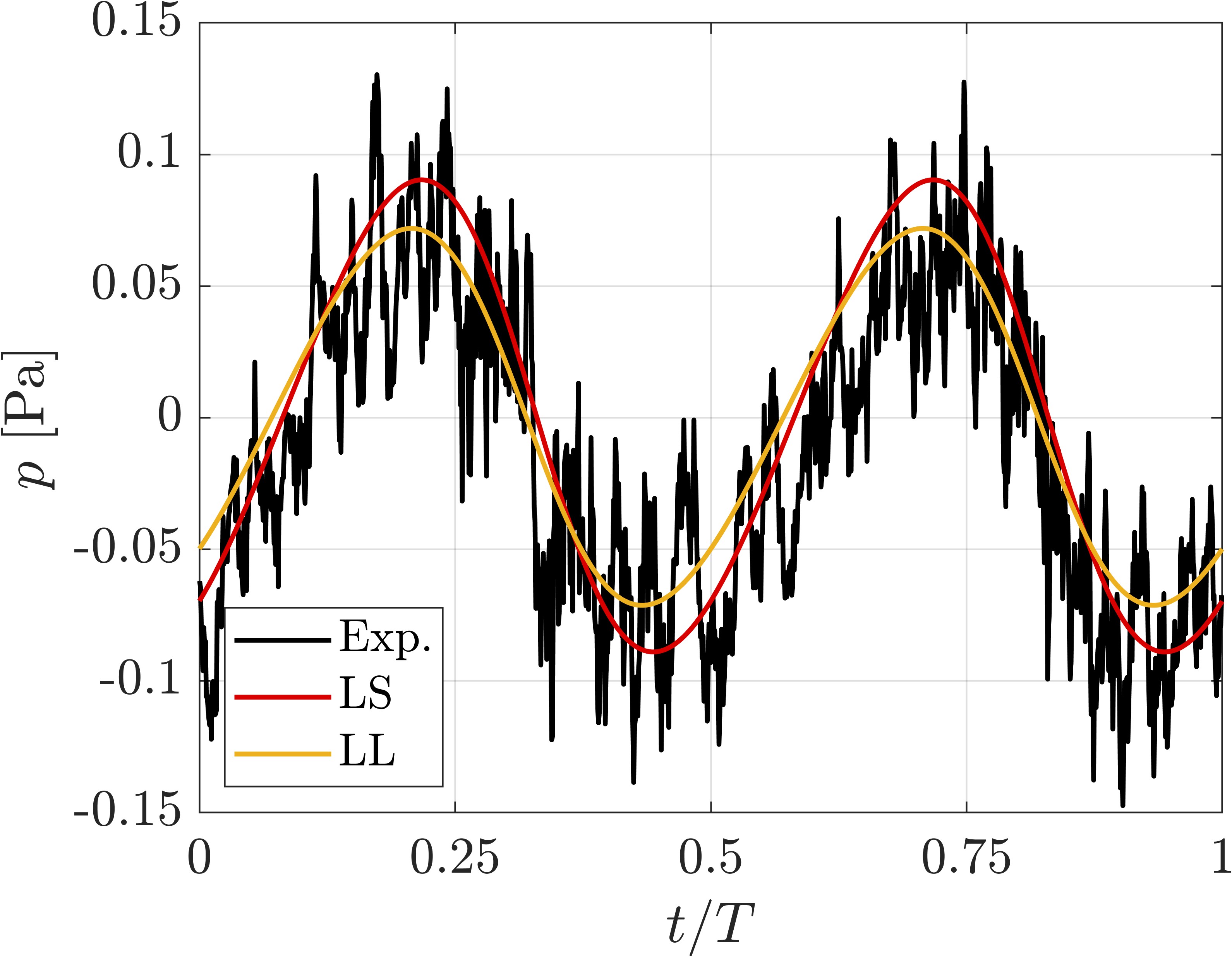} \label{fig:APC_time100}}\hfill
\subfloat[Narrowband $p_{dB,m}$ spectrum at the observer location $\theta_0=100^\circ$.]
{\includegraphics[width=0.45\textwidth]{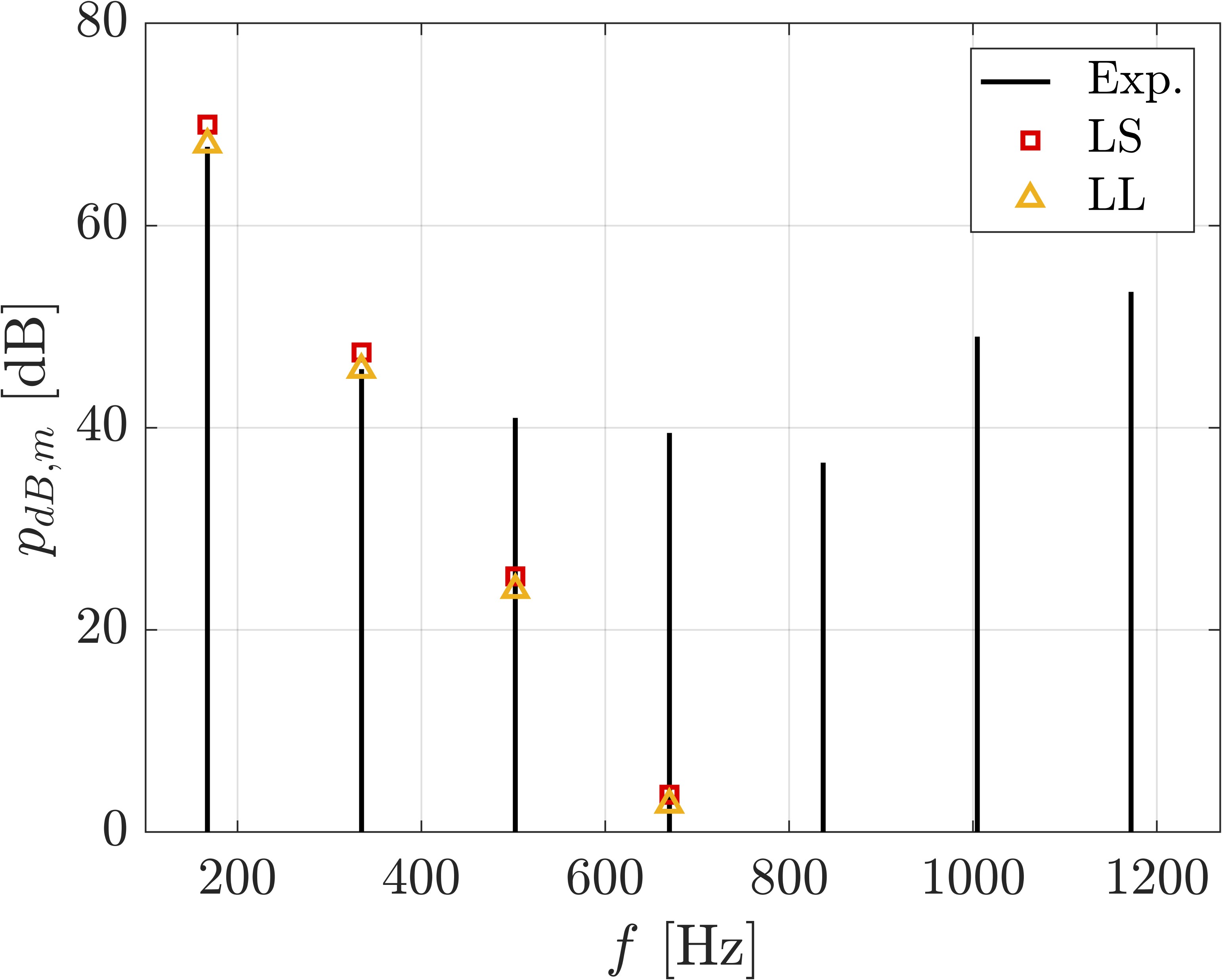}\label{fig:APC_spectrum100}}\hfill
\caption{Experimental vs \deleted[id=FF]{Lifting Line}\added{numerical} prediction for APC propeller. \inlinecomment{This figure has been updated in the revised manuscript.}}
\label{fig:APC_validation}
\end{figure}

\added{Figure~\ref{fig:APC_validation} shows the corresponding acoustic comparison, retaining only the first two non-zero multipole contributions in the numerical predictions. The first-BPF directivity is again well reproduced. The measured maximum is located around $\theta_0\approx 100^\circ$, and both numerical predictions recover the broad angular trend. The SPL errors in Table~\ref{tab:apc_validation_spl_bpf12} show that the LL approximation gives the closest first-BPF agreement for this case, with errors within about 1~dB over the microphone arc. The LS approximation slightly overpredicts the first-BPF level, with signed errors between about $-0.9$ and $-2.7$~dB. At the second BPF, both models show larger discrepancies, particularly at the lower and upper ends of the microphone arc. This confirms that the higher harmonic content is more sensitive to details of the aerodynamic loading and to experimental effects.}

\begin{table}
\centering
\begin{tabular}{cccc|ccc}
$\theta_0$ [$^\circ$] & \multicolumn{3}{c|}{$1^{\mathrm{st}}$ BPF} & \multicolumn{3}{c}{$2^{\mathrm{nd}}$ BPF} \\ 
  & Exp. [dB] & Exp.-LS [dB] & Exp.-LL [dB] & Exp. [dB] & Exp.-LS [dB] & Exp.-LL [dB] \\ \hline
60 & 65.60 & -0.85 & +0.62 & 47.65 & +6.08 & +6.94 \\ 
70 & 66.59 & -1.46 & +0.27 & 46.26 & +1.73 & +2.94 \\ 
80 & 67.35 & -1.87 & +0.03 & 44.54 & -1.91 & -0.40 \\ 
90 & 67.79 & -2.18 & -0.23 & 45.79 & -1.64 & +0.05 \\ 
100 & 68.11 & -2.17 & -0.26 & 46.89 & -0.56 & +1.20 \\ 
110 & 67.78 & -2.30 & -0.50 & 47.54 & +1.12 & +2.86 \\ 
120 & 66.57 & -2.67 & -1.00 & 47.46 & +3.28 & +4.94 \\ 
\end{tabular}
\caption{Experimental SPLs and signed numerical errors at the first and second BPF for the APC propeller. \inlinecomment{This table has been added in the revised manuscript.}}
\label{tab:apc_validation_spl_bpf12}
\end{table}

\deleted[id=FF]{The results confirm that the proposed formulation reproduces the dominant tonal field with high accuracy. In particular, the first-BPF level is captured remarkably well: the predicted angular distribution closely follows the measurements across the microphone arc with a maximum located around $\theta_0\approx 100^\circ$, and with only minor residual discrepancies. The maximum error is approximately $-0.91$~dB, while the mean error is as low as $-0.14$~dB. The single-microphone spectrum further shows that the model matches the lowest two tonal peaks well, while the discrepancy increases progressively at higher frequency. This behaviour is reflected in the time domain result. Overall, the comparison indicates that the lifting-line implementation provides a very accurate prediction of the first blade-passing tone and its angular behaviour, and that the remaining mismatches are consistent with the expected loss of fidelity at higher harmonics, where distributed loading, broadband mechanisms, and measurement noise increasingly contribute to the acoustic field. An analogous observation about convergence holds for these results, for which truncation to the first non-zero multipole order ($\ell_{max}-|m|=1$) likewise produces curves that are practically indistinguishable from the fully converged solution.}

\added{Overall, the two low-tip-Mach-number comparisons show that the proposed multipole formulation provides a consistent reconstruction of the dominant tonal field for both propellers. The first BPF is predicted with good accuracy in directivity and absolute level, and the main features of the coherent time-domain waveform are recovered. The second BPF remains reasonably represented, although with larger errors. For both propellers, retaining only the first few non-zero multipole orders is sufficient to reproduce the acoustic field at the microphone locations. This observation is consistent with the theoretical results discussed above and confirms that the acoustic field is governed by a small number of leading spherical multipoles.}

\subsection{\added{High tip-Mach number}}

\added{The previous comparisons concern propellers operating at relatively low tip Mach numbers. To assess the behaviour of the proposed multipole expansion under the more demanding conditions considered in the theoretical analysis, an additional benchmark is presented for a hovering rotor operating at \(M_t=0.8\). The comparison is based on measurements and reference calculations reported in the literature \citep{Boxwell1978,farassat1979,schmitz1980}. The experimental configuration was introduced by \citet{Boxwell1978} and consists of a two-bladed, \(1/7\)-scale model of the UH-1H helicopter main rotor, tested in an anechoic hover facility designed to limit aerodynamic recirculation. As summarised in table \ref{tab:boxwell-dataset}, the rotor has a diameter of \(D=2.090\,\mathrm{m}\), a constant-chord NACA 0012 section with \(c=0.076\,\mathrm{m}\), and the full-scale root-to-tip twist of \(10.9^\circ\). A root cut-out of $R_{\mathrm{hub}}=0.1R$ is assumed. The acoustic pressure was measured in the rotor plane at a distance \(r/D=1.5\) from the hub, where the high-speed impulsive-noise signature is particularly pronounced. \citet{Boxwell1978} investigated hover tip Mach numbers ranging from \(M_t=0.8\) to approximately unity and compared the measured in-plane waveforms with linear thickness-noise predictions. At \(M_t=0.8\), the linear theory reproduces the overall waveform reasonably well but underestimates the magnitude of the negative pressure peak by approximately a factor of two. The role of the neglected linear source terms was examined by \citet{farassat1979}, which included detailed blade-surface pressure and skin-friction distributions obtained from aerodynamic models. Their results showed that loading and skin-friction noise remain small compared with thickness noise over the range of conditions considered. This finding supports the interpretation that the remaining discrepancy is primarily associated with nonlinear flow effects, whose contribution was subsequently investigated by \citet{schmitz1980}. At \(M_t=0.8\), the quadrupole contribution was found to have a waveform similar in shape to that of the thickness contribution and to improve agreement with the measured pressure amplitude.}

\begin{table}
\centering
\small
\setlength{\tabcolsep}{7pt}
\begin{tabular}{lc}
\toprule
& \textbf{UH-1H Rotor} \\
\midrule
Number of blades $N$ & 2 \\
Tip radius $R$ [m] & 1.045 \\
Hub radius $R_{\mathrm{hub}}$ [m] & 0.105 \\
Mean chord length $c$ [m] & 0.076 \\
Twist angle $\beta$ [$^\circ$] & $10.9$ root-to-tip washout \\
Airfoil & NACA 0012 \\
\midrule
Rotational speed [rpm] & 2486 \\
Tip Mach $M_t$ [-] & 0.80 \\
\midrule
Microphone distance $r_0$ [m] & 3.14 \\
Microphone angle $\theta_0$ [$^\circ$] & 90 \\
$k r_0$ [-] & 2.40 \\
\bottomrule
\end{tabular}
\caption{Experimental parameters for the UH-1H model-rotor measurements reported by \citet{Boxwell1978}. \inlinecomment{This table has been added in the revised manuscript.}}
\label{tab:boxwell-dataset}
\end{table}

\added{The objective of the present comparison is therefore deliberately limited. The multipole formulation developed in this work is based on a linear surface-source description and does not include quadrupole volume sources or the nonlinear mechanisms associated with shock-wave generation. Moreover, only the thickness-noise contribution is considered, since it can be evaluated from the blade geometry and kinematics without requiring aerodynamic loading data and is the dominant linear contribution for this benchmark. The purpose is not to claim a complete physical prediction of transonic rotor noise, but to verify that the spherical multipole expansion correctly reproduces the established linear thickness-noise solution at \(M_t=0.8\). In future work, we plan to extend the spherical expansion to include the neglected quadrupole contribution.}

\begin{table}
\centering
\def~{\hphantom{0}}
\begin{tabular}{ccccc}
$\ell-|m|$ & $m=2$ & $m=4$ & $m=6$ & $m=8$ \\ \hline
0 & $9.891\times 10^{0}$ & $2.815\times 10^{1}$ & $4.580\times 10^{1}$ & $6.009\times 10^{1}$ \\ 
1 & $-$ & $-$ & $-$ & $-$ \\ 
2 & $2.149\times 10^{-1}$ & $1.245\times 10^{0}$ & $2.792\times 10^{0}$ & $4.429\times 10^{0}$ \\ 
3 & $-$ & $-$ & $-$ & $-$ \\ 
\end{tabular}
\caption{Leading thickness-noise multipoles $|A_{\ell m}^{\mathrm{LS}}| \added{[\mathrm{Pa}]}$ for the UH-1H validation case. \inlinecomment{This table has been added in the revised manuscript.}}
\label{tab:boxwell_validation_alm_ls}
\end{table}

\begin{table}
\centering
\def~{\hphantom{0}}
\begin{tabular}{ccccc}
$\ell-|m|$ & $m=2$ & $m=4$ & $m=6$ & $m=8$ \\ \hline
0 & $9.904\times 10^{0}$ & $2.826\times 10^{1}$ & $4.614\times 10^{1}$ & $6.083\times 10^{1}$ \\ 
1 & $-$ & $-$ & $-$ & $-$ \\ 
2 & $2.151\times 10^{-1}$ & $1.249\times 10^{0}$ & $2.811\times 10^{0}$ & $4.480\times 10^{0}$ \\ 
3 & $-$ & $-$ & $-$ & $-$ \\ 
\end{tabular}
\caption{Leading thickness-noise multipoles $|A_{\ell m}^{\mathrm{LL}}| \added{[\mathrm{Pa}]}$ for the UH-1H validation case. \inlinecomment{This table has been added in the revised manuscript.}}
\label{tab:boxwell_validation_alm_ll}
\end{table}

\begin{figure}
    \centering
    \subfloat[Time-domain pressure at the observer location $\theta_0=90^\circ$.]
    {\includegraphics[width=0.45\textwidth]{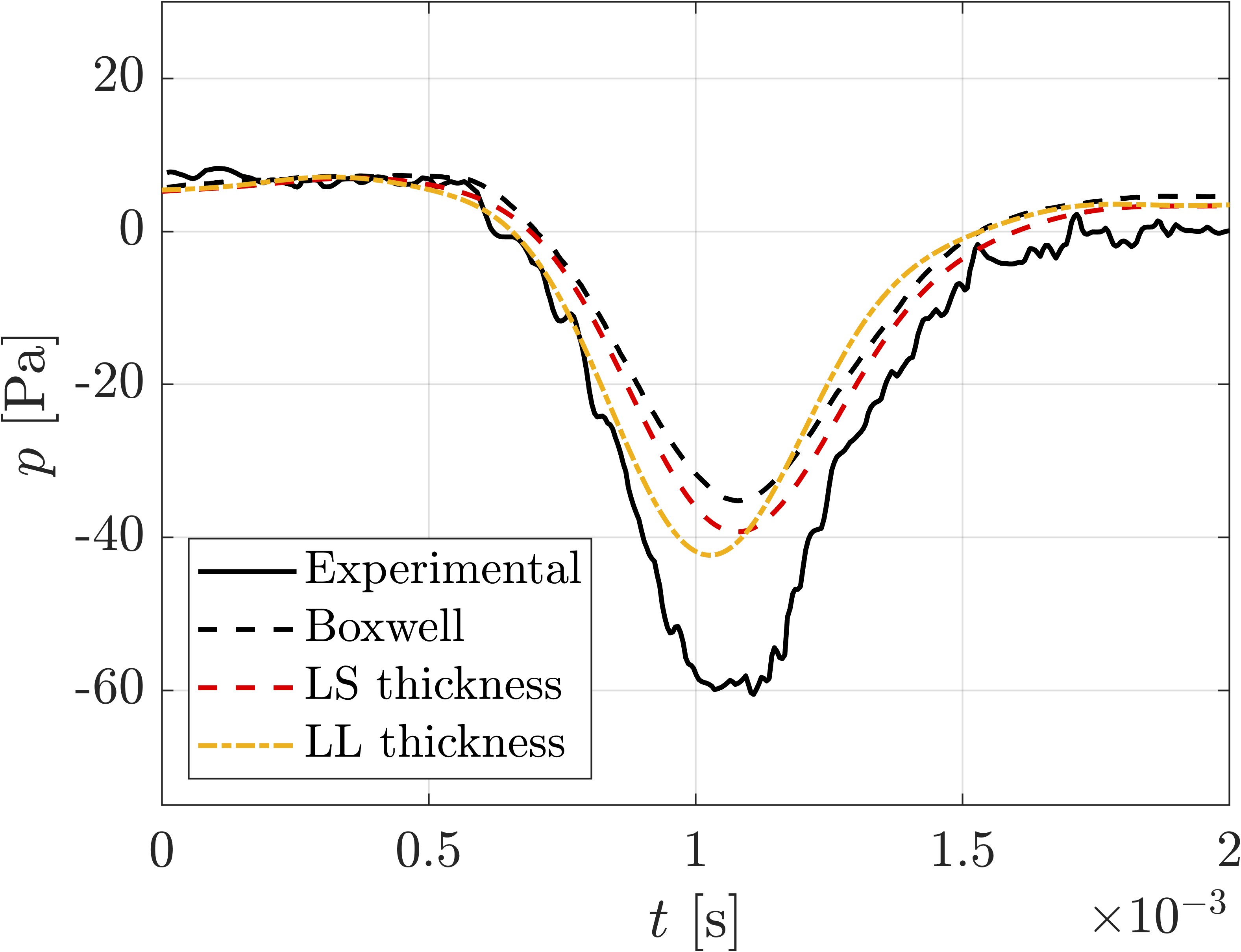} \label{fig:Boxwell_comparison}}\hfill
    \subfloat[Narrowband $p_{dB,m}$ spectrum at the observer location $\theta_0=90^\circ$.]
    {\includegraphics[width=0.45\textwidth]{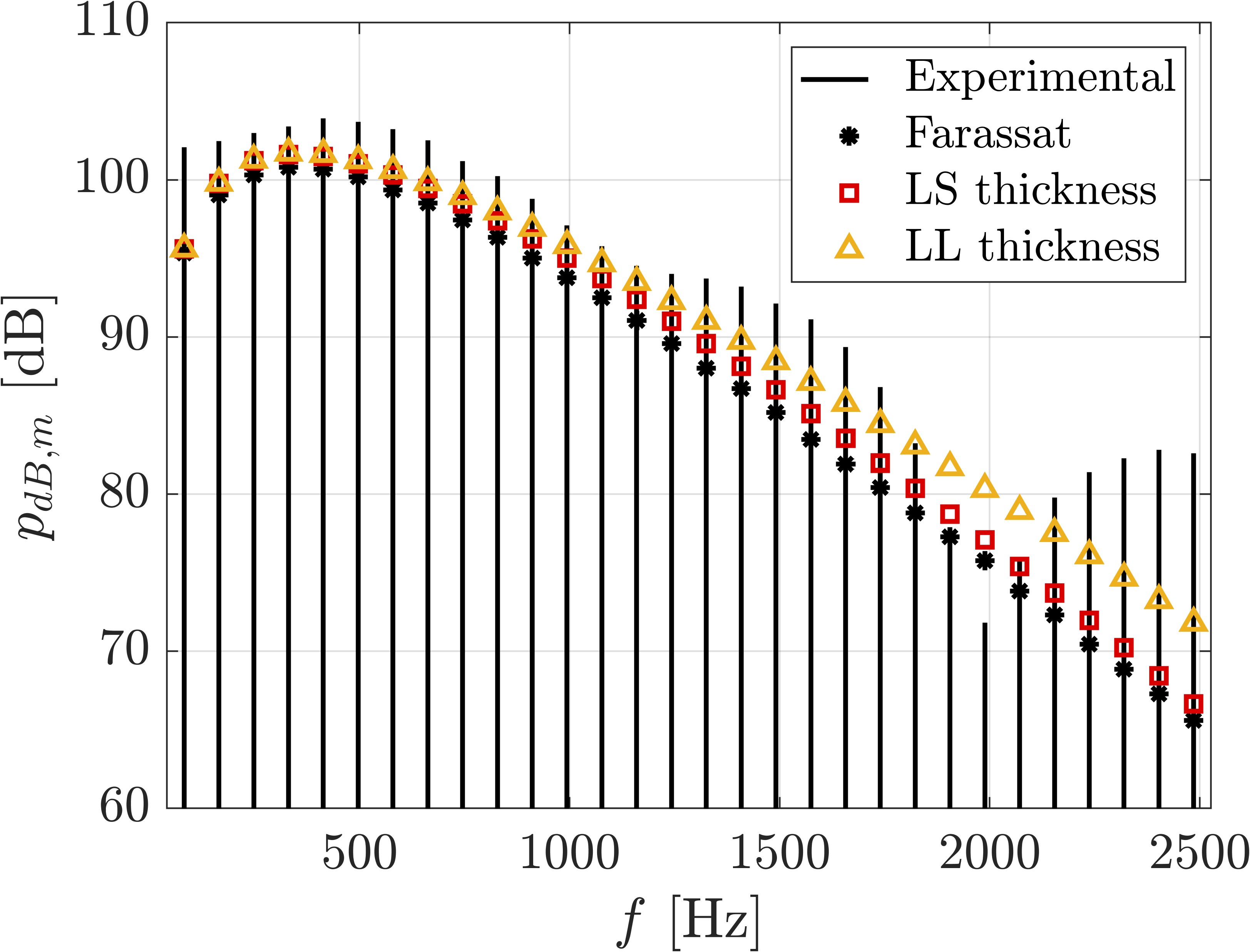}  \label{fig:Farassat_comparison}}
    \caption{Experimental vs numerical prediction for UH-1H rotor. Panel (a) compares the in-plane acoustic-pressure waveform with the measurements and the reference linear solution reported by \citet{Boxwell1978}. Panel (b) compares the corresponding in-plane spectrum with the measurements and the reference linear results reported by \citet{farassat1979}. \inlinecomment{This figure has been added in the revised manuscript.}}
    \label{fig:high_Mach_benchmark}
\end{figure}

\added{Figures \ref{fig:Boxwell_comparison} and \ref{fig:Farassat_comparison} compare the present predictions, retaining in this case only the leading thickness multipole $A_{mm}$ for each harmonic, with the experimental measurements and reference numerical results reported in the literature. The pressure waveform in figure \ref{fig:Boxwell_comparison} shows that the lifting-surface thickness-noise prediction closely follows the reference linear solution of \citet{Boxwell1978}. In particular, the overall temporal evolution of the pulse and the position of its minimum are reproduced well. The lifting-line approximation also captures the main waveform features, although small differences in the pulse depth and temporal evolution are visible. These discrepancies are expected because the lifting-line model collapses the chordwise source distribution, whereas the lifting-surface formulation retains the finite-chord phase effects that become increasingly important at high tip Mach numbers. In both cases, it was verified that increasing the number of retained multipoles progressively improves the agreement with the reference numerical solution.}

\added{The spectral comparison in figure \ref{fig:Farassat_comparison} leads to the same conclusion. The lifting-surface prediction remains close to the numerical thickness-noise results of \citet{farassat1979} over the range of blade-passing-frequency harmonics considered. The lifting-line results are also consistent at the lowest harmonics, but progressively depart from the lifting-surface and reference numerical solutions as the harmonic order increases. In particular, the lifting-line approximation increasingly overestimates the high-frequency levels. This behaviour is consistent with the results presented in Section \ref{sec:Comparison}.}

\begin{table}
\centering
\small
\begin{tabular}{cccccc}
BPF & $f$ [Hz] & Exp. [dB] & Exp.-Farassat [dB] & Exp.-LS [dB] & Exp.-LL [dB] \\ \hline
1 & 82.9 & 102.07 & +6.74 & +6.47 & +6.46 \\ 
2 & 165.7 & 102.47 & +3.43 & +2.70 & +2.67 \\ 
3 & 248.6 & 102.99 & +2.69 & +1.78 & +1.72 \\ 
4 & 331.4 & 103.39 & +2.59 & +1.78 & +1.67 \\ 
5 & 414.3 & 103.90 & +3.24 & +2.43 & +2.27 \\ 
6 & 497.1 & 103.69 & +3.51 & +2.68 & +2.46 \\ 
7 & 580.0 & 103.23 & +3.86 & +2.91 & +2.61 \\ 
8 & 662.8 & 102.52 & +4.01 & +3.06 & +2.68 \\ 
9 & 745.7 & 101.20 & +3.75 & +2.72 & +2.24 \\ 
10 & 828.5 & 100.24 & +3.91 & +2.84 & +2.27 \\ 
\end{tabular}
\caption{Experimental SPLs and signed numerical errors at the first ten BPFs for the UH-1H rotor. \inlinecomment{This table has been added in the revised manuscript.}}
\label{tab:boxwell_validation_harmonic_errors}
\end{table}

\added{Both comparisons also illustrate the limitations of the underlying linear source model. Although the predicted waveform shape is reasonably well captured at \(M_t=0.8\), linear thickness-noise solutions underestimate the magnitude of the measured negative-pressure peak by approximately a factor of two, consistently with the observations of \citet{Boxwell1978} and \citet{farassat1979}. The discrepancy between the linear predictions and the experimental data reflects physical contributions that are deliberately excluded from the present surface-source model. As discussed by \citet{schmitz1980}, nonlinear effects and quadrupole volume sources become increasingly important as the rotor approaches the transonic regime. The present benchmark consequently verifies that the spherical multipole expansion accurately reproduces the established linear thickness-noise solution at \(M_t=0.8\), while also making explicit the range of validity of the adopted acoustic model.}

\section{Conclusions}
\label{sec:Conclusions}

A spherical-multipole expansion has been developed for \added{the retained surface terms of an} \deleted[id=FF]{Goldstein’s} acoustic analogy in the particular case of rotating propellers. \added{ Mean-flow effects and volume quadrupole sources are excluded from the present study.} The main advantage of the formulation lies in the explicit separation between source and observer dependence: all source information is condensed into harmonic multipole coefficients, while the observer dependence is carried entirely by spherical harmonics and radial propagation factors. As a result, the source integrals are evaluated only once for each harmonic and multipole \deleted[id=FF]{order}\added{degree}, after which the acoustic field at arbitrary observer locations is recovered inexpensively, without repeated integration over the blade. This source/observer decoupling yields a clear computational advantage over formulations in which the observer coordinates remain embedded inside the source integrals. A further benefit of the multipole representation is that the radiated power can be obtained directly from the multipole amplitudes, thus providing a compact and physically meaningful measure of source strength. In the hovering subsonic regime considered here, the spherical expansion was found to converge rapidly with respect to the $\ell$-index. In particular, for each harmonic the dominant far-field radiation is generally captured by the first two non-zero multipoles, corresponding to the leading symmetric and antisymmetric contributions. This behaviour was demonstrated through coefficient decay, directivity decompositions, and reconstructed pressure fields\deleted[id=FF]{, and explains why accurate tonal predictions can be obtained with very small truncations of the series}. The resulting rapid convergence translates directly into computational savings with respect to alternative formulations available in the literature\added{, while remaining complementary to existing strategies for simplifying or accelerating the evaluation of acoustic radiation integrals.}

Two reduced descriptions of the surface integral were then introduced to clarify the physical content of the leading multipole coefficients. The first is a lifting-surface formulation, obtained by projecting the blade onto the plane of rotation. The underlying assumption is that the blade is made of thin sections operating at small incidence. The individual multipoles exhibit different symmetry properties with respect to the rotor plane: the antisymmetric family collects the \deleted[id=FF]{lift}\added{thrust}-driven loading contribution, whereas the symmetric family collects the thickness and \deleted[id=FF]{drag}\added{torque}-like contributions. \deleted[id=FF]{In the range of validity of this model, the dominant acoustic contributions are generally associated with lift and thickness, while drag usually remains secondary.} The formulation retains the chordwise phase effects, which become increasingly important as the tip Mach number increases, the harmonic spectrum broadens, and the time-domain waveform departs from a sinusoidal shape. The second reduced model is a lifting-line formulation, obtained through a high-aspect-ratio expansion in which the source information is retained through compact sectional moments. In this case, no small-incidence assumption is required. The leading contribution is a force term governed by sectional \deleted[id=FF]{lift}\added{thrust} and \deleted[id=FF]{drag}\added{torque}, whereas thickness enters only at a higher order. This approach loses the detailed chordwise information, but in return it is suited to more strongly twisted blades and entails a lower computational cost. The theoretical developments have been complemented by comparisons with \deleted[id=FF]{exact surface} integration \added{over the true blade geometry, and both}\deleted[id=FF]{, convergence analyses, and validation against two experimental datasets taken from the literature. Both} reduced formulations result\added{ed} less expensive than exact surface integration while retaining the relevant acoustic content. \deleted[id=FF]{The experimental comparisons confirm}\added{The experimental assessment considers three different propellers, selected to test the formulations over complementary geometries and operating conditions. Across these three configurations, the results show} that each formulation reproduces the tonal directivity and the blade-passing harmonics with good accuracy in their intended regimes. 

Overall, the spherical-multipole perspective provides both physical insight and computational efficiency for tonal propeller-noise prediction. It offers a unified framework for near- and far-field analysis, while clarifying the respective roles of \deleted[id=FF]{lift}\added{thrust}, \deleted[id=FF]{drag}\added{torque} and thickness in shaping the radiated field. These features make the present approach a promising basis for fast prediction, physical interpretation, and optimization of tonal propeller noise over a broad range of operating conditions.




\appendix

\section{Equatorial factors}\label{appA}
The factors \(B_{L_{\ell m}}=B_{T'_{\ell m}}\) and \(B_{D_{\ell m}}=B_{T_{\ell m}}\), which appear in the approximate expressions for the multipole coefficients, correspond to the values of the spherical harmonics and their \(\theta\)-derivative evaluated on the equatorial plane. Using particular results for the associated Legendre polynomials \citep[p.~334]{abramowitzstegun}, they can be expressed explicitly in terms of the indices \(\ell\) and \(m\) as
\begin{gather}
B_{L_{\ell m}} = B_{T'_{\ell m}} = C_{\ell m}  2^{m+1}  \left(-4\right)^{-\frac{1}{2}\left(\ell+m+1\right)} \frac{\left(\ell+m+1\right)!}{\left(\frac{1}{2}\left(\ell+m+1\right)\right)! \left(\frac{1}{2}\left(\ell-m-1\right)\right)!} \\
B_{D_{\ell m}}=B_{T_{\ell m}} = C_{\ell m}  2^{m} \left(-4\right)^{-\frac{1}{2}\left(\ell+m\right)} \frac{\left(\ell+m\right)!}{\left(\frac{1}{2}\left(\ell+m\right)\right)! \left(\frac{1}{2}\left(\ell-m\right)\right)!}
\end{gather}
where $C_{\ell m} = \sqrt{\frac{2\ell+1}{4\pi}\frac{\left(\ell-m\right)!}{\left(\ell+m\right)!}}$. The above expressions are understood to be used only for admissible parity combinations of $(\ell,m)$ (as imposed by the symmetry of the corresponding contribution), so that the half-integer factorial arguments are non-negative integers.

\section{Definition of far-field acoustic levels}\label{appB}

The multipole expansion provides the coefficients $A_{\ell m}$. For a given azimuthal harmonic index $m$, the acoustic pressure amplitude $\hat{p}_m$ and the radiated acoustic power $\Pi_m$ are obtained from Eqs. \eqref{eq:pressure_formula} and \eqref{eq:power formula} as series over the degree $\ell$ (with $\ell\ge |m|$). The complete field involves both positive and negative values of $m$; however, the contributions at $\pm m$ are complex conjugates and it is sufficient to evaluate $m\ge 0$ and account for symmetry when forming total levels.

The sound pressure level (SPL) associated with the $m$-th harmonic is defined as
\begin{equation}
    p_{dB,m}=20\log_{10}\!\left(\frac{\sqrt{2} \left|\hat{p}_m\right|}{p_{\mathrm{ref}}}\right),
\qquad
p_{\mathrm{ref}}=2\times 10^{-5}\ \mathrm{Pa}
\end{equation}
The overall SPL is obtained by summing the harmonic energies (Parseval's theorem)
\begin{equation}
p_{dB}=20\log_{10}\!\left(\frac{\sqrt{2\sum_{m=0}^{\infty}\left|\hat{p}_m\right|^2}}{p_{\mathrm{ref}}}\right)
\label{eq:SPL_calculation}
\end{equation}

Analogously, the sound power level associated with the $m$-th harmonic is
\begin{equation}
    \Pi_{dB,m}=10\log_{10}\!\left(\frac{2 \Pi_m}{\Pi_{\mathrm{ref}}}\right),
\qquad 
\Pi_{dB}=10\log_{10}\!\left(\frac{2\sum_{m=0}^{\infty}\Pi_m}{\Pi_{\mathrm{ref}}}\right),
\qquad
\Pi_{\mathrm{ref}}=10^{-12}\ \mathrm{W}
\label{eq:power_calculation}
\end{equation}

A further quantity of interest in the far field is the directivity. The directivity of the $m$-th harmonic is defined as the power radiated per unit solid angle in the direction $(\theta_0,\phi_0)$, normalised by the average power per unit solid angle of that harmonic. It can be written as
\begin{equation}
    D_m(\theta_0,\phi_0)
=
\frac{4\pi
\left|\displaystyle\sum_{\ell=|m|}^{\infty} (-i)^{\ell+1}\,A_{\ell m}\,Y_{\ell m}(\theta_0,\phi_0)\right|^2}
{\displaystyle\sum_{\ell=|m|}^{\infty} \left|A_{\ell m}\right|^2}
\label{m directivity}
\end{equation}
where the denominator is proportional to $\Pi_m$. The corresponding directivity for the total field is obtained by summing over the harmonics as in the power expression in Eq. \eqref{eq:power formula}, i.e.
\begin{equation}
    D(\theta_0,\phi_0)
=
\frac{4\pi\displaystyle\sum_{m=-\infty}^{\infty}\frac{1}{m^2}
\left|\sum_{\ell=|m|}^{\infty} (-i)^{\ell+1}\,A_{\ell m}\,Y_{\ell m}(\theta_0,\phi_0)\right|^2}
{\displaystyle\sum_{m=-\infty}^{\infty}\frac{1}{m^2}\sum_{\ell=|m|}^{\infty}\left|A_{\ell m}\right|^2}
\label{directivity definition}
\end{equation}
Owing to conjugate symmetry, the sums over $m$ may be restricted to $m>0$ when evaluating $D$ numerically. By construction, the directivity is normalised so that its average over the sphere is unity, i.e.
\begin{equation}
    \frac{1}{4\pi} \int^{2\pi}_{0} \int^{\pi}_{0} D(\theta_0,\phi_0)\,\sin{\theta_0} d\theta_0 d\phi_0=1
\end{equation}

\bibliographystyle{jfm}
\bibliography{jfm}

\end{document}